\documentclass[amsmath,showpacs,nofootinbib,12pt]{revtex4-2}
\usepackage{graphicx}
\usepackage{dcolumn}
\usepackage{bm}
\usepackage{subcaption}
\usepackage{color} 
\usepackage{slashed}
\usepackage{float}
\usepackage{amsfonts}
\begin{document}
\newcommand{\hs}{\hspace*{0.5cm}}
\newcommand{\vs}{\vspace*{0.5cm}}
\newcommand{\be}{\begin{equation}}
\newcommand{\ee}{\end{equation}}
\newcommand{\bea}{\begin{eqnarray}}
\newcommand{\eea}{\end{eqnarray}}
\newcommand{\ben}{\begin{enumerate}}
\newcommand{\een}{\end{enumerate}}
\newcommand{\bde}{\begin{widetext}}
\newcommand{\ede}{\end{widetext}}
\newcommand{\nn}{\nonumber}
\newcommand{\crn}{\nonumber \\}
\newcommand{\Tr}{\mathrm{Tr}}
\newcommand{\non}{\nonumber}
\newcommand{\noi}{\noindent}
\newcommand{\al}{\alpha}
\newcommand{\la}{\lambda}
\newcommand{\bet}{\beta}
\newcommand{\ga}{\gamma}
\newcommand{\va}{\varphi}
\newcommand{\om}{\omega}
\newcommand{\pa}{\partial}
\newcommand{\+}{\dagger}
\newcommand{\fr}{\frac}
\newcommand{\bc}{\begin{center}}
\newcommand{\ec}{\end{center}}
\newcommand{\Ga}{\Gamma}
\newcommand{\de}{\delta}
\newcommand{\De}{\Delta}
\newcommand{\ep}{\epsilon}
\newcommand{\varep}{\varepsilon}
\newcommand{\ka}{\kappa}
\newcommand{\La}{\Lambda}
\newcommand{\si}{\sigma}
\newcommand{\Si}{\Sigma}
\newcommand{\ta}{\tau}
\newcommand{\up}{\upsilon}
\newcommand{\Up}{\Upsilon}
\newcommand{\ze}{\zeta}
\newcommand{\ps}{\psi}
\newcommand{\Ps}{\Psi}
\newcommand{\ph}{\phi}
\newcommand{\vph}{\varphi}
\newcommand{\Ph}{\Phi}
\newcommand{\Om}{\Omega}
\newcommand{\AdrHEPC}{Phenikaa Institute for Advanced Study and Faculty of Fundamental Sciences, Phenikaa University, Yen Nghia, Ha Dong, Hanoi 12116, Vietnam}
\newcommand{\AdrH}{Institute of Physics, Vietnam Academy of Science and Technology, 10 Dao Tan, Ba Dinh, Hanoi 100000, Vietnam}

\title{Questions of flavor physics and neutrino mass from a flipped hypercharge} 

\author{Duong Van Loi}
\email{loi.duongvan@phenikaa-uni.edu.vn (corresponding author)}
\affiliation{\AdrHEPC} 
\author{N. T. Duy} 
\email{ntduy@iop.vast.vn}
\affiliation{\AdrH} 
\author{Nguyen Huy Thao}
\email{nguyenhuythao@hpu2.edu.vn}
\affiliation{Department of Physics, Hanoi Pedagogical University 2, Phuc Yen, Vinh Phuc, Vietnam} 
\author{Phung Van Dong} 
\email{dong.phungvan@phenikaa-uni.edu.vn}
\affiliation{\AdrHEPC} 
\date{\today}

\begin{abstract}
The flavor structure of quarks and leptons is not yet fully understood, but it hints a more fundamental theory of non-universal generations. We therefore propose a simple extension of the Standard Model by flipping (i.e., enlarging) the hypercharge $U(1)_Y$ to $U(1)_X\otimes U(1)_N$ for which both $X$ and $N$ depend on generations of both quark and lepton. By anomaly cancellation, this extension not only explains the existence of just three fermion generations as observed but also requires the presence of a right-handed neutrino per generation, which motivates seesaw neutrino mass generation. Furthermore, in its minimal version with a scalar doublet and two scalar singlets, the model naturally generates the measured fermion-mixing matrices while it successfully accommodates several flavor anomalies observed in the neutral meson mixings, $B$-meson decays, lepton-flavor-violating processes of charged leptons, as well as satisfying constraints from particle colliders.
\end{abstract}

\maketitle

\section{Introduction}

Although the Standard Model (SM) has been highly successful in describing observed phenomena, it leaves many striking features of the physics of our world unexplained. This work would focus on the issues relating to the number of fermion generations, the generation of neutrino masses, fermion mass hierarchies, and flavor mixing profiles \cite{ParticleDataGroup:2022pth}.

In the SM, the electroweak symmetry reveals a partial unification of weak and electromagnetic interactions, which is based upon the non-Abelian gauge group $SU(2)_L \otimes U(1)_Y$, where $Y$ is an Abelian charge, well known as hypercharge \cite{Glashow:1961tr, Weinberg:1967tq, Salam:1968rm, Glashow:1970gm}. The electric charge operator takes the form $Q = T_3 + Y$, in which $T_3$ is the third component of the $SU(2)_L$ weak isospin. The value of $T_3$ is quantized due to the non-Abelian nature of $SU(2)_L$. In contrast, the value of $Y$ is entirely arbitrary on the theoretical ground because the Abelian $U(1)_Y$ algebra is trivial. Indeed, the hypercharge is often chosen to describe observed electric charges, while it does not explain them. An interesting question relating to the nature of the SM hypercharge is whether the conventional choice of generation-universal hypercharge causes the SM to be unable to address the issues. The present work does not directly answer this question. Instead of that, we look for an extension of $U(1)_Y$ to generation-dependent Abelian factors, as general, which naturally solves the issues.

For this aim, we embed $U(1)_Y$ in $U(1)_X\otimes U(1)_N$ for which both $X$ and $N$ are generation-dependent but determining $Y=X+N$, as observed. It is clear that $X, N$ may be an intermediate new physics phase resulting from a GUT and/or string breaking. Additionally, anomaly cancellation fixes both the number of fermion generations and values of $X, N$. Interestingly, we find {\it for the first time} that both quark and lepton generations are not universal under a gauge charge as of $X, N$. We investigate the model with a minimal scalar content in detail, which is responsible for the small, nonzero neutrino masses \cite{Kajita:2016cak, McDonald:2016ixn}, the measured fermion-mixing matrices \cite{ParticleDataGroup:2022pth, Gonzalez-Garcia:2021dve}, and several flavor-physics anomalies, such as mass splittings in $K$- and $B$-meson systems \cite{ParticleDataGroup:2022pth, HFLAV:2022pwe}, $B$-meson decays \cite{HFLAV:2022pwe, LHCb:2021awg}, and lepton-flavor-violating (LFV) processes of charged leptons \cite{ParticleDataGroup:2022pth, SINDRUMII:2006dvw, BaBar:2009hkt, MEG:2016leq,Willmann:1998gd}.

A few recent studies have attempted to explain the observed fermion mass and mixing hierarchies by decomposing the SM hypercharge to family hypercharges, say $U(1)_Y\to U(1)_{Y_1}\otimes U(1)_{Y_2}\otimes U(1)_{Y_3}$ \cite{FernandezNavarro:2023rhv} or $U(1)_Y\to U(1)_{Y_{1,2}}\otimes U(1)_{Y_3}$ \cite{Davighi:2023evx}, similar to baryon and lepton numbers that can be decomposed to family baryon and lepton numbers, respectively. The new observation of this proposal is that each family hypercharge identifies a relevant fermion family; hence, the number of family hypercharges present in the theory explains the number of the observed fermion families. The compelling feature of this approach is that if the Higgs doublet(s) carry the only third family hypercharge, then the only third family Yukawa couplings are allowed at the renormalizable level; by contrast, the remaining Yukawa couplings are suppressed, arising only from non-renormalizable operators. Consequently, both the models successfully describe charged fermion mass and mixing hierarchies. However, the reason for the existence of the observed fermion families is not convincing yet. This is because, in both models, every anomaly is canceled separately within each family, as in the SM. Therefore, there is no reason why each family hypercharge contains only a fermion family (since various repeated fermion families may be allowed and assigned to the same family hypercharge); thus, the number of fermion families is arbitrary. Below, we present a novel model in which each fermion family is anomalous, and the anomaly cancellation restricts the number of fermion families to three. 

Let us emphasize the two features of the present work. First, we argue that the number of fermion generations is precisely three, as observed, which comes only from anomaly cancellation. This is quite different from the 3-3-1 model \cite{Singer:1980sw, Valle:1983dk, Pisano:1992bxx, Frampton:1992wt, Foot:1994ym, Dong:2006mg, Dong:2006cn, Dong:2007ee,VanDong:2023lbn} as well as our previous proposals \cite{Nam:2020twn, VanDong:2022cin, VanLoi:2023utt, VanLoi:2023pkt}, in which anomaly cancellation implies that the fermion generation number is an integer multiple of three, and then it is necessary to add the QCD asymptotic freedom condition to get the number of fermion generations equal to three. Second, in the present work, we consider the possibility that the first lepton generation (the third quark generation) carries Abelian charges different from the remaining lepton (quark) generations under the new gauge groups, $U(1)_X\otimes U(1)_N$. Consequently, the fermion-mixing matrices are recovered, appropriate to experiment \cite{ParticleDataGroup:2022pth, Gonzalez-Garcia:2021dve}, because necessary small mixings arise only from nonrenormalizable operators. Interestingly enough, flavor-changing neutral currents (FCNCs) appear at the tree level in both the quark and lepton sectors. 

The rest of this work is organized as follows. We present the new model in Sec. \ref{model}. We investigate the fermion mass spectra in Sec. \ref{fermionmass}. We diagonalize the gauge and scalar sectors in Sec. \ref{gaugescalar} to identify physical fields. We determine the interactions of fermions and gauge bosons in Sec. \ref{inter}. We examine flavor physics observables and compare them to experimental results in Sec. \ref{flaphys}. We discuss the collider bounds in Sec. \ref{collider}. Finally, we summarize our results and conclude this work in Sec. \ref{conclusion}.        

\section{\label{model}The model}
\subsection{Anomaly cancellation and generation number}
As mentioned, the model under consideration is based on gauge symmetry,
\be SU(3)_C\otimes SU(2)_L\otimes U(1)_X\otimes U(1)_N,\label{gaugesymmetry} \ee
in which the first two factors are exactly those of the SM, whereas the last two factors are flipped (i.e., extended) from the weak hypercharge symmetry $U(1)_Y$. The new gauge charges depend on flavors of both quarks and leptons as
\bea X &=&3z[Bi^{r^2(r-1)}+Li^{r(r-1)}], \label{newcharge}\\
N &=&Y-X,\eea
where $B(L)$ denotes normal baryon (lepton) number, $Y$ labels the hypercharge, $z$ is an arbitrary non-zero parameter, $i$ is the imaginary unit, and $r$ is a flavor index, $r=1,2,\cdots,N_f$. Notice that $X$ is Hermitian, since $i^{r(r-1)}=(-1)^{\fr{r(r-1)}{2}}$ is always real. Additionally, the charges $X$'s of quark and lepton generations determined by Eq. (\ref{newcharge}) are either the same or opposite in sign, leading to reduced degrees of freedom in the model. The electric charge operator is embedded in the gauge symmetry as 
\be Q=T_3+X+N  \ee
with $T_n \,(n=1,2,3)$ are the $SU(2)_L$ generators. The SM fermions transform under the gauge symmetry as follows: 
\bea l_{rL} &=&\left(\nu_{rL},e_{rL}\right)^T\sim \left({\bf 1}, {\bf 2}, 3z i^{r(r-1)},-1/2-3z i^{r(r-1)}\right),\label{SMfersb}\\
e_{rR} &\sim& \left({\bf 1}, {\bf 1}, 3zi^{r(r-1)},-1-3zi^{r(r-1)}\right),\\
q_{rL} &=&\left(u_{rL},d_{rL}\right)^T\sim \left({\bf 3}, {\bf 2}, zi^{r^2(r-1)}, 1/6-zi^{r^2(r-1)}\right),\\
u_{rR} &\sim& \left({\bf 3}, {\bf 1}, zi^{r^2(r-1)}, 2/3-zi^{r^2(r-1)}\right),\\
d_{rR} &\sim& \left({\bf 3}, {\bf 1}, zi^{r^2(r-1)}, -1/3-zi^{r^2(r-1)}\right).\label{SMferse}
 \eea

It is interesting that the charge $X$ defined by Eq. (\ref{newcharge}) is periodic in $r$ with period 4, i.e., with $r=1,2,3,4,5,6,7,8,\cdots$, then
\be X=z,z,-z,z,z,z,-z,z,\cdots \ee
for the quark generations and
\be X=3z,-3z,-3z,3z,3z,-3z,-3z,3z,\cdots \ee
for the lepton generations. Hence, we express the number of fermion generations as $N_f=4x-y$ with $x=1,2,3,\cdots$ and $y=0,1,2,3$. Take an example, $N_f=5$ then $x=2$ and $y=3$. Considering the anomaly $[SU(2)_L]^2U(1)_X$, we get 
\be[SU(2)_L]^2U(1)_X \sim \sum_{\mathrm{doublets}}X_{f_L}=\left\{\begin{array}{ll}6z(x-1) &\text{ if }y=1\\
6zx&\text{ if }y=0,2,3\end{array},\right.\ee
which implies that this anomaly is canceled if and only if $x=y=1$, or equivalently $N_f=3$ as observed.\footnote{The result $N_f=3$ is unique and independent of the QCD asymptotic freedom condition. This is quite different from the 3-3-1 model \cite{Singer:1980sw,Valle:1983dk,Pisano:1992bxx,Frampton:1992wt,Foot:1994ym, Dong:2006mg, Dong:2006cn, Dong:2007ee,VanDong:2023lbn} as well as our previous works \cite{Nam:2020twn,VanDong:2022cin,VanLoi:2023utt,VanLoi:2023pkt}.} Because of lepton and quark generation discrepancies, we conveniently use two kinds of generation indices, such as $\al,\bet=1,2$ for the first two quark generations, while $a,b=2,3$ for the last two lepton generations; generically, $n,m=1,2,3$ run over $N_f=3$.

With $N_f=3$ and the fermion content as in Eqs. (\ref{SMfersb})--(\ref{SMferse}), two anomalies $[\text{Gravity}]^2U(1)_X$ and $[U(1)_X]^3$ are not canceled yet, namely 
\bea
\left[\text{Gravity}\right]^2U(1)_X &\sim& \sum_{\mathrm{fermions}}(X_{f_L}-X_{f_R})=-3z,\\
\left[U(1)_X\right]^3 &\sim&\sum_{\mathrm{fermions}}(X^3_{f_L}-X^3_{f_R})=-27z^3.\eea
To cancel these anomalies, we introduce right-handed neutrinos $\nu_{pR}\sim ({\bf 1}, {\bf 1}, X_{\nu_{pR}},-X_{\nu_{pR}})$ for $p=1,2,\cdots,N_R$ into the theory as fundamental constituents, satisfying
\be \sum_{p=1}^{N_R} X_{\nu_{pR}} =-3z,\hs \sum_{p=1}^{N_R} X^3_{\nu_{pR}} = -27z^3.\label{eanomaly}  \ee
Solving the equations in Eq. (\ref{eanomaly}), as well as requiring that at least two right-handed neutrinos to be identically responsible for neutrino mass generation, we obtain a unique nontrivial solution, such as 
\be X_{\nu_{1R}} = 3z,\hs X_{\nu_{aR}} = -3z,\label{solution}\ee
which implies that the resulting right-handed neutrinos have the lepton number as usual.\footnote{The solution as obtained differs from that in the conventional $U(1)_{B-L}$ extension whose $(B-L)_{\nu_{nR}}=(-1,-1,-1)$ or $(B-L)_{\nu_{nR}}=(-4,-4,5)$ \cite{Montero:2007cd,VanDong:2023thb}.}

With presence of the three right-handed neutrinos, whose $X$ charges obey Eq. (\ref{solution}), it is easily checked that the remaining anomalies, including $[SU(3)_C]^2 U(1)_X$, $[SU(3)_C]^2 U(1)_N$, $[SU(2)_L]^2 U(1)_N$, $[\text{Gravity}]^2U(1)_N$, $[U(1)_N]^3$, $[U(1)_X]^2 U(1)_N$, and $U(1)_X[U(1)_N]^2$, are all canceled, independent of arbitrary $z$. 

\subsection{Minimal particle content and symmetry breaking}
The particle content of the model, including fermions and scalars, as well as their quantum numbers under the gauge symmetry, are listed in Table \ref{tab1}. In addition to the SM fermions, three right-handed neutrinos must be included as fundamental fermions to suppress the anomalies, as shown in the previous subsection. Concerning the scalar sector, we introduce two singlets $\chi_{1,2}$ and a doublet $\ph$ under $SU(2)_L$. The singlets $\chi_{1,2}$ are necessarily presented to break $U(1)_X\otimes U(1)_N$ down to the weak hypercharge symmetry $U(1)_Y$, provide the Majorana masses for right-handed neutrinos, and recover the mixing matrices in quark and lepton sectors. Of course, the scalar doublet $\ph$ that is identified to the SM-Higgs doublet must be used to break $SU(2)_L\otimes U(1)_Y$ down to the electromagnetic symmetry $U(1)_Q$ and generate the masses for ordinary charged fermions, as well as Dirac masses for neutrinos. 

The scheme of symmetry breaking is given by \bc
\begin{tabular}{c} 
$SU(3)_C\otimes SU(2)_L\otimes U(1)_X\otimes U(1)_N$\\
$\downarrow \La_{1,2}$\\
$SU(3)_C\otimes SU(2)_L\otimes U(1)_Y$\\
$\downarrow v$\\
$SU(3)_C\otimes U(1)_Q$
\end{tabular}\ec Here, the scalar fields develop the vacuum expectation values (VEVs), such as
\be \langle \ph\rangle = \begin{pmatrix}0\\ \fr{v}{\sqrt2}\end{pmatrix},\hs \langle\chi_1\rangle = \fr{\La_1}{\sqrt2}, \hs \langle\chi_2\rangle=\fr{\La_2}{\sqrt2},\ee
satisfying $v=246 \text{ GeV}$ and $\La_{1,2}\gg v$ for consistency with the SM.

\begin{table}[h!]
\bc
\begin{tabular}{l|cccccccc}
\hline\hline
Multiplets & $SU(3)_C$ & $SU(2)_L$ & $U(1)_X$ & $U(1)_N$ \\ \hline 
$l_{1L}=(\nu_{1L},e_{1L})^T$ & $\bf 1$ & $\bf 2$ & $3z$ & $-1/2-3z$\\
$\nu_{1R}$ & $\bf 1$ & $\bf 1$ & $3z$ & $-3z$\\
$e_{1R}$ & $\bf 1$ & $\bf 1$ & $3z$ & $-1-3z$\\
$l_{aL}=(\nu_{aL},e_{aL})^T$ & $\bf 1$ & $\bf 2$ & $-3z$ & $-1/2+3z$\\
$\nu_{aR}$ & $\bf 1$ & $\bf 1$ & $-3z$ & $3z$\\
$e_{aR}$ & $\bf 1$ & $\bf 1$ & $-3z$ & $-1+3z$\\
$q_{\al L}=(u_{\al L},d_{\al L})^T$ & $\bf 3$ & $\bf 2$ & $z$ & $1/6-z$\\
$u_{\al R}$ & $\bf 3$ & $\bf 1$ & $z$ & $2/3-z$\\
$d_{\al R}$ & $\bf 3$ & $\bf 1$ & $z$ & $-1/3-z$\\
$q_{3L}=(u_{3L},d_{3L})^T$ & $\bf 3$ & $\bf 2$ & $-z$ & $1/6+z$\\
$u_{3 R}$ & $\bf 3$ & $\bf 1$ & $-z$ & $2/3+z$\\
$d_{3 R}$ & $\bf 3$ & $\bf 1$ & $-z$ & $-1/3+z$\\
$\ph=(\ph_1^+,\ph_2^0)^T$ & $\bf 1$ & $\bf 2$  & $0$ & $1/2$\\
$\chi_1$ & $\bf 1$ & $\bf 1$ & $2z$ & $-2z$\\
$\chi_2$ & $\bf 1$ & $\bf 1$ & $6z$ & $-6z$\\
\hline\hline
\end{tabular}
\caption[]{\label{tab1}Matter content in the model, where $\al=1,2$ and $a=2,3$ are generation indices, while $z$ is an arbitrarily nonzero parameter.}
\ec
\end{table}

Notice that the scalar content introduced above is minimal. Alternatively, a generic model can be constructed by introducing two new scalar doublets, namely $\ph'\sim({\bf 1},{\bf 2},2z,1/2-2z)$ and $\ph''\sim({\bf 1},{\bf 2},6z,1/2-6z)$, in addition to the usual doublet $\ph$, while the scalar singlet $\chi_2$ must be retained for breaking $U(1)_X\otimes U(1)_N\to Y(1)_Y$ as well as providing Majorana right-handed neutrino masses (note that $\chi_1$ can be omitted). This would produce renormalizable Yukawa couplings by $\phi',\phi''$ instead of the non-renormalizable ones (see below), which recover a complete mixing in the quark and lepton sectors at tree level. Additionally, such a model presents phenomenological aspects of interest that differ from the current model and should be published elsewhere.

\section{\label{fermionmass}Fermion mass}
The Yukawa Lagrangian for quarks and leptons in the current model is given by
\bea\mathcal{L}_{\mathrm{Yukawa}}&=&y^d_{\al\bet}\bar{q}_{\al L}\ph d_{\bet R}+y^d_{33}\bar{q}_{3L}\ph d_{3R}+\frac{y^d_{\al 3}}{M}\bar{q}_{\al L}\ph\chi_1 d_{3R}+\frac{y^d_{3\bet}}{M}\bar{q}_{3L}\ph\chi_1^* d_{\bet R}\crn
&&+y^u_{\al\bet}\bar{q}_{\al L}\tilde{\ph} u_{\bet R}+y^u_{33}\bar{q}_{3L}\tilde{\ph} u_{3R}+\frac{y^u_{\al 3}}{M}\bar{q}_{\al L}\tilde{\ph}\chi_1 u_{3R}+\frac{y^u_{3\bet}}{M}\bar{q}_{3L}\tilde{\ph}\chi_1^* u_{\bet R}\crn
&&+y^e_{11}\bar{l}_{1L}\ph e_{1R}+y^e_{ab}\bar{l}_{aL}\ph e_{bR}+\fr{y^e_{1b}}{M}\bar{l}_{1L}\ph\chi_2 e_{bR}+\fr{y^e_{a1}}{M}\bar{l}_{aL}\ph\chi_2^* e_{1R}\crn
&&+y^\nu_{11}\bar{l}_{1L}\tilde{\ph} \nu_{1R}+y^\nu_{ab}\bar{l}_{aL}\tilde{\ph} \nu_{bR}+\fr{y^\nu_{1b}}{M}\bar{l}_{1L}\tilde{\ph}\chi_2 \nu_{bR}+\fr{y^\nu_{a1}}{M}\bar{l}_{aL}\tilde{\ph}\chi_2^* \nu_{1R}\crn
&&+\fr 1 2 f^\nu_{11}\bar{\nu}^c_{1R}\chi_2^*\nu_{1R}+\fr 1 2 f^\nu_{ab}\bar{\nu}^c_{aR} \chi_2\nu_{bR}+F^\nu_{1b}\bar{\nu}^c_{1R} \nu_{bR} +\mathrm{H.c.},\label{yukawa}
\eea
where $\tilde{\ph} = i\sigma_2\ph^*$ with $\sigma_2$ is the second Pauli matrix, $M$ is a new physics scale that defines the effective interactions, and the couplings $y$ and $f$ are dimensionless. The bare mass $F$ connects $\nu_{1R}$ and $\nu_{2,3R}$, possibly obtaining a value ranging from zero to $M$. 

\subsection{Charged fermion mass}
From terms in the first three lines of Eq. (\ref{yukawa}), we obtain the mass matrices for charged fermions, which are given by
\bea [M_q]_{\al\beta} &=& -y^q_{\al\beta}\frac{v}{\sqrt2}, \hs [M_q]_{33}=-y^q_{33}\frac{v}{\sqrt2}, \\
\left[M_q\right]_{\al 3} &=& -y^q_{\al 3}\fr{v\La_1}{2M}, \hs [M_q]_{3\beta}= -y^q_{3\beta}\frac{v\La_1}{2M},\\
\left[M_e\right]_{11} &=& -y^e_{11}\frac{v}{\sqrt2},\hs [M_e]_{ab} = -y^e_{ab}\frac{v}{\sqrt2},\\
\left[M_e\right]_{1b} &=& -y^e_{1b}\fr{v\La_2}{2M}, \hs [M_e]_{a1}= -y^e_{a1}\frac{v\La_2}{2M},\eea
where $q=u,d$. Notice that the small mixing between the first two and third quark generations can be induced by either $y_{\al 3}^q,y_{3\beta}^q<y_{\al\beta}^q,y_{33}^q$ or $\La_1<M$, while between the first and last two lepton generations can be understood by either $y_{1b}^e,y_{a1}^e<y_{11}^q,y_{ab}^e$ or $\La_2<M$. By applying biunitary transformations, we can diagonalize these mass matrices separately, and then get the realistic masses of the up quarks $u,c,t$, the down quarks $d,s,b$, as well as the charged leptons $e,\mu,\tau$, such as 
\bea V_{u_L}^\dag M_u V_{u_R} &=& M_u^D = \text{diag}(m_u,m_c,m_t),\\
V_{d_L}^\dag M_d V_{d_R} &=& M_d^D = \text{diag}(m_d,m_s,m_b),\\
V_{e_L}^\dag M_e V_{e_R} &=& M_e^D = \text{diag}(m_e,m_\mu,m_\tau), \eea
where $V_{u_{L,R}}$, $V_{d_{L,R}}$ and $V_{e_{L,R}}$ are unitary matrices, linking gauge states, $u=(u_1,u_2,u_3)^T$, $d=(d_1,d_2,d_3)^T$, $e=(e_1,e_2,e_3)^T$, to mass eigenstates, $u'=(u,c,t)^T$, $d'=(d,s,b)^T$, $e'=(e,\mu,\tau)^T$, respectively, 
\be u_{L,R} = V_{u_{L,R}}u'_{L,R}, \hs d_{L,R} = V_{d_{L,R}}d'_{L,R}, \hs e_{L,R} = V_{e_{L,R}}e'_{L,R}. \ee
The Cabibbo-Kobayashi-Maskawa (CKM) matrix is then given by $V=V_{u_L}^\dag V_{d_L}$.

\subsection{Neutrino mass}
In the current model, neutrinos have both Dirac and Majorana mass terms, and their total mass matrix takes a specific form,
\be \mathcal{L}_{\text{Yukawa}}\supset - \fr 1 2 (\begin{array}{cc} \bar{\nu}_L^c& \bar{\nu}_R\end{array}) \left(\begin{array}{cc} 0&M_D\\ M_D^T&M_M\end{array}\right)\left(\begin{array}{cc} \nu_L\\ \nu_R^c\end{array}\right)+\mathrm{H.c.}, \label{neu}\ee
where $\nu_{L,R}=(\nu_1,\nu_2,\nu_3)_{L,R}^T$ are related to gauge states, $M_D$ and $M_M$ are respectively the Dirac and Majorana mass matrices, 
\bea 
\left[M_D\right]_{11} &=& -y^\nu_{11}\frac{v}{\sqrt2},\hs [M_D]_{ab} = -y^\nu_{ab}\frac{v}{\sqrt2},\\
\left[M_D\right]_{1b} &=& -y^\nu_{1b}\fr{v\La_2}{2M}, \hs [M_D]_{a1}= -y^\nu_{a1}\frac{v\La_2}{2M},\\
\left[M_M\right]_{11} &=& -f^\nu_{11}\fr{\La_2}{\sqrt2},\hs [M_M]_{ab} = -f^\nu_{ab}\fr{\La_2}{\sqrt2},\\
\left[M_M\right]_{1b} &=& -F^\nu_{1b},\hs [M_M]_{a1} = -F^\nu_{1a}.  \eea

Supposing $M>\La_2\gg v$, i.e. $M_M\gg M_D$, the total mass matrix of neutrinos in Eq. (\ref{neu}) can be diagonalized via a transformation as
\be \left(\begin{array}{cc} \nu_L\\ \nu_R^c\end{array}\right) \simeq \left(\begin{array}{cc} 1&\kappa^*\\-\kappa^T&1\end{array}\right)\left(\begin{array}{cc} V_{\nu_L}&0\\ 0&V_{\nu_R}^*\end{array}\right)\left(\begin{array}{cc} \nu'_L\\ {\nu'}_R^c\end{array}\right), \ee
where $\kappa$ is the $\nu_L$-$\nu_R$ mixing element, $\kappa = M_DM_M^{-1}\sim v/(\La_2, M)$, while $\nu'_{L,R}=(\nu'_1,\nu'_2,\nu'_3)_{L,R}^T$ are related to mass eigenstates, connecting to $\nu_{L,R}$ via unitary matrices $V_{\nu_{L,R}}$ as 
\be \nu_L\simeq V_{\nu_L}\nu'_L,\hs \nu_R\simeq V_{\nu_R}\nu'_R.\ee
Then, the mass eigenvalues are approximately given by
\bea \text{diag}(m_1,m_2,m_3) &\simeq& -V_{\nu_L}^TM_DM_M^{-1}M_D^TV_{\nu_L},\\
\text{diag}(M_1,M_2,M_3)&\simeq& V_{\nu_R}^\dag M_M V_{\nu_R}^*, \eea
in which $m_{1,2,3}\sim v^2/\La_2$ are appropriately small, identified with the observed neutrino masses, whereas $M_{1,2,3}\sim \La_2$ are the sterile neutrino masses, being at the new physics scale. Note that the Pontecorvo-Maki-Nakagawa-Sakata (PMNS) matrix can be written as $U=V_{e_L}^\dag V_{\nu_L}$. Note also that $F$ only contributes to right-handed neutrino mixing but does not set the seesaw scale. 

\section{\label{gaugescalar}Gauge and scalar sectors}
\subsection{Gauge sector}
The gauge bosons acquire masses via the scalar kinetic term $\sum_{S=\phi,\chi_1,\chi_2}(D^\mu \langle S\rangle)^\dag(D_\mu \langle S\rangle)$ when the gauge symmetry breaking occurs. The covariant derivative takes the form 
\be D_\mu = \pa_\mu + i g_s t_p G_{p\mu} + i g T_n A_{n\mu}+ i g_X X B_\mu + i g_N N C_\mu, \label{deriva}\ee
where $(g_s,g,g_X,g_N)$, $(t_p, T_n, X, N)$, and $(G_p, A_n, B, C)$ are coupling constants, generators, and gauge bosons of the ($SU(3)_C$, $SU(2)_L$, $U(1)_X$, $U(1)_N$) groups, respectively. Identifying the charged gauge bosons as $W^\pm_\mu = (A_{1\mu}\mp iA_{2\mu})/\sqrt2$, we obtain
\be \mathcal{L}\supset \frac{g^2v^2}{4}W^{\mu+}W_\mu^- +\fr 1 2 (A^\mu_3\, B^\mu\, C^\mu)M_0^2 (A_{3\mu}\, B_\mu\, C_\mu)^T, \ee
where 
\be M_0^2=\left(\begin{array}{ccc} \fr 1 4 g^2v^2 & 0 & -\fr 1 4 gg_Nv^2 \\
0 & 4g_X^2z^2(\La_1^2+9\La_2^2) & -4g_Xg_Nz^2(\La_1^2+9\La_2^2) \\
-\fr 1 4 gg_Nv^2 & -4g_Xg_Nz^2(\La_1^2+9\La_2^2) & \fr 1 4 g_N^2[16z^2(\La_1^2+9\La_2^2)+v^2]\end{array}\right). \ee
Hence, the boson $W$ is a physical field by itself with mass $m_W^2=g^2v^2/4$, which is identified to the SM $W$ boson, thus $v=246\text{ GeV}$, as expected.

Concerning the neutral gauge bosons, the mass-squared matrix $M_0^2$ always has a zero eigenvalue (i.e. photon mass) with corresponding eigenstate (i.e. photon field),
\be A = \frac{g_Xg_N A_3+gg_N B+gg_X C}{\sqrt{g_X^2g_N^2+g^2g_N^2+g^2g_X^2}}. \ee
From here, the interaction of the photon with fermions can be calculated \cite{Dong:2005wgt}. Identifying the coefficient of these interaction vertices with the electromagnetic coupling constant, we get the sine of the Weinberg's angle as $s_W=g_Xg_N/\sqrt{g_X^2g_N^2+g^2g_N^2+g^2g_X^2}$, and thus the hypercharge coupling to be $g_Y=g_Xg_N/\sqrt{g_X^2+g_N^2} = g_X s_\theta = g_N c_\theta$, where the angle $\theta$ is defined by $t_\theta=g_N/g_X$. We rewrite the photon field,
\be A = s_W A_3+c_W(s_\theta B+c_\theta C). \ee 
Hence, we define the SM $Z$ boson orthogonal to the photon $A$ and a new gauge boson $Z'$ orthogonal to both $A$ and $Z$, such as
\bea Z &=& c_WA_3-s_W(s_\theta B+c_\theta C),\\
Z' &=& c_\theta B-s_\theta C. \eea

In the new basis ($A,Z,Z'$), the photon $A$ is decoupled as a physical field, whereas two states $Z$ and $Z'$ still mix by themselves via a $2\times 2$ symmetric submatrix with the elements
\bea m_Z^2 &=& \frac{g^2v^2}{4c_W^2}, \hs m_{ZZ'}^2 = \frac{g^2v^2}{4c_W^2}s_Wt_\theta,\\
m_{Z'}^2 &=& \frac{g^2t_W^2}{s_{2\theta}^2}[16z^2(\La_1^2+9\La_2^2)+s_\theta^4v^2]. \eea
Diagonalizing this submatrix, we get two physical fields,
\be Z_1 = c_\va Z - s_\va Z',\hs Z_2 = s_\va Z + c_\va Z',\label{ZZp}  \ee
and two corresponding masses,
\bea m^2_{Z_1} &=& \fr 1 2 \left[m^2_{Z}+m^2_{Z'}-\sqrt{(m^2_Z-m^2_{Z'})^2+4m^4_{ZZ'}}\right]\simeq m^2_Z-\fr{m^4_{ZZ'}}{m^2_{Z'}},\\
m^2_{Z_2} &=& \fr 1 2 \left[m^2_{Z}+m^2_{Z'}+\sqrt{(m^2_Z-m^2_{Z'})^2+4m^4_{ZZ'}}\right] \simeq m^2_{Z'}, \eea
where the approximations apply due to $v\ll \La_{1,2}$. Also, the mixing angle $\va$ in Eq. (\ref{ZZp}) is given by
\be t_{2\va}=\fr{2m^2_{ZZ'}}{m^2_{Z'}-m^2_{Z}}\simeq \fr{s_\theta^3c_\theta v^2}{8s_W z^2 (\La_1^2+9\La_2^2)}.\ee
It is easy to see that the $Z$-$Z'$ mixing is small as suppressed by $v^2/\La_{1,2}^2$. Additionally, the field $Z_1$ has a mass approximating that of the SM, and thus, it is called the SM $Z$-like boson, whereas the field $Z_2$ is a new heavy gauge boson with mass at $\La_{1,2}$ scale.

It is noteworthy that the present model contains two Abelian gauge groups $U(1)_{X, N}$, in which the SM fermions have both non-zero $U(1)_X$ and $U(1)_N$ charges. Consequently, a non-zero gauge kinetic mixing between two relevant gauge bosons, i.e. $\mathcal{L}\supset -\fr 1 2 \ep_0 B_{\mu\nu}C^{\mu\nu}$, can arise at the one-loop level, given that this mixing vanishes at a high-energy scale due to some grand unification. Therefore, the $Z$-$Z'$ mixing is not only given by the mass mixing discussed above but also induced by the gauge kinetic mixing. Additionally, this kinetic mixing is easily computed by generalizing the result in \cite{Bauer:2022nwt} to be $\ep_0=\frac{g_Xg_N}{24\pi^2}\sum_fX_f(N_{f_L}+N_{f_R})\ln\frac{m_r}{m_f}$, where $f$ runs over every fermion of the SM with mass $m_f$, and $m_r$ is a renormalization scale. Thus, we estimate $\ep_0\sim \frac{10^{-4}}{24\pi^2}\left(\frac{g_X}{0.1}\right)\left(\frac{g_N}{0.1}\right)\left(\frac{z}{0.01}\right)\ln\left[10^{14}\left(\frac{m_r}{10^{16}\text{ GeV}}\right)\left(\frac{10^2\text{ GeV}}{m_f}\right)\right]\sim 10^{-5}$. This kinetic mixing effect is radically smaller than that from the tree-level mass mixing, since $\varphi\sim 10^{-3}\gg \ep_0\sim 10^{-5}$, taking $\La_{1,2}\gtrsim\mathcal{O}(10)$ TeV (see below). Hence, the gauge kinetic mixing is negligible and suppressed. 

\subsection{Scalar sector}
The current model's scalar sector contains a doublet $\ph$ and two singlets $\chi_{1,2}$ under $SU(2)_L$. Thus, the scalar potential has a simple form as
\bea V &=& \mu_1^2 \phi^\dagger \phi + \mu_2^2 \chi_1^* \chi_1 + \mu_3^2 \chi_2^* \chi_2 + (\la\chi_1^3\chi_2^*+\mathrm{H.c.})\crn
&&+ \la_1(\phi^\dagger \phi)^2 + \la_2 (\chi_1^* \chi_1)^2 + \la_3 (\chi_2^* \chi_2)^2\crn
&&+\la_4(\phi^\dagger \phi)(\chi_1^* \chi_1) + \la_5(\phi^\dagger \phi)(\chi_2^* \chi_2)+ \la_6 (\chi_1^* \chi_1)(\chi_2^* \chi_2),\eea
where the couplings $\la$'s are dimensionless, whereas $\mu$'s have a mass dimension. The necessary conditions for this scalar potential to be bounded from below and yielding a desirable vacuum structure are
\bea \mu_{1,2,3}^2 &<& 0,\hs |\mu_1|\ll |\mu_{2,3}|, \hs \la_{1,2,3} > 0,\\
\la_4 &>& -2\sqrt{\la_1\la_2},\hs \la_5>-2\sqrt{\la_1\la_3}, \hs \la_6>-2\sqrt{\la_2\la_3}.\eea

To obtain the physical scalar spectrum, we expand the scalar fields around their VEVs, such as 
\bea
\phi &=&\left(\begin{array}{c} \phi^+_1 \\ \frac{1}{\sqrt2}(v+S_1+iA_1) \end{array}\right), \\ 
\chi_1 &=&\frac{1}{\sqrt2}(\La_1+S_2+iA_2),\hs \chi_2=\frac{1}{\sqrt2}(\La_2+S_3+iA_3),
\eea
and then substitute them into the scalar potential. By using the potential minimum conditions given by
\bea
2\la_1v^2+\la_4\La_1^2+\la_5\La_2^2+2\mu_1^2 &=& 0,\\
2\la_2\La_1^2+\la_4v^2+\la_6\La_2^2+3\la\La_1\La_2+2\mu_2^2 &=& 0,\\
\la\La_1^3+(2\la_3 \La_2^2+\la_5v^2+\la_6\La_1^2+2\mu_3^2)\La_2 &=& 0,
\eea
we get the mass-squared matrix for $CP$-even scalar sector as
\be M_S^2=\left(\begin{array}{ccc} 2\la_1v^2 & \la_4v\La_1 & \la_5v\La_2 \\
\la_4v\La_1 & \fr 1 2 (4\la_2\La_1+3\la\La_2)\La_1 & \fr 1 2 (2\la_6\La_1\La_2+3\la\La_1^2) \\
\la_5v\La_2 & \fr 1 2 (2\la_6\La_1\La_2+3\la\La_1^2) & \fr{1}{2\La_2}(4\la_3\La_2^3-\la\La_1^3)\end{array}\right). \ee
Because of the condition, $v\ll\La_{1,2}$, the first row and first column of $M_S^2$ consist of elements much smaller than those of the rest. Therefore, the matrix $M_S^2$ can be diagonalized by using the seesaw approximation to separate the light state ($S_1$) from the heavy states ($S_{2,3}$). Labeling the new basis as ($H,H_1,H_2$), for which $H$ is decoupled as a physical field, we have
\be H\simeq S_1-\ep_1 S_2-\ep_2 S_3 \ee
with a corresponding mass
\be m^2_H\simeq 2\la_1v^2-(\ep_1 \la_4\La_1+\ep_2 \la_5\La_2)v, \ee
while the remaining states $H_1\simeq \ep_1 S_1 + S_2$ and $H_2\simeq \ep_2 S_1 + S_3$ mix by themselves via a submatrix as 
\be \mathcal{M}^2\simeq \fr 1 2\left(\begin{array}{cc} 
(4\la_2\La_1+3\la\La_2)\La_1 & (2\la_6\La_2+3\la\La_1)\La_1 \\
 (2\la_6\La_2+3\la\La_1)\La_1 & \fr{1}{\La_2}(4\la_3\La_2^3-\la\La_1^3)\end{array}\right). \ee
Above, the mixing parameters are given by
\bea \ep_1 &=& \frac{[\la(\la_4\La_1^3+3\la_5\La_1\La_2^2)-2(2\la_3\la_4-\la_5\la_6)\La_2^3]v}{2[3\la^2\La_1^3\La_2+\la(\la_2\La_1^4-3\la_3\La_2^4+3\la_6\La_1^2\La_2^2)-(4\la_2\la_3-\la_6^2)\La_1\La_2^3]},\\
\ep_2 &=& \frac{[3\la(\la_4\La_1^2-\la_5\La_2^2)-2(2\la_2\la_5-\la_4\la_6)\La_1\La_2]v\La_2}{2[3\la^2\La_1^3\La_2+\la(\la_2\La_1^4-3\la_3\La_2^4+3\la_6\La_1^2\La_2^2)-(4\la_2\la_3-\la_6^2)\La_1\La_2^3]}, \eea
which are small as suppressed by $v/\La_{1,2}$.

Diagonalizing the submatrix $\mathcal{M}^2$, we get two physical fields,
 \be \mathcal{H}_1 = c_\xi H_1 - s_\xi H_2, \hs \mathcal{H}_2 = s_\xi H_1 + c_\xi H_2,\ee
 with corresponding masses
 \bea
m^2_{\mathcal{H}_{1,2}} &=& \frac{1}{4\La_2}\left\{4\la_3\La_2^3-\la\La_1^3+(4\la_2\La_1+3\la\La_2)\La_1\La_2\right.\crn
&&\left.\mp\sqrt{[4\la_3\La_2^3-\la\La_1^3-(4\la_2\La_1+3\la\La_2)\La_1\La_2]^2+4(2\la_6\La_2+3\la\La_1)^2\La_1^2\La_2^2}\right\}, \eea
where the mixing angle $\xi$ is given by 
\be t_{2\xi} = \fr{2(2\la_6\La_2+3\la\La_1)\La_1\La_2}{4\la_3\La_2^3-\la\La_1^3-(4\la_2\La_1+3\la\La_2)\La_1\La_2}. \ee
The Higgs boson $H$ has a mass in weak scale like the SM Higgs boson, so $H$ is called the SM-like Higgs boson, whereas $\mathcal{H}_{1,2}$ are the new Higgs bosons, heavy in $\La_{1,2}$ scale.

The $CP$-odd scalars, $A_{1,2,3}$, mix by themselves via a mass-squared matrix 
\be M_A^2 = \fr{\la}{2}\left(\begin{array}{ccc}
0 & 0 & 0 \\
0 & -9\La_1\La_2 & 3\La_1^2 \\
0 & 3\La_1^2 & -\La_1^3/\La_2
\end{array}\right). \ee
This matrix has exactly two zero eigenvalues corresponding to two eigenstates,
\be G_{Z_1} = A_1,\hs G_{Z_2} =\frac{\La_1A_2+3\La_2A_3}{\sqrt{\La_1^2+9\La_2^2}},  \ee
which are the Goldstone bosons associated with the neutral gauge bosons, $Z_1$ and $Z_2$, respectively. The remaining eigenstate labeled $\mathcal{A}$ is a physical pseudoscalar orthogonal to $G_{Z_2}$, heavy at the $\La_{1,2}$ scale, namely 
\be \mathcal{A}= \frac{3\La_2A_2-\La_1A_3}{\sqrt{\La_1^2+9\La_2^2}}, \hs m^2_\mathcal{A}= -\frac{\la(\La_1^2+9\La_2^2)\La_1}{2\La_2}.\ee
Here, the requirement of positive squared mass implies the parameter $\la$ to be negative.

Concerning the charged scalars, we obtain a massless eigenstate, $G_W^\pm \equiv \ph_1^\pm$, identical to the Goldstone boson eaten by the SM $W$ boson.

\section{\label{inter} Fermion-gauge boson interaction}
We now consider the interaction of gauge bosons with fermions, which results from the fermion kinetic term, i.e., $\sum_F\bar{F}i\gamma^\mu D_\mu F$, where $F$ runs over fermion multiplets in the model. For convenience, we rewrite the covariant derivative in Eq. (\ref{deriva}) in the new form of
\bea D_\mu &=& \pa_\mu + ig_s t_p G_{p\mu} + i gs_W Q A_\mu + i g (T_+W^+_\mu + \mathrm{H.c.})  \crn
&&+ \fr{ig}{c_W}\left[c_\va (T_3-s^2_W Q)-s_\va\fr{s_W}{s_\theta c_\theta}(X-s^2_\theta Y)\right]Z_{1\mu} \crn
&&+ \fr{ig}{c_W}\left[s_\va (T_3-s^2_W Q)+c_\va\fr{s_W}{s_\theta c_\theta}(X-s^2_\theta Y)\right]Z_{2\mu},\eea 
where $T_\pm=(T_1\pm i T_2)/\sqrt{2}$ are the weight-raising and lowering operators of the $SU(2)_L$ group. Notice that $Q$, $T_3$, and $Y$ are universal for every flavor of neutrinos, charged leptons, up-type quarks, and down-type quarks, but $X$ is not. Consequently, both $Z_1$ and $Z_2$ flavor-change when interacting with fermions, in which the flavor-changing effect associated with $Z_1$ results from the $Z$--$Z'$ mixing to be small, whereas the flavor change associated with $Z_2$ is dominant, even for $\va=0$.

It is easily checked that the interaction of gluons and photon with fermions is similar to the SM, while the interaction of the $W$ boson with fermions is modified by the PMNS matrix,
\be \mathcal{L} \supset -\frac{g}{\sqrt2}(\bar{\nu}_{iL}\gamma^\mu U_{ij}^\dag e_{jL} + \bar{u}_{iL}\gamma^\mu V_{ij}d_{jL})W^+_\mu + \mathrm{H.c.},\ee
where $i,j=1,2,3$ are mass eigenstate indexes, i.e., $\nu_{iL}=\{\nu'_{1L},\nu'_{2L},\nu'_{3L}\}$, $e_i=\{e,\mu,\tau\}$, $u_i=\{u,c,t\}$, and $d_i=\{d,s,b\}$. 

For the interaction of $Z_{1,2}$ with fermions, using the unitary condition of mixing matrices,
\be V_{\nu_{L,R}}^\dag V_{\nu_{L,R}}=V_{e_{L,R}}^\dag V_{e_{L,R}}=V_{u_{L,R}}^\dag V_{u_{L,R}}=V_{d_{L,R}}^\dag V_{d_{L,R}}=1,\ee
we obtain a flavor-conserving part, given in the form of
\bea \mathcal{L} &\supset& -\fr{g}{2c_W}\left\{C_{1L}^{Z_I}\bar{\nu}'_{1L}\gamma^\mu\nu'_{1L}+C_{2L}^{Z_I}(\bar{\nu}'_{2L}\gamma^\mu\nu'_{2L}+\bar{\nu}'_{3L}\gamma^\mu\nu'_{3L})\right.\crn
&& \left.+ C_{R}^{Z_I}(\bar{\nu}'_{1R}\gamma^\mu\nu'_{1R}-\bar{\nu}'_{2R}\gamma^\mu\nu'_{2R}-\bar{\nu}'_{3R}\gamma^\mu\nu'_{3R})+ \bar{f}\ga^\mu [g^{Z_I}_V(f)- g^{Z_I}_A(f)\ga_5]f \right\}Z_{I\mu},\eea
where $I=1,2$, and $f$ denotes the physical charged fermions in the model. Additionally, the flavor-conserving couplings are given by
\bea 
C_{1L}^{Z_1} &=& c_\va - s_\va \fr{s_W}{s_\theta c_\theta}(6z+s_\theta^2),\hs C_{2L}^{Z_1} = c_\va + s_\va \fr{s_W}{s_\theta c_\theta}(6z-s_\theta^2),\\
C_{R}^{Z_1} &=& -s_\va \fr{6s_Wz}{s_\theta c_\theta},\\
g_V^{Z_1}(f) &=& c_\va[T_3(f_L)-2Q(f)s_W^2]-s_\va \fr{s_W}{s_\theta c_\theta}\{[T_3(f_L)-2Q(f)]s_\theta^2+2X(f)\},\\
g_A^{Z_1}(f) &=& T_3(f_L)(c_\va-s_\va s_W t_\theta),\\
C_{1L,2L}^{Z_2} &=& C_{1L,2L}^{Z_1}|_{c_\va\to s_\va, s_\va\to -c_\va},\hs C_{R}^{Z_2} = C_{R}^{Z_1}|_{s_\va\to -c_\va},\\
g_{V,A}^{Z_2}(f) &=& g_{V,A}^{Z_1}(f)|_{c_\va\to s_\va, s_\va\to -c_\va}.
\eea 
More specifically, we show the flavor-conserving couplings of $Z_{1,2}$ with the charged fermions in Tables \ref{tab2} and \ref{tab3}, respectively. It is easy to see that the $Z_1$ couplings with to the fermions are identical to those of the SM $Z$ boson in the limit $\va\to 0$.
\begin{table}[h]
\bc
\begin{tabular}{c|cc}
\hline\hline
$f$ & $g^{Z_1}_V(f)$ & $g^{Z_1}_A(f)$ \\
\hline 
$e$ & $c_\va\left(2s^2_W-\fr 1 2\right)-3s_\va s_W\left(\fr{1}{2}t_\theta+\fr{2z}{s_\theta c_\theta}\right)$ & $\fr 1 2\left(s_\va s_W t_\theta - c_\va\right)$\\
$\mu,\tau$ & $c_\va\left(2s^2_W-\fr 1 2\right)-3s_\va s_W\left(\fr{1}{2}t_\theta-\fr{2z}{s_\theta c_\theta}\right)$ & $\fr 1 2\left(s_\va s_W t_\theta - c_\va\right)$\\
$u,c$ & $c_\va\left(\fr 1 2 -\fr 4 3 s^2_W\right)+s_\va s_W\left(\fr 5 6 t_\theta-\fr{2z}{s_\theta c_\theta}\right)$ & $\fr 1 2\left(c_\va-s_\va s_W t_\theta\right)$\\
$t$ & $c_\va\left(\fr 1 2 -\fr 4 3 s^2_W\right)+s_\va s_W\left(\fr 5 6 t_\theta+\fr{2z}{s_\theta c_\theta}\right)$ & $\fr 1 2\left(c_\va-s_\va s_W t_\theta\right)$\\
$d,s$ & $c_\va\left(\fr 2 3 s^2_W-\fr 1 2\right)-s_\va s_W\left(\fr 1 6 t_\theta+\fr{2z}{s_\theta c_\theta}\right)$ & $\fr 1 2\left(s_\va s_W t_\theta-c_\va\right)$\\
$b$ & $c_\va\left(\fr 2 3 s^2_W-\fr 1 2\right)-s_\va s_W\left(\fr 1 6 t_\theta-\fr{2z}{s_\theta c_\theta}\right)$ & $\fr 1 2\left(s_\va s_W t_\theta-c_\va\right)$\\
\hline\hline
\end{tabular}
\caption[]{\label{tab2} Flavor-conserving couplings of $Z_1$ with the charged fermions.}
\ec
\end{table}
\begin{table}[h]
\bc
\begin{tabular}{c|cc}
\hline\hline
$f$ & $g^{Z_2}_V(f)$ & $g^{Z_2}_A(f)$ \\
\hline 
$e$ & $s_\va\left(2s^2_W-\fr 1 2\right)+3c_\va s_W\left(\fr{1}{2}t_\theta+\fr{2z}{s_\theta c_\theta}\right)$ & $-\fr 1 2\left(c_\va s_W t_\theta+s_\va\right)$\\
$\mu,\tau$ & $s_\va\left(2s^2_W-\fr 1 2\right)+3c_\va s_W\left(\fr{1}{2}t_\theta-\fr{2z}{s_\theta c_\theta}\right)$ & $-\fr 1 2\left(c_\va s_W t_\theta+s_\va\right)$\\
$u,c$ & $s_\va\left(\fr 1 2 -\fr 4 3 s^2_W\right)-c_\va s_W\left(\fr 5 6 t_\theta-\fr{2z}{s_\theta c_\theta}\right)$ & $\fr 1 2\left(c_\va s_W t_\theta+s_\va\right)$\\
$t$ & $s_\va\left(\fr 1 2 -\fr 4 3 s^2_W\right)-c_\va s_W\left(\fr 5 6 t_\theta+\fr{2z}{s_\theta c_\theta}\right)$ & $\fr 1 2\left(c_\va s_W t_\theta+s_\va\right)$\\
$d,s$ & $s_\va\left(\fr 2 3 s^2_W-\fr 1 2\right)+c_\va s_W\left(\fr 1 6 t_\theta+\fr{2z}{s_\theta c_\theta}\right)$ & $-\fr 1 2\left(c_\va s_W t_\theta+s_\va\right)$\\
$b$ & $s_\va\left(\fr 2 3 s^2_W-\fr 1 2\right)+c_\va s_W\left(\fr 1 6 t_\theta-\fr{2z}{s_\theta c_\theta}\right)$ & $-\fr 1 2\left(c_\va s_W t_\theta+s_\va\right)$\\
\hline\hline
\end{tabular}
\caption[]{\label{tab3} Flavor-conserving couplings of $Z_2$ with the charged fermions.}
\ec
\end{table}

To obtain flavor-changing part, we look at fermion-$Z_{1,2}$ interactions induced by $X$-charge, namely 
\bea \mathcal{L} &\supset& \frac{zgt_W}{s_\theta c_\theta}(\bar{l}_{L,R}\gamma^\mu T_ll_{L,R}+\bar{q}_{L,R}\gamma^\mu T_qq_{L,R})(s_\va Z_{1\mu}-c_\va Z_{2\mu}),\eea
where we have denoted $T_l=\text{diag}(3,-3,-3)$, $T_q=\text{diag}(1,1,-1)$, and $l=\nu,e$, while $q=u,d$. Changing to the mass basis via transformations $l_{L,R}=V_{l_{L,R}}l'_{L,R}$ and $q_{L,R}=V_{q_{L,R}}q'_{L,R}$, we obtain
\bea \mathcal{L} &\supset& \frac{zgt_W}{s_\theta c_\theta}(\bar{l}'_{L,R}\gamma^\mu V_{l_{L,R}}^\dag T_lV_{l_{L,R}}l'_{L,R}+\bar{q}'_{L,R}\gamma^\mu V_{q_{L,R}}^\dag T_qV_{q_{L,R}}q'_{L,R})(s_\va Z_{1\mu}-c_\va Z_{2\mu})\crn
&\supset& \frac{2zgt_W}{s_\theta c_\theta}(3[V_{l_L}^*]_{1i}[V_{l_L}]_{1j}\bar{l}_{iL}\gamma^\mu l_{jL}-[V_{q_L}^*]_{3i}[V_{q_L}]_{3j}\bar{q}_{iL}\gamma^\mu q_{jL})(s_\va Z_{1\mu}-c_\va Z_{2\mu})+(L\to R)\crn
&\equiv & (\Gamma_{ij}^{l_L}\bar{l}_{iL}\gamma^\mu l_{jL}+\Gamma_{ij}^{q_L}\bar{q}_{iL}\gamma^\mu q_{jL})(s_\va Z_{1\mu}-c_\va Z_{2\mu})+(L\to R),\label{flavor}\eea
which give rise to flavor-changing interactions for $i\neq j$. Here, we have labeled
\be \Gamma_{ij}^{l_L} = \frac{6gzt_W}{s_\theta c_\theta}[V_{l_L}^*]_{1i}[V_{l_L}]_{1j},\hs \Gamma_{ij}^{q_L} = -\frac{2gzt_W}{s_\theta c_\theta}[V_{q_L}^*]_{3i}[V_{q_L}]_{3j}. \label{fcnc} \ee

\section{\label{flaphys} Flavor phenomenologies}
To explain some flavor anomalies based on flavor-changing interactions in the current model, we first perform some assumptions for related parameters. It has been previously mentioned that the CKM and PMNS matrices are determined as $V=V_{u_L}^\dag V_{d_L}$ and $U=V_{e_L}^\dag V_{\nu_L}$, respectively. For the sake of simplicity, in this section, we align the lepton mixing to the charged lepton sector, i.e., $V_{\nu_L}=1$ and $U=V_{e_L}^\dag$. Similarly, for the quark sector, we align the quark mixing to the down quark sector, i.e., $V_{u_L}=1$ and $V=V_{d_L}$. That said, we focus solely on studying the flavor-changing of down quarks. It is noted that $V_{u_R,d_R}$ are completely arbitrary on the experimental side, i.e. they are not fixed by the current experiment, similar to those of the SM. Therefore, we choose $V_{u_R}=1$, while we parameterize the right-handed down-type quark mixing matrix $V_{d_R}$ through three Euler's angles $\theta^{d_R}_{ij}$ and a CP-violating phase $\delta ^{d_R}$ in the same way that we do so for the CKM and PMNS matrices, namely 
\be 
V_{d_R}=\left(\begin{array}{ccc}
	c^{d_R}_{12}c^{d_R}_{13} & s^{d_R}_{12}c^{d_R}_{13}&s^{d_R}_{13}e^{-i\delta^{d_R}}
	\\
	-s^{d_R}_{12}c_{23}^{d_R}-c_{12}^{d_R}s^{d_R}_{13}s_{23}^{d_R}e^{i\delta^{d_R}}&c^{d_R}_{12}c^{d_R}_{23}-s_{12}^{d_R}s^{d_R}_{13}s_{23}^{d_R}e^{i\delta^{d_R}}&c_{13}^{d_R}s_{23}^{d_R}\\
	s^{d_R}_{12}s_{23}^{d_R}-c_{12}^{d_R}s^{d_R}_{13}c_{23}^{d_R}e^{i\delta^{d_R}}&-c^{d_R}_{12}s_{23}^{d_R}-s_{12}^{d_R}s^{d_R}_{13}c_{23}^{d_R}e^{i\delta^{d_R}}&c^{d_R}_{13}c^{d_R}_{23}
\end{array}\right),
\ee  
where $c^{d_R}_{ij}\equiv\cos\theta^{d_R}_{ij}$ and $s^{d_R}_{ij}\equiv\sin\theta^{d_R}_{ij}$. Since $V_{d_R}$ has not been determined, as mentioned, the mixing angles $\theta^{d_R}_{ij}$ and the CP phase $\delta^{d_R}$ are free. To reduce the degrees of freedom, we assume that there is a relation among $\theta^{d_R}_{ij}$ following the Euler's angles of CKM matrix $\theta^{\text{CKM}}_{ij}$ according to one of the following four scenarios,   
\bea 
\fr{s_{13}^{d_R}}{s_{12}^{d_R}}&=&\fr{s^{\text{CKM}}_{13}}{s^{\text{CKM}}_{12}}, \hs \fr{s_{23}^{d_R}}{s_{12}^{d_R}}=\fr{s^{\text{CKM}}_{23}}{s^{\text{CKM}}_{12}} \hs \text{(Normal relation -- QNR)}, \label{quark_relation1}\\
\fr{s_{13}^{d_R}}{s_{12}^{d_R}}&=&\fr{s^{\text{CKM}}_{12}}{s^{\text{CKM}}_{13}}, \hs \fr{s_{23}^{d_R}}{s_{12}^{d_R}}=\fr{s^{\text{CKM}}_{12}}{s^{\text{CKM}}_{23}} \hs \text{(Inverted relation -- QIR)} ,\\ 
\fr{s_{13}^{d_R}}{s_{12}^{d_R}}&=&\fr{s^{\text{CKM}}_{13}}{s^{\text{CKM}}_{12}}, \hs \fr{s_{23}^{d_R}}{s_{12}^{d_R}}=\fr{s^{\text{CKM}}_{12}}{s^{\text{CKM}}_{23}} \hs \text{(Mixed relation -- QMR1)}, \\
\fr{s_{13}^{d_R}}{s_{12}^{d_R}}&=&\fr{s^{\text{CKM}}_{12}}{s^{\text{CKM}}_{13}}, \hs \fr{s_{23}^{d_R}}{s_{12}^{d_R}}=\fr{s^{\text{CKM}}_{23}}{s^{\text{CKM}}_{12}} \hs \text{(Mixed relation -- QMR2)}, \label{quark_relation2}
\eea 
in which $s^{\text{CKM}}_{ij}\equiv\sin\theta^{\text{CKM}}_{ij}$ \cite{Chau:1984fp}. Hence, for each the assumed relation, the matrix $V_{d_R}$ contains only two free parameters, $s_{12}^{d_R}$ and $\delta^{d_R}$. Notice that the Euler's angles of the CKM matrix can be defined via the Wolfenstein parameters $\la,A,\bar{\rho},\bar{\eta}$ \cite{Wolfenstein:1983yz,Buras:1994ec,Charles:2004jd}, i.e., 
\bea s^{\text{CKM}}_{12}=\la,\hs s^{\text{CKM}}_{23}= A\la^2, \hs s^{\text{CKM}}_{13}=A\la^3\sqrt{\bar{\rho}^2+\bar{\eta}^2}/(1-\la^2/2).
\eea  
Similarly, although we have imposed $U=V_{e_L}^\dag$, the right-handed charged lepton mixing matrix $V_{e_R}$ is still arbitrary on the experimental side. Thus, we can parameterize it via three Euler's angles $\theta^{e_R}_{ij}$ and a CP phase $\delta^{e_R}$ in the same way above and assume that there are four different scenarios of relation among $\theta^{e_R}_{ij}$ following the mixing angles of PMNS matrix $\theta^{\text{PMNS}}_{ij}$, such as 
	\bea 
\fr{s_{12}^{e_R}}{s_{23}^{e_R}}&=&\fr{s^{\text{PMNS}}_{12}}{s^{\text{PMNS}}_{23}}, \hs \fr{s_{13}^{e_R}}{s_{23}^{e_R}}=\fr{s^{\text{PMNS}}_{13}}{s^{\text{PMNS}}_{23}} \hs \text{(Normal relation -- LNR)},\label{lepton_relation1}\\ 
\fr{s_{12}^{e_R}}{s_{23}^{e_R}}&=&\fr{s^{\text{PMNS}}_{23}}{s^{\text{PMNS}}_{12}}, \hs \fr{s_{13}^{e_R}}{s_{23}^{e_R}}=\fr{s^{\text{PMNS}}_{23}}{s^{\text{PMNS}}_{13}}\hs \text{(Inverted relation -- LIR)},\\
\fr{s_{12}^{e_R}}{s_{23}^{e_R}}&=&\fr{s^{\text{PMNS}}_{23}}{s^{\text{PMNS}}_{12}}, \hs \fr{s_{13}^{e_R}}{s_{23}^{e_R}}=\fr{s^{\text{PMNS}}_{13}}{s^{\text{PMNS}}_{23}} \hs \text{(Mixed relation -- LMR1)},\\
\fr{s_{12}^{e_R}}{s_{23}^{e_R}}&=&\fr{s^{\text{PMNS}}_{12}}{s^{\text{PMNS}}_{23}}, \hs \fr{s_{13}^{e_R}}{s_{23}^{e_R}}=\fr{s^{\text{PMNS}}_{23}}{s^{\text{PMNS}}_{13}} \hs \text{(Mixed relation -- LMR2)}, \label{lepton_relation2}
\eea
where $s^{e_R}_{ij}\equiv\sin\theta^{e_R}_{ij}$ and $s^{\text{PMNS}}_{ij}\equiv\sin\theta^{\text{PMNS}}_{ij}$. In this work, we take the best-fit values of neutrino oscillation data with normal ordering hierarchy, given in Ref. \cite{Gonzalez-Garcia:2021dve}. Therefore, for each the above relation, the matrix $V_{e_R}$ contains only $s_{23}^{e_R}$ and $\delta^{e_R}$ as free parameters. Furthermore, in the limit $v \ll \La_{1,2}$, we have $t_{2\varphi}\sim v^2/\La_{1,2}^2 \ll 1$, hence we can neglect the $Z$--$Z'$ mixing. For the VEVs $\La_{1,2}$, we assume that $\La_1=k\La_2$ where $k$ is a dimensionless coefficient. Consequently, our model leaves six free parameters $z$, $k,\La_2$, $s^{d_R}_{12},s^{e_R}_{23}$, and $\theta$. Numerical values of the relevant common SM parameters are listed in Table \ref{commonSMvals}, while those of known input parameters associated with quark and lepton flavors are listed in Tables \ref{input-par} and \ref{input-par-lepton}, respectively. 

We would like to note that the new scalars $\mathcal{H}_{1,2}$ and $\mathcal{A}$ also induce flavor-violating interactions, in addition to the new gauge boson $Z'$. However, these flavor-violating interactions are proportional to $m_{u,d,e}/\La_{1,2}\ll 1$, and thus, significantly smaller compared to those caused by the $Z'$ gauge boson. Therefore, the following analysis will only focus on flavor phenomenologies from the $Z'$ gauge boson. 

\begin{table}[h!]
	\begin{centering}
		\begin{tabular}{|c|c|}
			\hline
			Parameters & Values   \tabularnewline
			\hline 
			$\al_{\text{em}} $ & $1/137 $ \cite{ParticleDataGroup:2022pth} \tabularnewline
			$m_W $ & $80.377\ \text{GeV}$  \cite{ParticleDataGroup:2022pth} 
			\tabularnewline
			$m_Z $ & $91.1876 \ \text{GeV}$ \cite{ParticleDataGroup:2022pth} \tabularnewline
			$G_F$ & $1.1663788 \times 10^{-5} \ \text{GeV}^{-2}$ \cite{ParticleDataGroup:2022pth} \tabularnewline	
			$s_W^2$ & $0.23121$ \cite{ParticleDataGroup:2022pth} \tabularnewline	
			\hline 
		\end{tabular}
		\par
		\protect\caption{\label{commonSMvals}Common SM parameters. }
	\end{centering}	
\end{table}

 \begin{table}[h!]
	\begin{centering}
		\begin{tabular}{|c|c|c|c|}
			\hline
			Parameters & Values  & Parameters & Values \tabularnewline
			\hline 
			$ f_{K}$ & $155.7(3) \ \text{MeV}$  \cite{FlavourLatticeAveragingGroupFLAG:2021npn} &  $m_{K}$ & $497.611(13)  \ \text{MeV} $ \cite{ParticleDataGroup:2022pth}
			\tabularnewline
			$f_{B_s}$ & $230.3(1.3) \ \text{MeV}$ \cite{FlavourLatticeAveragingGroupFLAG:2021npn}	 &  $ m_{B_s}$ & $5366.88(11) \ \text{MeV}$  \cite{ParticleDataGroup:2022pth}\tabularnewline
			$ f_{B_d}$ & $190.0(1.3) \ \text{MeV}$  \cite{FlavourLatticeAveragingGroupFLAG:2021npn} &  $m_{B_d}$ & $5279.65(12)  \ \text{MeV} $ \cite{ParticleDataGroup:2022pth} \tabularnewline
			$ m_{u}$ & $2.14(8) \ \text{MeV}$  \cite{FlavourLatticeAveragingGroupFLAG:2021npn} &  $m_{d}$ & $4.70(5)  \ \text{MeV} $ \cite{FlavourLatticeAveragingGroupFLAG:2021npn}\tabularnewline
			$\bar{m}_{c} \ (3 \ \text{GeV})$ & $0.988(11)  \ \text{GeV} $ \cite{FlavourLatticeAveragingGroupFLAG:2021npn} & $ m_{s}$ & $93.40(57) \ \text{MeV}$  \cite{FlavourLatticeAveragingGroupFLAG:2021npn} \tabularnewline
			$m_{t}$ & $172.69(30) \  \text{GeV} $ \cite{ParticleDataGroup:2022pth} & $\bar{m}_{b}(\bar{m}_b)$ & $4.196(14)  \ \text{GeV} $ \cite{UTfit:2022hsi} \tabularnewline
			$N(E_{\ga})$ & $3.3\times 10^{-3} $ \cite{Misiak:2020vlo}& $C_7^{\text{SM}}(\mu_b=2.0 \ \text{GeV})$ & $-0.3636 $ \cite{Misiak:2006ab,Czakon:2006ss,Misiak:2020vlo} \tabularnewline
			$C_9^{\text{SM}}(\mu_b=5.0 \ \text{GeV})$ & $4.344  $\ \cite{Beneke:2017vpq} &	$C_{10}^{\text{SM}}(\mu_b=5.0 \ \text{GeV})$ & $-4.198  $\  \cite{Beneke:2017vpq} \tabularnewline
			$y_s$ & $0.0645(3) $ \cite{HFLAV:2022pwe}& 	$\la $ & $0.22519(83) $ \cite{UTfit:2022hsi} \tabularnewline
			$A $ & $0.828(11) $ \cite{UTfit:2022hsi}  &
			$\bar{\rho} $ & $0.1609(95) $ \cite{UTfit:2022hsi} \tabularnewline
			$\bar{\eta} $ & $0.347(10) $ \cite{UTfit:2022hsi} &&
			\tabularnewline  
			
			\hline 
		\end{tabular}
		\par
			\protect\caption{\label{input-par}Numerical values of known input parameters for quark flavors.}
	\end{centering}
\end{table} 

\begin{table}[h!]
	\begin{centering}
		\begin{tabular}{|c|c|c|c|}
			\hline
			Parameters & Values  & Parameters & Values \tabularnewline
			\hline 
			$ m_{e}$ & $5\times 10^{-4} \ \text{GeV}$  \cite{ParticleDataGroup:2022pth} &  $(s_{12}^{\text{PMNS}})^2$ & $0.304^{+0.012}_{-0.012} $ \cite{Gonzalez-Garcia:2021dve}
			\tabularnewline
			$m_{\mu}$ & $0.105 \ \text{GeV}$ \cite{ParticleDataGroup:2022pth}	 &  $(s_{23}^{\text{PMNS}})^2$ & $0.450^{+0.019}_{-0.016} $ \cite{Gonzalez-Garcia:2021dve}
			\tabularnewline
			$ m_{\tau}$ & $1.776 \ \text{GeV}$  \cite{FlavourLatticeAveragingGroupFLAG:2021npn} &  $(s_{13}^{\text{PMNS}})^2$ & $0.02246^{+0.00062}_{-0.00062} $ \cite{Gonzalez-Garcia:2021dve} 
			\tabularnewline
			$ \Ga_{\mu}$ & $3\times 10^{-19} \ \text{GeV}$  \cite{ParticleDataGroup:2022pth} &  $\delta_{\text{CP}}^{\text{PMNS}} (^{\circ})$ & $230^{+36}_{-25} $ \cite{Gonzalez-Garcia:2021dve}
			\tabularnewline
			$\Ga_{\tau} $ & $2.27\times 10^{-12} \ \text{GeV} $ \cite{ParticleDataGroup:2022pth} && 
			\tabularnewline
			\hline 
		\end{tabular}
		\par
		\protect\caption{\label{input-par-lepton}Numerical values of known input parameters for lepton flavors.}
	\end{centering}
\end{table}

\subsection{Quark flavor phenomenologies}
This subsection focuses on flavor phenomenologies in the quark sector with controllable theoretical uncertainties. Because the quark generations are not universal under $U(1)_X\otimes U(1)_N$, the model predicts flavor-changing processes in the quark sector associated with the new gauge boson $Z'$. These processes occur at the tree level for $K,B_s$, and $B_d$ meson oscillations or at both tree and loop levels for the quark transitions $b\to s e_I^+ e_I^-$ with $e_I=\{e_1, e_2\}=\{e,\mu\}$, such as branching ratio of $B_s\to \mu^+\mu^-$, branching ratio of inclusive decay BR$(\bar{B}\to X_s\gamma)$, and ratios $R_{K,K^*}=\text{BR}(B^{+,0}\to K^{+,0*}\mu^+\mu^-)/\text{BR}(B^{+,0}\to K^{+,0*}e^+e^-)$. 

\begin{figure}[H]
	\centering
	\begin{tabular}{c}
	\includegraphics[width=10cm]{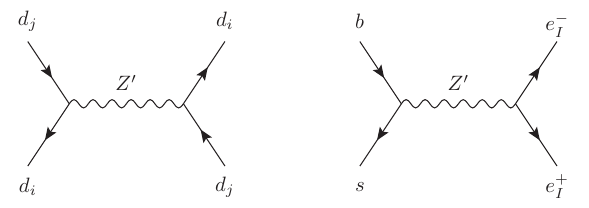}
	\end{tabular}
	\caption{\label{tree_diagrams}Tree-level diagrams induced by new gauge boson $Z'$ for meson mixings (left) and $b\to s e_I^+e_I^-$ transitions (right), where $d_{i,j}=\{d,s,b\}$ and $d_i\neq d_j$, $e_I=\{e,\mu\}$.}
\end{figure}

\begin{figure}[h!]
	\centering
	\begin{tabular}{c}
	\includegraphics[width=13cm]{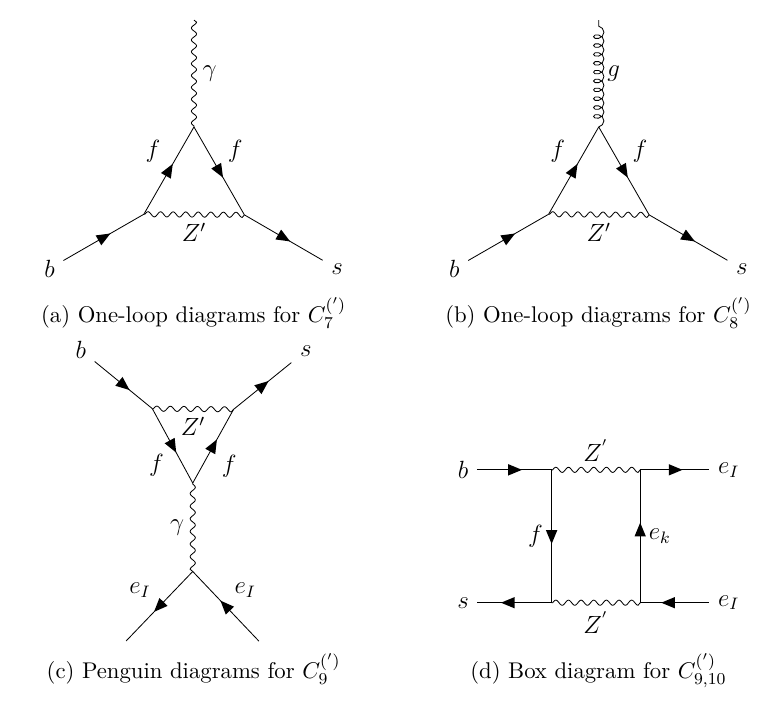}
	\end{tabular}
	\caption{\label{loop_diagrams}Radiative contributions to $C^{(')}_{7,8,9,10}$, where $k=e,\mu,\tau$ and $f=d,s,b$.}
\end{figure}

The effective Hamiltonian relevant for the above processes can be written as \cite{Buchalla:1995vs}
\bea \mathcal{H}^{\text{quark}}_{\text{eff}} &=& \sum_{X=K,B_s,B_d}(C_X^2\mathcal{O}_X^2+C^{'2}_X\mathcal{O}_X^{'2}+2C_XC^{'}_X\mathcal{O}_X\mathcal{O}^{'}_X)\crn 
&&-\fr{4G_F}{\sqrt{2}}V^{*}_{ts}V_{tb}\sum_{Y=7,8,9,10}(C_Y\mathcal{O}_Y+C_Y^{'}\mathcal{O}_Y^{'}),\eea
where $G_F$ is the Fermi constant and $V_{ts,tb}$ are the CKM matrix elements. The first summation contains contributions to meson mixing systems $\bar{d}_id_j\to d_j\bar{d_j}$ with $d_{i,j}=\{d_1,d_2,d_3\}=\{d,s,b\}$ and $d_i\neq d_j$, while the second summation relevant to the $b\to s e_I^+ e_I^-$ observables. The primed operators $\mathcal{O}^{'}_{X,Y}$ are chirally flipped counterparts $P_L \leftrightarrow P_R$ of unprimed operators $\mathcal{O}_{X,Y}$, they are defined as
\bea 
\mathcal{O}^{(')}_K &=&\bar{s}\ga^{\mu}P_{L(R)}d , \hs \mathcal{O}^{(')}_{B_s}=\bar{s}\ga^{\mu}P_{L(R)}b ,\hs \mathcal{O}^{(')}_{B_d}=\bar{d}\ga^{\mu}P_{L(R)}b , \\  
\mathcal{O}^{(')}_7 &=&\fr{e}{16\pi^2}m_b(\bar{s}\sigma^{\mu\nu}P_{R(L)} b)F_{\mu\nu}, \hs \mathcal{O}^{(')}_8=\fr{g_s}{16\pi^2}m_b(\bar{s}\sigma^{\mu\nu}T^a P_{R(L)} b)G^a_{\mu\nu}, \\ 
\mathcal{O}^{(')}_{9}&=&\fr{e^2}{16\pi^2}(\bar{s}\ga^{\mu}P_{L(R)}b)(\bar{e}_I\ga_{\mu}e_I),\hs \mathcal{O}^{(')}_{10}=\fr{e^2}{16\pi^2}(\bar{s}\ga^{\mu}P_{L(R)} b)(\bar{e}_I\ga_{\mu}\ga_5 e_I), \eea
where $P_{L,R}=\fr 1 2 (1\mp\gamma_5)$. The operators $\mathcal{O}^{(')}_{7,8}$ contribute mainly to BR$(\bar{B}\to X_s\ga)$, whereas $\mathcal{O}^{(')}_{9,10}$ dominate the BR$(B_s\to e_I^+e_I^-)$ and the ratios $R_{K,K^*}$. The new physics contributions to the Wilson coefficients (WCs) $C^{(')\text{NP}}_{X,Y}$ can come from either the tree level or the quantum level (loop, penguin, and box diagrams) or both ones. Generally, we can decompose the new physics contributions as $C^{(')\text{NP}}_{X,Y}=C^{(')\text{tree}}_{X,Y}+C^{(')\text{loop}}_{X,Y}+C^{(')\text{penguin}}_{X,Y}+C^{(')\text{box}}_{X,Y}$, where the contribution of each style of diagrams is indicted by the superscripts. For the tree-level contributions as described by Feynman diagrams in Fig. \ref{tree_diagrams}, we obtain 
\bea C_K^{\text{tree}} &=& -\fr{2gzt_W}{s_\theta c_\theta m_{Z'}}[V^*_{d_L}]_{31}[V_{d_L}]_{32}, \hs C_K^{'\text{tree}}=-\fr{2gzt_W}{s_\theta c_\theta m_{Z'}}[V^*_{d_R}]_{31}[V_{d_R}]_{32}, \label{tree1}\\ 
C_{B_s}^{\text{tree}} &=& -\fr{2gzt_W}{s_\theta c_\theta m_{Z'}}[V^*_{d_L}]_{32}[V_{d_L}]_{33},\hs C_{B_s}^{'\text{tree}}=-\fr{2gzt_W}{s_\theta c_\theta m_{Z'}}[V^*_{d_R}]_{32}[V_{d_R}]_{33}, \\
C_{B_d}^{\text{tree}} &=&-\fr{2gzt_W}{s_\theta c_\theta m_{Z'}}[V^*_{d_L}]_{31}[V_{d_L}]_{33}, \hs C_{B_d}^{'\text{tree}}=-\fr{2gzt_W}{s_\theta c_\theta m_{Z'}}[V^*_{d_R}]_{31}[V_{d_R}]_{33},\label{tree3} \\ 
C^{\text{tree},e_I}_{9} &=& \Ga^{d_L}_{23}\fr{m_W^2}{c_W V^*_{ts}V_{tb}}\fr{1}{g}\fr{(4\pi)^2}{e^2}\fr{g_V^{Z_2}(e_I)}{m_{Z'}^2},\hs C^{'\text{tree},e_I}_{9} = \Ga^{d_R}_{23}\fr{m_W^2}{c_W V^*_{ts}V_{tb}}\fr{1}{g}\fr{(4\pi)^2}{e^2}\fr{g_V^{Z_2}(e_I)}{m_{Z'}^2}, \\ 
 C^{\text{tree},e_I}_{10} &=& -\Ga^{d_L}_{23}\fr{m_W^2}{c_W V^*_{ts}V_{tb}}\fr{1}{g}\fr{(4\pi)^2}{e^2}\fr{g_A^{Z_2}(e_I)}{m_{Z'}^2}, \hs C^{'\text{tree},e_I}_{10}=-\Ga^{d_R}_{23}\fr{m_W^2}{c_W V^*_{ts}V_{tb}}\fr{1}{g}\fr{(4\pi)^2}{e^2}\fr{g_A^{Z_2}(e_I)}{m_{Z'}^2}.\label{WCs_quark1} \eea
For the quantum-level contributions, they are obtained from the one-loop, penguin, and box diagrams that contain gauge boson $Z'$, down quarks $f=d,s,b$, and charged leptons $k=e,\mu,\tau$ to be internal lines, given in Fig. \ref{loop_diagrams}. We use 't Hooft gauge $\zeta=1$ for calculating these diagrams. With the diagrams (a) and (b), we calculate on-shell, i.e., $q^2=0$, $p_s^2=m_s^2$, and $p_b^2=m_b^2$. Because $m_s\ll m_b$, we set the $s$ quark mass to be zero, $m_s=0$, and keep the mass of $b$ quark at the linear order, i.e. $m_b^2=0$. Additionally, we calculate in the limit $m_f^2/m_{Z'}^2,m_k^2/m_{Z'}^2\ll 1$ since $m_{f,k}\sim \mathcal{O}(1) \ \text{GeV} \ll m_{Z'}\sim \mathcal{O}(1)$ TeV, for simplicity. It is important to note that under this limit other loop diagrams with unphysical Goldstone boson $\phi_{Z'}$ are suppressed by factors $m_f^2/m_{Z'}^2,m_k^2/m_{Z'}^2\ll 1$, hence we can safely ignore the box diagrams with the Goldstone boson $\phi_{Z'}$ and keep only the ones with physical gauge boson $Z'$. We have the expressions for these contributions at the scale $\mu_{Z'}=m_{Z'}$ as      
	\bea
C^{\text{loop}}_{7} (\mu_{Z'}) &\simeq& -\fr{2m_W^2}{9m_{Z'}^2V^{*}_{ts}V_{tb}}\fr{1}{g^2}\sum_{i=1,2,3}\left(\Ga^{*d_L}_{i2}\Ga^{d_L}_{i3}-3\fr{m_{d_i}}{m_b}\Ga^{*d_L}_{i2}\Ga^{d_R}_{i3}\right) ,\label{loop1}\\ 
C^{'\text{loop}}_{7} (\mu_{Z'}) &\simeq& - \fr{2m_W^2}{9m_{Z'}^2V^{*}_{ts}V_{tb}}\fr{1}{g^2}\sum_{i=1,2,3}\left(\Ga^{*d_R}_{i2}\Ga^{d_R}_{i3}-3\fr{m_{d_i}}{m_b}\Ga^{*d_R}_{i2}\Ga^{d_L}_{i3}\right) ,\\ 
C^{\text{loop}}_{8} (\mu_{Z'}) &\simeq& -3C^{\text{loop}}_{7}  (\mu_{Z'}), \hs C^{'\text{loop}}_{8}  (\mu_{Z'})\simeq -3C^{'\text{loop}}_{7} (\mu_{Z'}) ,\\ 
C_9^{\text{penguin},e_I,\ga}(\mu_{Z'}) &\simeq& -\fr{m_W^2}{m_{Z'}^2}\fr{1}{g^2}\sum_{i=1,2,3}\fr{\Ga^{*d_L}_{i2}\Ga^{d_L}_{i3}}{V_{ts}^*V_{tb}}\left(-\fr{1}{27}+\fr{2}{9}\ln{\frac{m^2_{d_i}}{m^2_{Z'}}}\right), \\ 
C_9^{'\text{penguin},e_I,\ga}(\mu_{Z'}) &\simeq& -\fr{m_W^2}{m_{Z'}^2}\fr{1}{g^2}\sum_{i=1,2,3}\fr{\Ga^{*d_R}_{i2}\Ga^{d_R}_{i3}}{V_{ts}^*V_{tb}}\left(-\fr{1}{27}+\fr{2}{9}\ln{\frac{m^2_{d_i}}{m^2_{Z'}}}\right),\label{pen2} \\ 
C^{\text{box},e_I}_{9}(\mu_{Z'}) &\simeq& -\fr{m_W^2}{4s_W^2 m_{Z'}^2} \fr{1}{g^4}\sum_{i=1,2,3}\sum_{j=1,2,3}\fr{\Ga^{*d_L}_{i2}\Ga^{d_L}_{i3}(|\Ga^{e_L}_{jI}|^2+|\Ga^{e_R}_{jI}|^2)}{V^{*}_{ts}V_{tb}},\\
C^{'\text{box},e_I}_{9}(\mu_{Z'}) &\simeq& -\fr{m_W^2}{4s_W^2 m_{Z'}^2}\fr{1}{g^4}\sum_{i=1,2,3}\sum_{j=1,2,3}\fr{\Ga^{*d_R}_{i2}\Ga^{d_R}_{i3}(|\Ga^{e_L}_{jI}|^2+|\Ga^{e_R}_{jI}|^2)}{V^{*}_{ts}V_{tb}}, \\
C^{\text{box},e_I}_{10}(\mu_{Z'}) &\simeq& -\fr{m_W^2}{4s_W^2 m_{Z'}^2} \fr{1}{g^4}\sum_{i=1,2,3}\sum_{j=1,2,3}\fr{\Ga^{*d_L}_{i2}\Ga^{d_L}_{i3}(|\Ga^{e_L}_{jI}|^2-|\Ga^{e_R}_{jI}|^2)}{V^{*}_{ts}V_{tb}},\\
C^{'\text{box},e_I}_{10}(\mu_{Z'}) &\simeq& -\fr{m_W^2}{4s_W^2 m_{Z'}^2}\fr{1}{g^4}\sum_{i=1,2,3}\sum_{j=1,2,3}\fr{\Ga^{*d_R}_{i2}\Ga^{d_R}_{i3}(|\Ga^{e_L}_{jI}|^2-|\Ga^{e_R}_{jI}|^2)}{V^{*}_{ts}V_{tb}}. \label{WCs_quark2}
\eea 
It should be noted that the penguin diagrams with off-shell SM $Z$-like boson do not give the contributions to WCs in the limit $m_{d_i}^2/m_{Z'}^2 \to 0$; thus, we do not include these diagrams in our calculation.

Next, we determine the new physics contributions to each observable in terms of WCs. Firstly, for the meson mixings, we can decompose the contributions to meson mass differences as $\Delta m_{K,B_s,B_d}=\Delta m^{\text{SM}}_{K,B_s,B_d}+\Delta m^{\text{NP}}_{K,B_s,B_d}$, where the SM contributions $\Delta m^{\text{SM}}_{K,B_s,B_d}$ are shown in the second column in Table \ref{SM_exp data}, and the new physics contributions $\Delta m^{\text{NP}}_{K,B_s,B_d}$ are estimated by \cite{Gabbiani:1996hi,Langacker:2000ju}
\bea
\Delta m_K^{\text{NP}} &\simeq& \fr{2}{3} \mathrm{Re}\left\{(C^\text{tree}_K)^2-\left[\fr 3 2+ \left(\fr{m_K}{m_d+m_s}\right)^2\right]C^\text{tree}_KC^{'\text{tree}}_K+(C^{'\text{tree}}_K)^2\right\}m_K f^2_K, \label{ptdt16}\\
\Delta m_{B_s}^{\text{NP}} &\simeq& \frac{2}{3} \mathrm{Re}\left\{ (C^\text{tree}_{B_s})^2-\left[\fr 3 2+ \left(\fr{m_{B_s}}{m_s+m_b}\right)^2\right]C^\text{tree}_{B_s}C^{'\text{tree}}_{B_s}+(C^{'\text{tree}}_{B_s})^2\right\}m_{B_s} f^2_{B_s},
\label{ptdt17}\\
\Delta m_{B_d}^{\text{NP}} &\simeq& \fr{2}{3} \mathrm{Re}\left\{ (C^\text{tree}_{B_d})^2-\left[\fr 3 2+ \left(\fr{m_{B_d}}{m_d+m_b}\right)^2\right]C^\text{tree}_{B_d}C^{'\text{tree}}_{B_d}+(C^{'\text{tree}}_{B_d})^2\right\}m_{B_d} f^2_{B_d}\label{ptdt18}.\eea 
Note that the SM $Z$-like boson also contributes to meson mass differences due to the mixing of $Z$--$Z'$. However, these contributions are proportional with $s_{\varphi}^4$, so we ignore them. 
\begin{table}[h!]
	\begin{centering}
	\begin{tabular}{|c|c|c|}
			\hline
			Observables & SM predictions  & Experimental values  \tabularnewline
			\hline 
			$\Delta m_K$ & $0.467\times 10^{-2}   \  \text{ps}^{-1}$ \cite{ParticleDataGroup:2022pth}& $0.5293(9)\times 10^{-2}  \  \text{ps}^{-1}$ \cite{ParticleDataGroup:2022pth} \tabularnewline
			$\Delta m_{B_s}$ & $18.77(86) \  \text{ps}^{-1}$ \cite{Lenz:2019lvd} &  $17.765(6)  \  \text{ps}^{-1}$ \cite{HFLAV:2022pwe} \tabularnewline
			$\Delta m_{B_d}$ & $0.543(29)\ \text{ps}^{-1} $ \cite{Lenz:2019lvd}&	$0.5065(19)  \  \text{ps}^{-1}$  \cite{HFLAV:2022pwe} \tabularnewline
			$\text{BR}(B_s\to \mu^+\mu^-)$ & $(3.66 \pm 0.14 ) \times 10^{-9} $ \cite{Beneke:2019slt} &	$(3.45\pm 0.29) \times 10^{-9}$  \cite{LHCb:2021awg} \tabularnewline 
			$\text{BR}(\bar{B}\to X_s \gamma)$ & $(3.40 \pm 0.17) \times 10^{-4}$ \cite{Misiak:2020vlo} &	$(3.49\pm 0.19) \times 10^{-4}$  \cite{HFLAV:2022pwe}  \tabularnewline
			$R_K$ & $1.00\pm 0.01$ \cite{Bordone:2016gaq}&	$0.949^{+0.042}_{-0.041}\pm0.022$ \cite{LHCb:2021awg}			\tabularnewline	
			$R_{K^{*}}$ & $1.00\pm 0.01$ \cite{Bordone:2016gaq}&	$1.027^{+0.072+ 0.027}_{-0.068-0.026}$ \cite{LHCb:2021awg}			
			\tabularnewline
			\hline 
		\end{tabular}
		\par
	\end{centering}	
	\caption{\label{SM_exp data} The SM predictions and experimental values for flavor-changing observables related to quark sectors.}
\end{table}
 
For the branching ratio BR$(B_s\to \mu^+\mu^-)$, we have the following formula \cite{Mohanta:2005gm},
\be
\mathrm{BR}(B_s \to \mu^+\mu^-)=\frac{\tau_{B_s}}{16\pi^3}\al_{\text{em}}^2 G_F^2 f_{B_s}^2|V_{tb}V^*_{ts}|^2m_{B_s}m_{\mu}^2\sqrt{1-\fr{4m_{\mu}^2}{m^2_{B_s}}}|C_{10}^\mu-C^{'\mu}_{10}|^2,
\ee
where $\tau_{B_s}$ is the lifetime of $B_s$ meson, $\al_{\text{em}}$ is the fine-structure constant, and the WCs are defined as $C_{10}^\mu=C_{10}^{\text{SM}}+C_{10}^{\text{NP},\mu}$, $C^{'\mu}_{10}=C_{10}^{'\text{NP},\mu}$ with $C_{10}^{\text{SM}}$ is the SM WC and given in Table \ref{input-par} and $C_{10}^{(')\text{NP},\mu}=C_{10}^{(')\text{tree},\mu}+C_{10}^{(')\text{box},\mu}$. Due to the effect of $B_s$--$\bar{B}_s$ oscillations, the available experimental value relates to theoretical prediction as \cite{DeBruyn:2012wj}
\be
\mathrm{BR}(B_s \to \mu^+\mu^-)_{\text{exp}}\simeq \fr{1}{1-y_s}\mathrm{BR}(B_s \to \mu^+\mu^-),
\ee
where $y_s=\frac{\Delta \Ga_{B_s}}{2 \Ga_{B_s}}$ and the value of $y_s$ is presented in Table \ref{input-par}.

The branching ratio for the decay $\bar{B}\to X_s\gamma$ is given as \cite{Gambino:2001ew,Buras:2011zb}
\be
\text{BR}(\bar{B}\to X_s\ga)=\fr{6\al_{\text{em}}}{\pi C}\left|\fr{V_{ts}^*V_{tb}}{V_{cb}}\right|^2 \left[|C_7(\mu_b)|^2+|C^{'}_7(\mu_b)|^2+N(E_{\gamma})\right]\text{BR}(\bar{B}\rightarrow X_c e\bar{\nu}) , \label{bra1}
\ee
where $N(E_{\gamma})$ is a non-perturbative contribution which amounts around $4\%$ of the branching ratio. We compute the leading order contribution to $N(E_{\gamma})$ followed the Eq. (3.8) in Ref. \cite{Misiak:2020vlo} and then obtain $N(E_{\ga})\simeq 3.3\times 10^{-3}$. Additionally, $C$ is the semileptonic phase-space factor, $C=|V_{ub}/V_{cb}|^2\Ga(\bar{B}\rightarrow X_c e\bar{\nu}_e)/\Ga(\bar{B}\rightarrow X_u e\bar{\nu}_e)$, and BR$(\bar{B}\rightarrow X_c e\bar{\nu})$ is the branching ratio for semi-leptonic decay. It is necessary to consider the QCD corrections to complete the calculation for this branching ratio. The WCs $C_{7}^{(')}(\mu_b)$ are evaluated at the matching scale $\mu_b=2$ GeV by running down from the higher scale $\mu_{Z'}$ via the renormalization group equations. Its expression can be split as
\bea C_{7}(\mu_b)=C_{7}^{\text{SM}}(\mu_b)+C_{7}^{\text{NP}}(\mu_b),\hs  C^{'}_{7}(\mu_b)=C_{7}^{'\text{NP}}(\mu_b), \eea 
where $C_{7}^{\text{SM}}(\mu_b)$ is the SM WC and have been calculated up to next-to-next-leading order of QCD corrections with the result shown in Table \ref{input-par}. Otherwise, for NP contribution, we have the result at leading order \cite{Buras:2011zb} as 
\bea 
C^{(')\text{NP}}_{7}(\mu_{b})=\ka_7C^{(')\text{loop}}_7(\mu_{Z'})+\ka_8C_{8}^{(')\text{loop}}(\mu_{Z'})+\Delta_{Z'}^{(')\text{current}}(\mu_{b}),
\eea
where the last term stems from the mixing of neutral current-current operators generated by $Z'$ and the dipole operators $\mathcal{O}_{7,8}$. Besides, the coefficients $\ka_{7,8}$ are called NP magic numbers, and their numerical values are given in Ref. \cite{Buras:2011zb}. 

Lepton flavor universality violating (LFUV) observables $R_{K,K^*}$ in the range of squared dilepton mass $q^2=[1.1,6.0] \ \text{GeV}^2$ are defined in terms of new physics WCs $C^{(')\text{NP},e_I}_{9,10}$, given in \cite{Cornella:2021sby},
\bea 
\fr{R_K}{R_{K}^{\text{SM}}} &=&\left\{1+0.24\text{Re}[C^{\text{NP},\mu}_{9}+C^{'\text{NP},\mu}_{9}]-0.26\text{Re}[C^{\text{NP},\mu}_{10}+C^{'\text{NP},\mu}_{10}]+0.03\left(|C^{\text{NP},\mu}_9+C^{'\text{NP},\mu}_{9}|^2\right.\right.\crn
&&\left.\left.+|C^{\text{NP},\mu}_{10}+C^{'\text{NP},\mu}_{10}|^2\right)\right\}\left\{1+0.24\text{Re}[C^{\text{NP},e}_{9}+C^{'\text{NP},e}_{9}]-0.26\text{Re}[C^{\text{NP},e}_{10}+C^{'\text{NP},e}_{10}]\right.\crn
&&\left.+0.03\left(|C^{\text{NP},e}_9+C^{'\text{NP},e}_{9}|^2+|C^{\text{NP},e}_{10}+C^{'\text{NP},e}_{10}|^2\right)\right\}^{-1},\\
\fr{R_{K^*}}{R_{K^*}^{\text{SM}}} &=&\left\{1+0.18\text{Re}[C^{\text{NP},\mu}_{9}-C^{'\text{NP},\mu}_{9}]-0.29\text{Re}[C^{\text{NP},\mu}_{10}-C^{'\text{NP},\mu}_{10}]+0.03\left(|C^{\text{NP},\mu}_9-C^{'\text{NP},\mu}_{9}|^2\right.\right.\crn
&&\left.\left.+|C^{\text{NP},\mu}_{10}-C^{'\text{NP},\mu}_{10}|^2\right)\right\}\left\{1+0.18\text{Re}[C^{\text{NP},e}_{9}-C^{'\text{NP},e}_{9}]-0.29\text{Re}[C^{\text{NP},e}_{10}-C^{'\text{NP},e}_{10}]\right.\crn
&&\left.+0.03\left(|C^{\text{NP},e}_9-C^{'\text{NP},e}_{9}|^2+|C^{\text{NP},e}_{10}-C^{'\text{NP},e}_{10}|^2\right)\right\}^{-1}.
 \eea 
We also need to take into account QCD corrections here. At the leading order, the $C^{('),e_I}_{9,10}$ are shifted by $\ep \simeq \fr{\al_s}{4\pi}\ln{(m_{Z'}/m_b)}$ where $\al_s$ is the strong coupling at scale $m_{Z'}$. This effect of QCD corrections modifies the value of WCs by around a few percent with $m_{Z'}\sim \mathcal{O}(1) \ \text{TeV} \gg m_b $.  However, the effect of QCD correction is insignificant in the ratios $R_{K,K^*}$ because they are small and canceled between the numerator and the denominator of these ratios. Therefore, in this work, we ignore the effect of QCD corrections in $R_{K,K^*}$ and BR$(B_s\to \mu^+\mu^-)$. 
  
All observables mentioned above should be compared with the experimental values in the last column in Table \ref{SM_exp data}. It is important to note that the central values of SM prediction and the measurement results of these observables are very close. However, the uncertainties in SM prediction are quite large, especially in meson mass differences, compared to experimental ones. Therefore, it is better to consider the ratio between SM and respective experimental values on each observable since the uncertainties can be canceled via the numerator and the denominator of these ratios. Hence, we obtain constraints for $B^0_{s,d}$--$\bar{B}^0_{s,d}$ meson systems as 
\be
 \fr{(\Delta m_{B_d})_{\mathrm{SM}}}{(\Delta m_{B_d})_{\mathrm{exp}}}
 =1.0721(1\pm0.0535), \hs 
\fr{(\Delta m_{B_s})_{\mathrm{SM}}}{(\Delta m_{B_s})_{\mathrm{exp}}} =1.0566(1\pm0.0458),  \ee
which are equivalent  
\bea \fr{(\Delta m_{B_d})_{\text{NP}}}{(\Delta m_{B_d})_{\mathrm{exp}}}\in[-0.1295,-0.0147],   \hs \fr{(\Delta m_{B_s})_{\text{NP}}}{(\Delta m_{B_s})_{\mathrm{exp}}}\in[-0.105,-0.0082] . \label{constraint2}  \eea
However, in $K^0$--$\bar{K}^0$ meson system, the lattice QCD calculations for long-distance effect are not well-controlled. Therefore, we assume the present theory contributes about 30\% to $\Delta m_K$, it reads    
\bea 
\fr{(\Delta m_{K})_{\text{SM}}}{(\Delta m_{K})_{\text{exp}}}=1(1\pm0.3), \eea 
and then translates to the following constraint
\bea  \fr{(\Delta m_{K})_{\text{NP}}}{(\Delta m_{K})_{\text{exp}}}\in [-0.3,0.3], 
\label{constraint1}\eea
in agreement with \cite{Buras:2015kwd}. For the branching ratios BR$(B_s\to \mu^+\mu^-)$ and BR$(\bar{B}\to X_s\gamma)$, we have constraints as
\bea \fr{\text{BR}(B_s\to \mu^+\mu^-)_{\text{exp}}}{\text{BR}(B_s\to \mu^+\mu^-)_{\text{SM}}}&=&\fr{1}{1-y_s}\fr{|C^{\mu}_{10}-C^{'\mu}_{10}|^2}{|C_{10}^{\text{SM}}|^2}\crn 
&=&0.9426(1\pm 0.0924), \label{Bsmm_constraint}\\
\fr{\text{BR}(\bar{B}\to X_s\gamma)_{\text{exp}}}{\text{BR}(\bar{B}\to X_s\gamma)_{\text{SM}}} &=& 1+\fr{|C_7^{\text{NP}}|^2+|C_7^{'\text{NP}}|^2+2C_7^{\text{SM}}\text{Re}[C^{\text{NP}}_7]}{|C_7^{\text{SM}}|^2+N(E_{\ga})}\crn 
&=& 1.0265(1\pm0.0739) \label{bsga_constraint}.
\eea

\subsection{Lepton flavor phenomenologies }
\begin{figure}[h!]
	\centering
	\begin{tabular}{c}
\includegraphics[width=13.0cm]{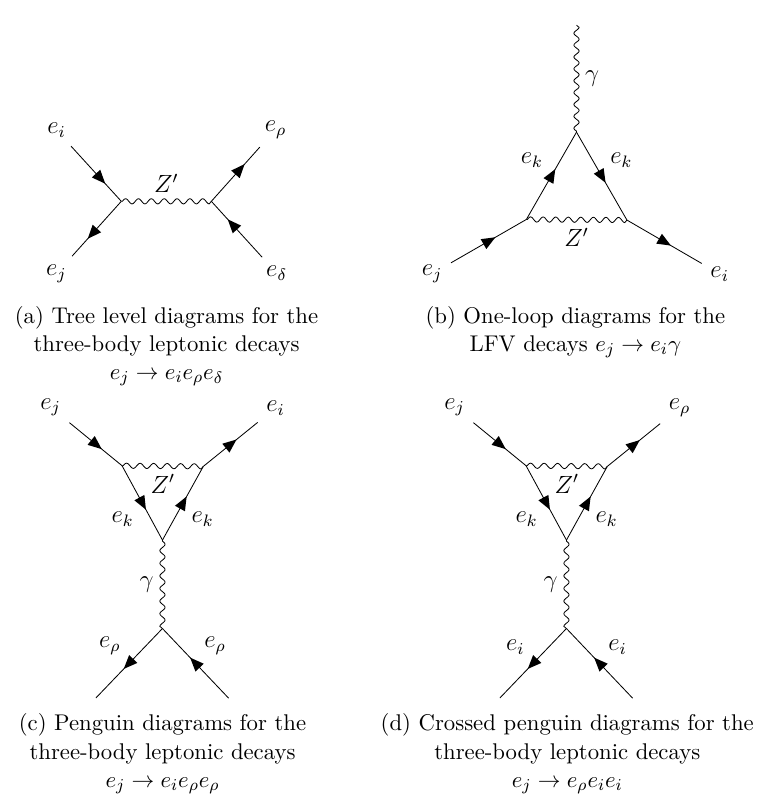}
	\end{tabular}
	\caption{\label{l3lp}Feynman diagrams for three-body leptonic and LFV decays.}
\end{figure}

For the lepton flavor violating (LFV) decays $e_j \to e_i\gamma$ with $e_{i,j}=\{e_1,e_2,e_3\}=\{e,\mu,\tau\}$ and $e_i\neq e_j$, we have the following the effective Hamiltonian contributing by new neutral gauge boson $Z'$ at the one-loop level 
\bea
\mathcal{H}^{\text{lepton}}_{\text{eff}}=C^{ij}_L\bar{e}_i\sigma_{\mu\nu}P_L e_jF^{\mu\nu}+(L\to R), \label{leptonfcnc}\eea 
where the coefficients $C^{ij}_{L,R}$ are obtained by calculating one-loop diagrams containing the SM charged leptons $e_k =\{e_1,e_2,e_3\}=\{e,\mu,\tau\}$ and new neutral gauge boson $Z'$ as internal lines, see subfigure (b) of Fig. \ref{l3lp}. Here we calculate these diagrams in the limit $m_{e_k}^2/ m_{Z'}^2\ll 1$ and keeping the masses of external leptons $m_{e_{i,j}}$, similar to the quark flavor section. We obtain the expressions for these coefficients as 
\bea C_{L}^{ij} &=& \fr{e}{3(4\pi)^2m_{Z'}^2}\sum_{k=1}^3(m_{e_j}\Ga^{*e_R}_{ki}\Ga^{e_R}_{kj}-3m_{e_k}\Ga^{*e_R}_{ki}\Ga^{e_L}_{kj}+m_{e_i}\Ga^{*e_L}_{ki}\Ga^{e_L}_{kj}), \\ 
C_{R}^{ij} &=& \fr{e}{3(4\pi)^2m_{Z'}^2}\sum_{k=1}^3(m_{e_j}\Ga^{*e_L}_{ki}\Ga^{e_L}_{kj}-3m_{e_k}\Ga^{*e_L}_{ki}\Ga^{e_R}_{kj}+m_{e_i}\Ga^{*e_R}_{ki}\Ga^{e_R}_{kj}),\label{WCs_lepton}\eea 
where $\Ga_{ij}^{e_{L,R}}$ are the LFV couplings given in Eq. (\ref{fcnc}). The branching ratios of the LFV decays are determined by \cite{Crivellin:2018qmi}
\be \text{BR}(e_j\to e_j \gamma)=\fr{(m_{e_j}^2-m_{e_i}^2)^3}{4\pi m_{e_j}^3\Ga_{e_j}}(|C^{ij}_L|^2+|C^{ij}_R|^2), \ee
where $\Ga_{e_j}$ is the total decay width of decaying lepton $e_j$. 

Besides, the effective Hamiltonian in Eq. (\ref{leptonfcnc}) also contributes to branching ratios of three-body leptonic decays such as $\tau\to 3\mu (3e)$, $\tau \to e\mu\mu(ee\mu)$, and $\mu \to 3e$. There are three contributions to these observables, including the tree-level shown in subfigure (a) of Fig. \ref{l3lp} with the following operators 
\bea \mathcal{O}_{ab,\al\bet}^{L(R)L(R)}=(\bar{e}_i\ga_{\mu}P_{L(R)}e_j)(\bar{e}_{\rho}\ga^{\mu}P_{L(R)}e_{\delta}), \label{tree_ope}
 \eea
where $e_{i,j}=\{e_2,e_3\}=\{\mu,\tau\}$, $i\neq j$, and $e_{\rho,\delta}=\{e_1,e_2\}=\{e,\mu\}$. Note that these operators are also generated by the SM $Z$-like boson but suppressed due to small $Z$--$Z'$ mixing. This setup also does not allow the LFV decays of $Z$ boson, namely $Z\to e_ie_j$. Besides the tree level, the dipole operators in Eq. (\ref{leptonfcnc}) also generate the three-body decays via penguin diagrams, as shown in subfigures (c) and (d) of Fig. \ref{l3lp}. Furthermore, there are one-loop contributions that arise from the mixing of tree-level operators defined in Eq. (\ref{tree_ope}) with ``hidden" operators that do not trigger flavor violating decays at the tree level but do so in QED penguin diagrams, such as $\mathcal{O}^{L(R),L(R)}_{e\mu,\mu\mu(\tau\tau)},\mathcal{O}^{L(R),L(R)}_{e\tau,\tau\mu(\tau\tau)},\mathcal{O}^{L(R),L(R)}_{\mu\tau,\tau\tau}$ \cite{Buras:2021btx}. The branching ratios of three-body leptonic decays, including all mentioned contributions, were explicitly given in Ref. \cite{Buras:2021btx}.  

On the other hand, for the lepton flavor conversing (LFC) observables including the electron and muon anomalous magnetic moments $\Delta a_{e,\mu}$ and the electric dipole moments $d_{e,\mu}$, we have the following formulas \cite{Crivellin:2018qmi},
\bea \Delta a_{e_I} &=& -\fr{4m_{e_I}}{e}\text{Re}[C_R^{II}],\label{lfc1} \\ 
d_{e_I} &=& -2\text{Im}[C_R^{II}],\hs I=1,2.  \label{lfc}\eea 

The LFV couplings of $Z'$ also cause a transition of muonium (Mu: $\mu^+e^-$) into antimuonium ($\overline{\text{Mu}}$: $\mu^-e^+$), which resembles the $K^0$-$\bar{K}^0$ mixing in the quark sector. The effective Lagrangian for this process can be written as
\be
\mathcal{L}_{\text{Mu}-\overline{\text{Mu}}}=-\sum_{i=1}^3\fr{\mathcal{G}_i}{\sqrt{2}}\mathcal{Q}_i, 
\ee 
where the coefficients and corresponding operators are given by 
\bea 
\mathcal{Q}_1 &=&(\bar{\mu}\ga_{\mu}(1-\ga_{5})e)(\bar{\mu}\ga^{\mu}(1-\ga_{5)}e),\hs \mathcal{G}_1=\fr{|\Ga^{l_L}_{\mu e}|^2}{4\sqrt{2}m_{Z'}^2},\crn
\mathcal{Q}_2 &=&(\bar{\mu}\ga_{\mu}(1+\ga_{5})e)(\bar{\mu}\ga^{\mu}(1+\ga_{5})e), \hs \mathcal{G}_2=\fr{|\Ga^{l_R}_{\mu e}|^2}{4\sqrt{2}m_{Z'}^2},\crn
\mathcal{Q}_3 &=&(\bar{\mu}\ga_{\mu}(1+\ga_{5})e)(\bar{\mu}\ga^{\mu}(1-\ga_{5})e) , \hs \mathcal{G}_3=\fr{\Ga^{l_L}_{\mu e}\Ga^{l_R,*}_{\mu e}}{2\sqrt{2}m_{Z'}^2}.
\eea 
Additionally, there are operators $\mathcal{Q}_{4,5}=(\bar{\mu}(1\mp\ga_{5})e)(\bar{\mu}(1\mp\ga_{5})e)$ contributing by neutral Higgs bosons, however these contributions are negligible, compared to the ones of $Z'$. In the presence of an external magnetic field $B$, the time-integrated probability of the Mu-to-$\overline{\text{Mu}}$ transition is given by \cite{Fukuyama:2021iyw}
\be
P(\text{Mu}\to \overline{\text{Mu}}) =2\tau^2\left(|c_{0,0}|^2|\mathcal{M}^B_{0,0}|^2+|c_{1,0}|^2|\mathcal{M}^B_{1,0}|^2+\sum_{m=\pm 1}|c_{1,m}|^2\fr{|\mathcal{M}_{1,m}|^2}{1+(\tau\Delta E)^2}\right),
\ee
where $\tau\simeq 2.2\times 10^{-6}~ s$ is the Mu lifetime, $|c_{F,m}|^2$ denotes the population of Mu$(F,m)$ state, and $\mathcal{M}^B_{F,m}$ is the amplitude of the Mu$(F,m)\to\overline{\text{Mu}}(F,m)$ transition.\footnote{In practice, the state of the produced Mu is a mixture of four states labeled by the magnitude of total angular momentum $F$ and the $z$-component of total angular momentum $m$, i.e., $(F,m)=(0,0),(1,0)$, and $(1,\pm 1)$.} Additionally,  $\Delta E$ is the energy splitting between $(1,1)$ and $(1,-1)$ states. Notice that the transition probability for $(1,\pm 1)$ states is suppressed for $B\gtrsim \mathcal{O}(10^{-6})$ Tesla. In this case, the total transition probability reads 
\bea
P(\text{Mu}\to \overline{\text{Mu}}) &\simeq & \fr{2.572\times 10^{-5}}{G_F^2}\left(|c_{0,0}|^2\left|\mathcal{G}_3-\fr{\mathcal{G}_1+\mathcal{G}_2-0.5\mathcal{G}_3}{\sqrt{1+X^2}}\right|^2\right . \crn && \left . +|c_{1,0}|^2\left|\mathcal{G}_3+\fr{\mathcal{G}_1+\mathcal{G}_2-0.5\mathcal{G}_3}{\sqrt{1+X^2}}\right|^2\right),
\eea 
where $X$ denotes the magnetic flux density. The experimental result reported by the PSI experiment for the Mu-to-$\overline{\text{Mu}}$ transition under $B=0.1$ Tesla is $P(\text{Mu}\to \overline{\text{Mu}}) <8.3\times 10^{-11}$ \cite{Willmann:1998gd}. Taking $X=6.31\times B/\text{Tesla}$, $|c_{0,0}|^2=0.32$, and $|c_{1,0}|^2=0.18$, the experimental result is decoded as 
\be
0.64\left|\mathcal{G}_1+\mathcal{G}_2-1.68\mathcal{G}_3\right|^2 +0.36\left|\mathcal{G}_1+\mathcal{G}_2+0.68\mathcal{G}_3\right|^2<9\times 10^{-6}G_F^2.  
\ee 

The LFV couplings of $Z'$ also contribute to the muon-to-electron conversion in a muonic atom. Specifically, we focus on the coherent conversion processes in which the nucleus's initial and final states are the same and the non-photonic processes at the tree level mediated by $Z'$, which are described by the following effective Lagrangian \cite{Kuno:1999jp,Kitano:2002mt,Cirigliano:2009bz}
\be
\mathcal{L}_{\mu\to e}=-\sum_{q=u,d}(C^q_{VL}\bar{e}\ga_{\nu}P_L\mu+C^q_{VR}\bar{e}\ga_{\nu}P_R\mu)\bar{q}\ga^{\nu}q+\mathrm{H.c.},
\ee 
where the coefficients $C^q_{VL,VR}$ are
\be 
C^q_{VL}=\fr{gg^{Z'}_V(q)\Ga^{l_L}_{e\mu}}{2c_Wm_{Z'}^2}, \hs C^q_{VR}=\fr{gg^{Z'}_V(q)\Ga^{l_R}_{e\mu}}{2c_Wm_{Z'}^2}.
\ee  
Here, the operators involving $\bar{q}\gamma^\nu\gamma_5q$ are omitted since they do not contribute to the coherent conversion processes. To evaluate the conversion rate, it is appropriate to use the effective Lagrangian at the nucleon level, such as
\be \mathcal{L}_{\mu\to e}=-\sum_{\mathcal{N}=p,n}(C^\mathcal{N}_{VL}\bar{e}\ga_{\nu}P_L\mu +C^\mathcal{N}_{VR}\bar{e}\ga_{\nu}P_R\mu)\bar{\psi}_N\ga^{\nu}\psi_N+\mathrm{H.c.}\ee
with $\psi_N$ to be the nucleon fields, and
\bea C^p_{VL}&=&2C^u_{VL}+C^d_{VL},\hs C^p_{VR}=2C^u_{VR}+C^d_{VR},\\
C^n_{VL}&=&C^u_{VL}+2C^d_{VL},\hs C^n_{VR}=C^u_{VR}+2C^d_{VR}.\eea
The $\mu\to e$ conversion branching ratio in a target of atomic nuclei $N$ can then be written as \cite{Kitano:2002mt,Cirigliano:2009bz} 
\be
\text{BR}(\mu~N\to e~N)=4m_{\mu}^5\left[|C^p_{VL}V^p_N+C^n_{VL}V^n_N|^2+|C^p_{VR}V^p_N+C^n_{VR}V^n_N|^2\right]/\Ga^{N}_{\text{capt}},
\ee
where $\Ga_{\text{capt}}^N$ is the total capture rate, $V^{p,n}_N$ are related to the overlap integrals between the lepton wave functions and the nucleon densities, depending on the nature of the target $N$. For instance, we consider the $\mu \to e$ conversion captured by Au nuclei, we have $V^p_{\text{Au}}=0.0974$ and $V^n_{\text{Au}}=0.146$ \cite{Kitano:2002mt}, $\Ga_{\text{capt}}^{\text{Au}}\simeq 8.7\times 10^{-18}$ GeV \cite{Suzuki:1987jf}, and the current experimental limit $\text{BR}(\mu~\text{Au}\to e~\text{Au})\leq 7.0\times 10^{-13}$ \cite{SINDRUMII:2006dvw}.

All predicted observables above should be compared with experimental results listed in Table \ref{lepton_exp}. It is straightforward to recognize that the contribution of the $Z'$ gauge boson with a mass at several TeVs implied by the collider searches (discussed below) to the anomalous magnetic moments, especially for $\Delta a_{\mu}$, be quite suppressed in comparison with other leptonic observables. Indeed, from Eqs. (\ref{WCs_lepton}) and (\ref{lfc1}), it is easy to see that $\Delta a_{\mu}$ is proportional with a factor, $-\fr{4m_{\mu}}{e}\fr{e}{48\pi^2m_{Z'}^2}\sim \mathcal{O}(10^{-11}-10^{-10})$, while the internal terms, Re$[m_{e_k}\Ga^{e_{L(R)}}_{k2}\Ga^{*e_{L(R)}}_{k2}]\sim \mathcal{O}(10^{-1}-10^0)$. Therefore, our model predicts $\Delta a_{\mu}\sim \mathcal{O}(10^{-12}-10^{-11})$,  remarkably smaller than experimental result $\Delta a_{\mu}^{\text{exp}}\sim \mathcal{O}(10^{-9})$ \cite{Muong-2:2023cdq}. In the following numerical analysis, we will investigate the branching ratios of LFV, the three-body leptonic decays, the electric dipole moments, the muonium-to-antimuonium transition, and the muon-to-electron conversion. 
	\begin{table}[h!]
	\begin{centering}
		\begin{tabular}{|c|c|c|c|}
			\hline
			LFV Observables & Experimental limits  & LFC Observables & Experimental limits  
			\tabularnewline
			\hline 
			BR$ (\mu \to e\ga)$ & $\leq 4.2\times 10^{-13}$  \cite{SINDRUMII:2006dvw,BaBar:2009hkt,MEG:2016leq} &  $\Delta a_e^{\text{Cs}}$ & $-0.88(36)\times 10^{-12} $ \cite{ParticleDataGroup:2022pth}
			\tabularnewline
			BR$(\tau\to e\ga)$ & $\leq 3.3\times 10^{-8}$ \cite{SINDRUMII:2006dvw,BaBar:2009hkt,MEG:2016leq} 	 &  $ \Delta a_{e}^{\text{Rb}}$ & $0.48(30)\times 10^{-12}$  \cite{ParticleDataGroup:2022pth}
			\tabularnewline
			BR$ (\tau\to \mu\ga)$ & $\leq 4.4\times 10^{-8}$  \cite{SINDRUMII:2006dvw,BaBar:2009hkt,MEG:2016leq}  &  $\Delta a_{\mu}$ & $249(48)\times 10^{-11}$ \cite{Muong-2:2023cdq} \tabularnewline
			BR$ (\mu^-\to e^-e^+e^-)$ & $\leq 1.0\times 10^{-12}$  \cite{ParticleDataGroup:2022pth} &  $|d_e|$ & $< 1.1\times 10^{-29}e$ cm  \cite{ACME:2018yjb}\tabularnewline
			BR$(\tau^-\to e^-e^+e^-)$ & $\leq 1.4\times 10^{-8} $ \cite{ParticleDataGroup:2022pth} & $ |d_{\mu}|$ & $<1.9\times 10^{-19}e$ cm  \cite{Muong-2:2008ebm} 
			\tabularnewline
			BR$(\tau^-\to e^-\mu^+\mu^-)$ & $\leq 1.6\times 10^{-8} $\cite{ParticleDataGroup:2022pth}  & & \tabularnewline
			BR$(\tau^-\to \mu^-e^+\mu^-)$ & $\leq 9.8\times 10^{-9} $ \cite{ParticleDataGroup:2022pth} & & 
			\tabularnewline
			BR$(\tau^-\to e^-\mu^+e^-)$ & $\leq 8.4\times 10^{-9} $ \cite{ParticleDataGroup:2022pth}& & \tabularnewline
			BR$(\tau^-\to \mu^-\mu^+\mu^-)$ & $\leq 1.1\times 10^{-8} $ \cite{ParticleDataGroup:2022pth}& &  \tabularnewline
			BR$(\mu~\text{Au}\to e~\text{Au})$& $\leq 7.0\times 10^{-13}$ \cite{SINDRUMII:2006dvw} &&\tabularnewline
			$P(\text{Mu}\to \overline{\text{Mu}})$&$<8.3\times 10^{-11}$ \cite{Willmann:1998gd}& &\tabularnewline
		\hline 
		\end{tabular}
		\par
			\protect\caption{\label{lepton_exp} Experimental results for leptonic flavor observables.}
	\end{centering}
\end{table} 
 
\subsection{Numerical results}
In this subsection, we will use the values of known input parameters from Tables \ref{commonSMvals}, \ref{input-par}, and \ref{input-par-lepton} for our numerical study. For the lepton flavor phenomenologies, we randomly seed the free parameters $z,s_{23}^{e_R},k$, $\La_2$, and $\theta$ in ranges as
\be z\in [-1,1],  \hs s_{23}^{e_R}\in [0,1], \hs \theta \in [0,\pi/2], \hs k\in[1,10], \hs \La_2 \in [1,50] \ \text{TeV}. \label{lepton_interval}\ee
Besides, we also compare the results of four relation scenarios of lepton mixing angles shown in Eqs. (\ref{lepton_relation1})--(\ref{lepton_relation2}). 

\begin{figure}[h!]
	\begin{subfigure}{0.49\textwidth}
	\includegraphics[width=\textwidth]{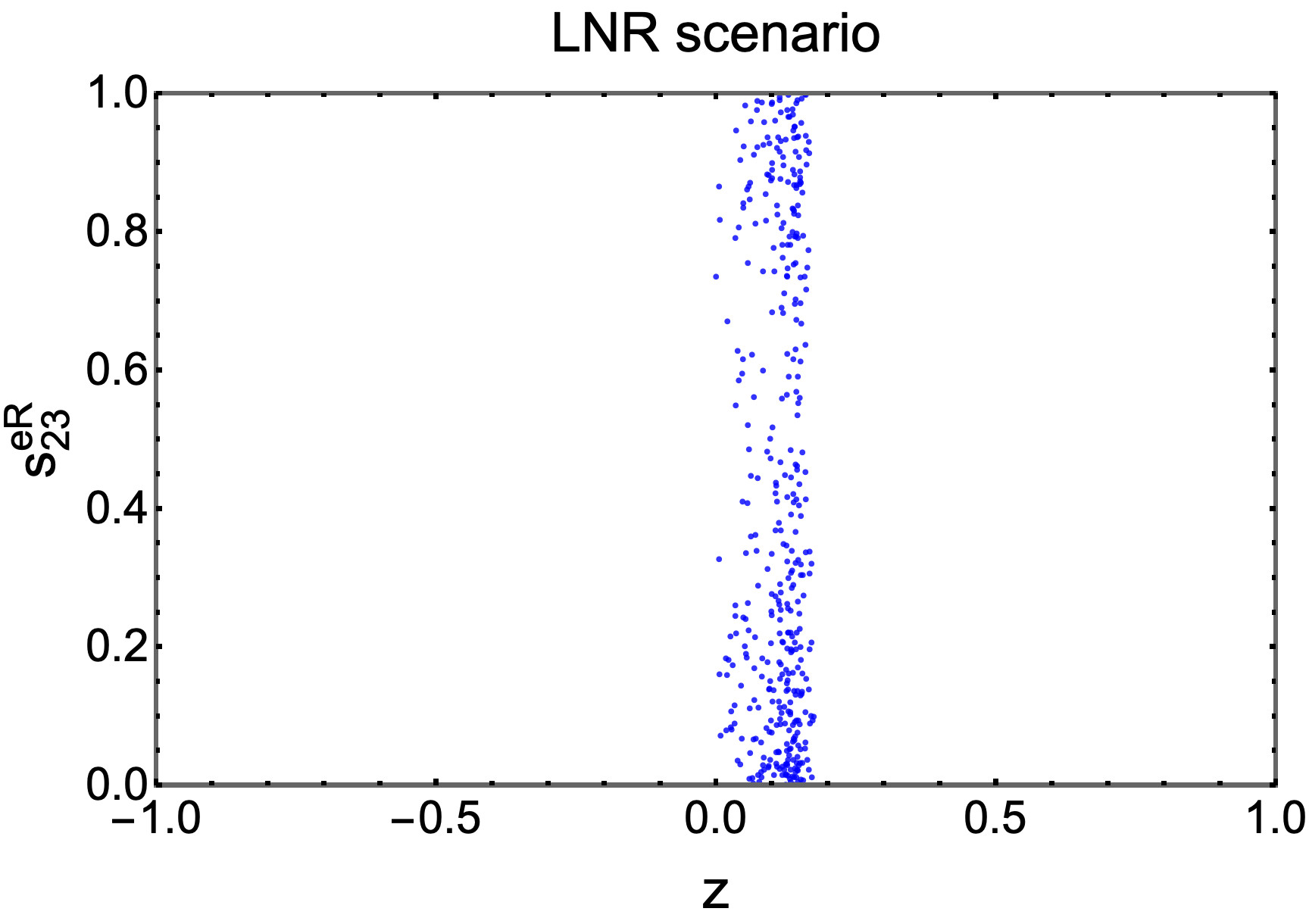}
	\end{subfigure}
	\hfill
	\begin{subfigure}{0.49\textwidth}
	\includegraphics[width=\textwidth]{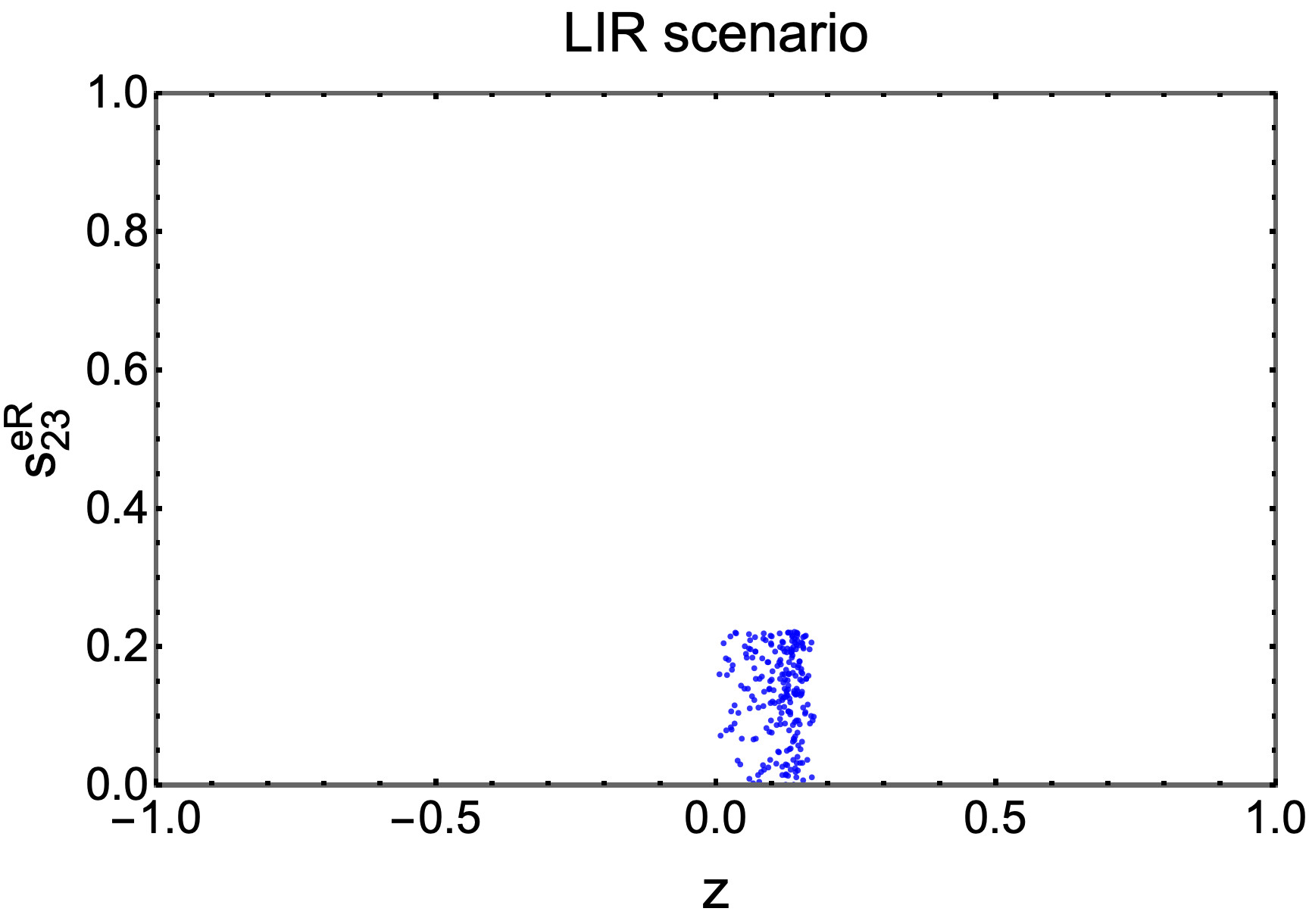}
	\end{subfigure}
	\medskip
	\begin{subfigure}{0.49\textwidth}
	\includegraphics[width=\textwidth]{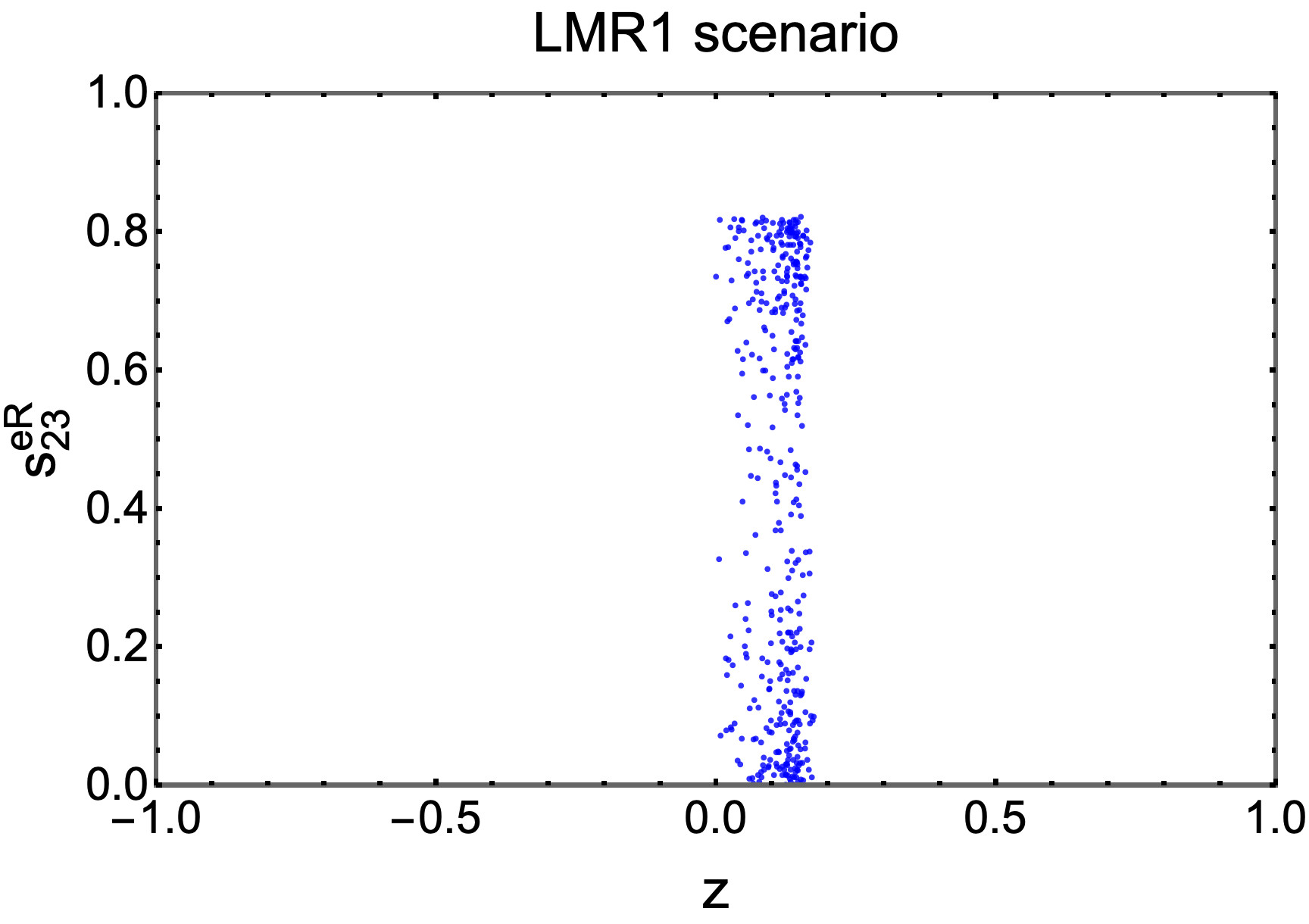}
	\end{subfigure}
	\hfill
	\begin{subfigure}{0.49\textwidth}
	\includegraphics[width=\textwidth]{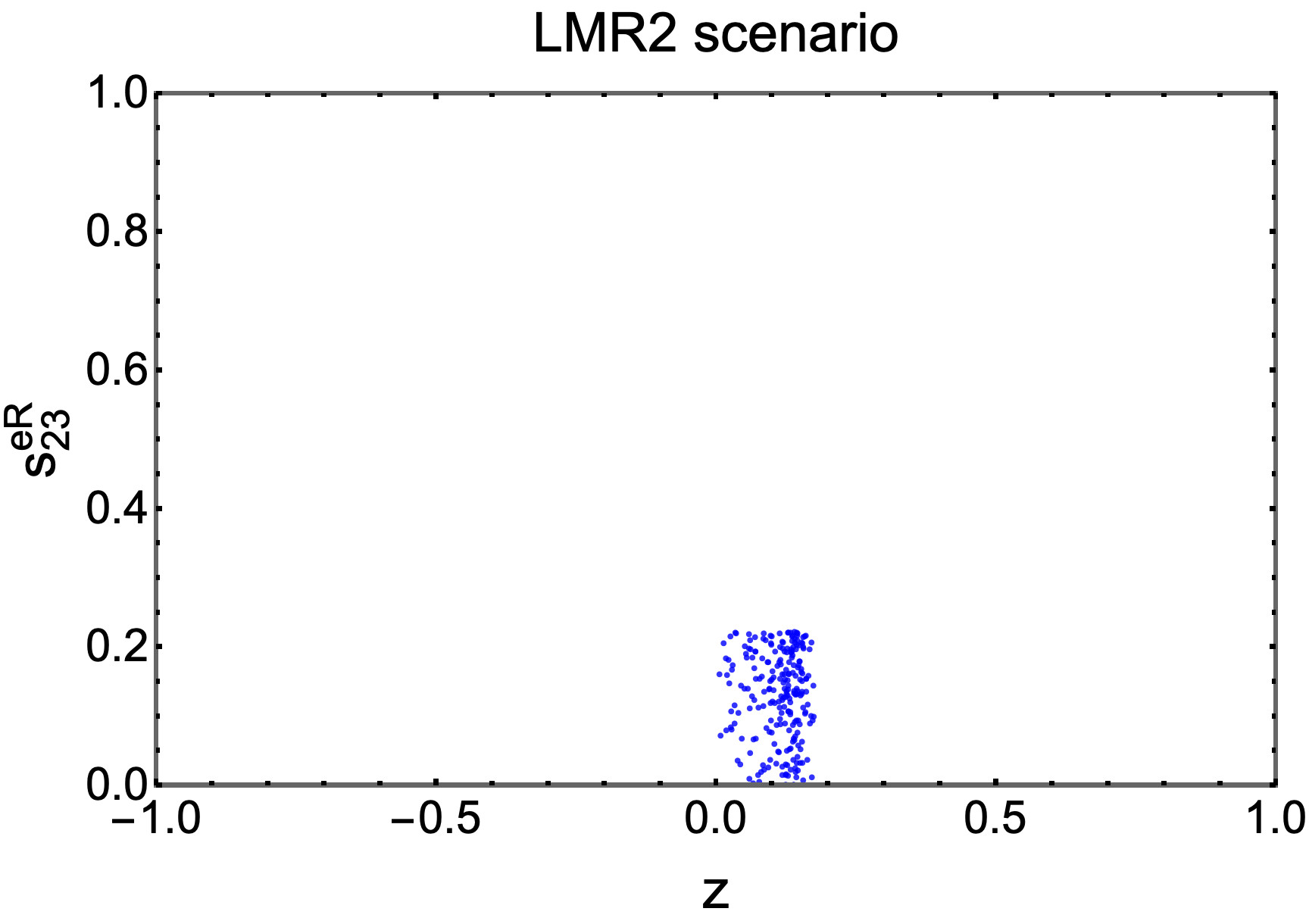}
	\end{subfigure}
	\caption{The correlations between mixing angle $s_{23}^{e_R}$ with charge parameter $z$ in four relation scenarios of lepton mixing angles. }
	\label{correlation_s23_vs_z_Lamda2big}
\end{figure}

We first obtain the correlation between mixing angle $s_{23}^{e_R}$ and charge parameter $z$ satisfying all constraints of leptonic observables within four relation scenarios of lepton mixing angles as in Fig. \ref{correlation_s23_vs_z_Lamda2big}. It is noteworthy that all the relation scenarios potentially fulfill the constraints, and the viable range of $z$ is $2.41 (6.55)\times 10^{-4}\lesssim z\lesssim 0.175 $ for the LNR and LMR1 (LIR and LMR2) scenarios. In addition, the whole range $s_{23}^{e_R}$ pleases the constraints in the LNR scenario. In contrast, the remaining scenarios accept only a partial range of $s_{23}^{e_R}$, namely, $0 \lesssim s_{23}^{e_R}\lesssim 0.82$ for the LMR1 scenario and  $0 \lesssim s_{23}^{e_R}\lesssim 0.22$ for the LIR and LMR2 scenarios. Notice that the LIR and LMR2 scenarios have $s_{13}^{e_R}=s_{23}^{e_R} s_{23}^{\text{PMNS}}/s_{13}^{\text{PMNS}}\simeq 4.476s_{23}^{e_R}$ \cite{Gonzalez-Garcia:2021dve}, and therefore $s_{23}^{e_R}$ in these two scenarios is constrained by an additional condition of $s_{13}^{e_R}\leq 1$. Furthermore, the LIR and LMR2 scenarios have an inverse relationship between $s_{12}^{e_R}$ and $s_{23}^{e_R}$. Therefore, the nearly identical panels of these scenarios also illustrate that the leptonic observables do not significantly rely on $s_{12}^{e_R}$, but primarily on $s_{23}^{e_R}$. This behavior is also applied to the LNR and LMR1 scenarios since they have the same $s_{13}^{e_R}$ whereas $s_{12}^{e_R}$ is changed, but the result is not modified remarkably.  

\begin{figure}[h!]
	\begin{subfigure}{0.49\textwidth}
	\includegraphics[width=\textwidth]{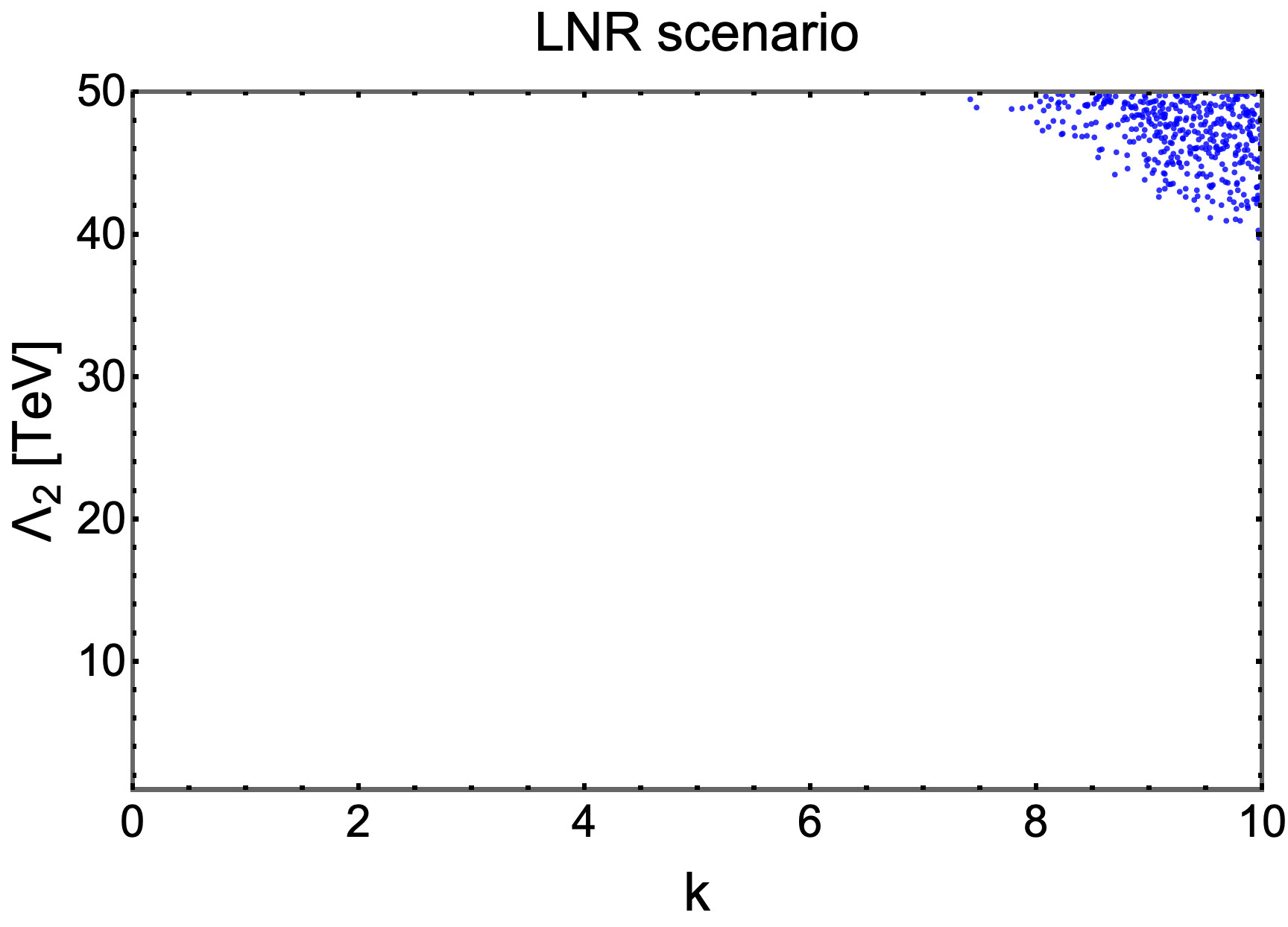}
	\end{subfigure}
	\hfill
	\begin{subfigure}{0.49\textwidth}
	\includegraphics[width=\textwidth]{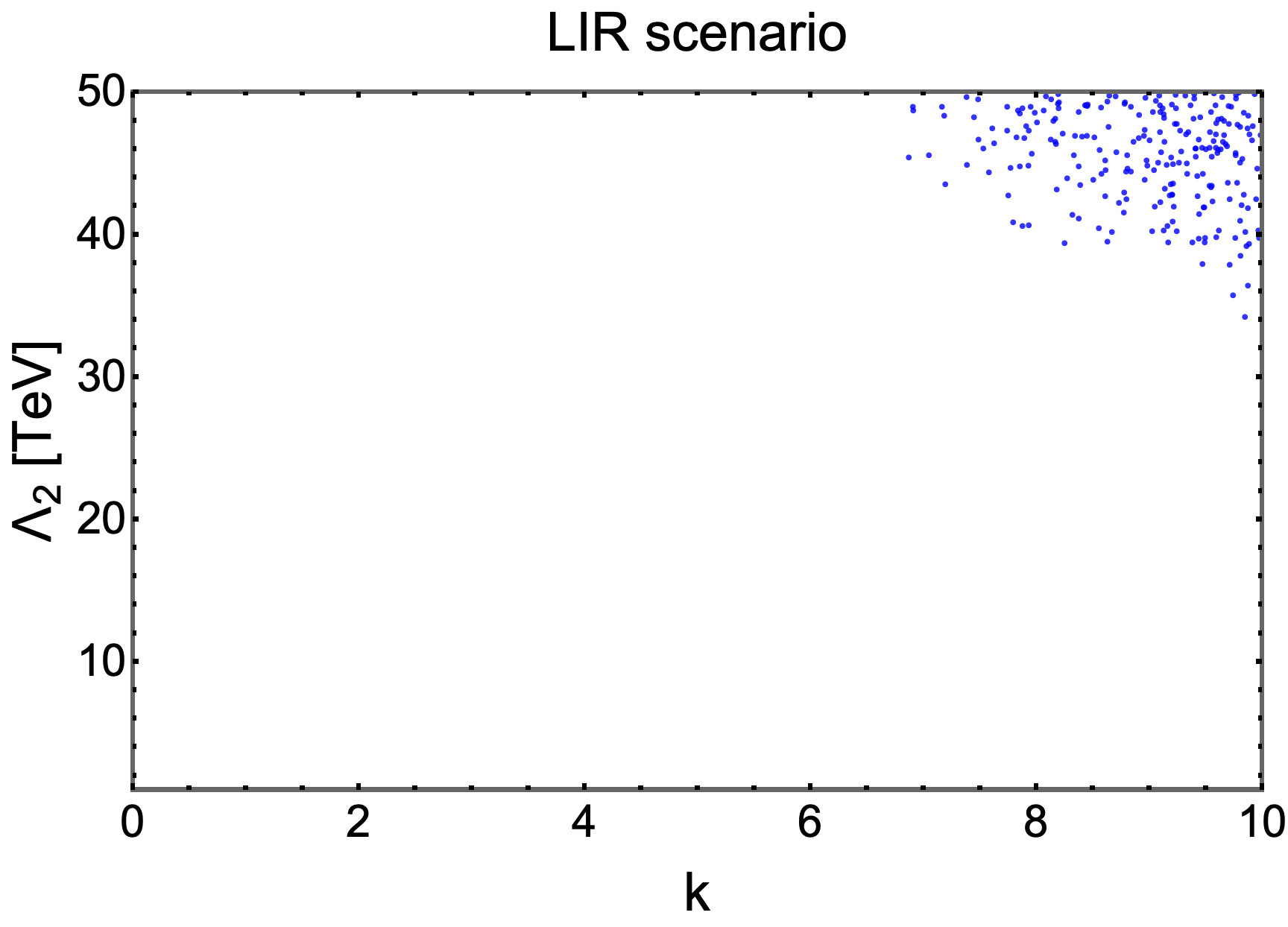}
	\end{subfigure}
	\medskip
	\begin{subfigure}{0.49\textwidth}
	\includegraphics[width=\textwidth]{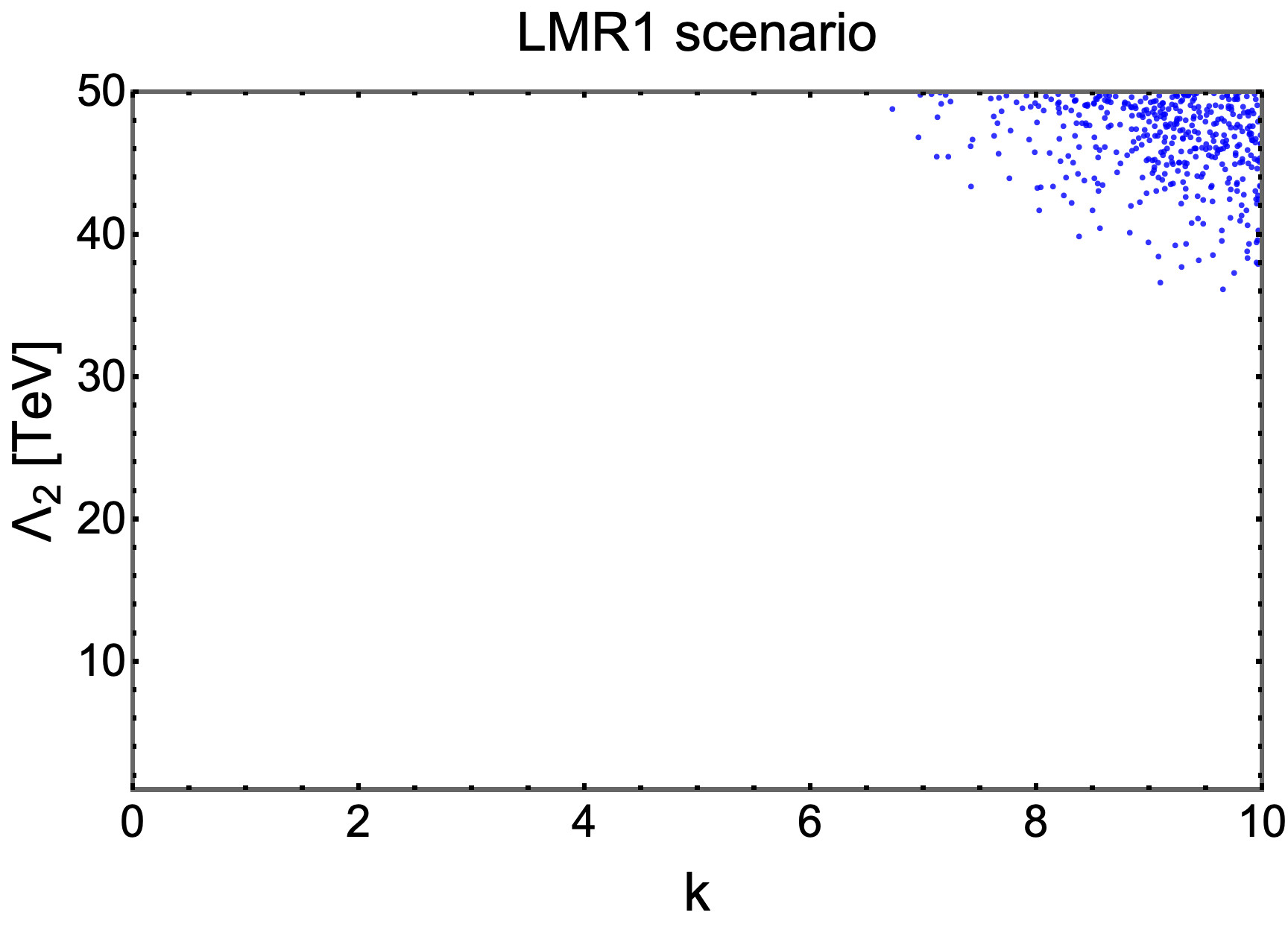}
	\end{subfigure}
	\hfill
	\begin{subfigure}{0.49\textwidth}
	\includegraphics[width=\textwidth]{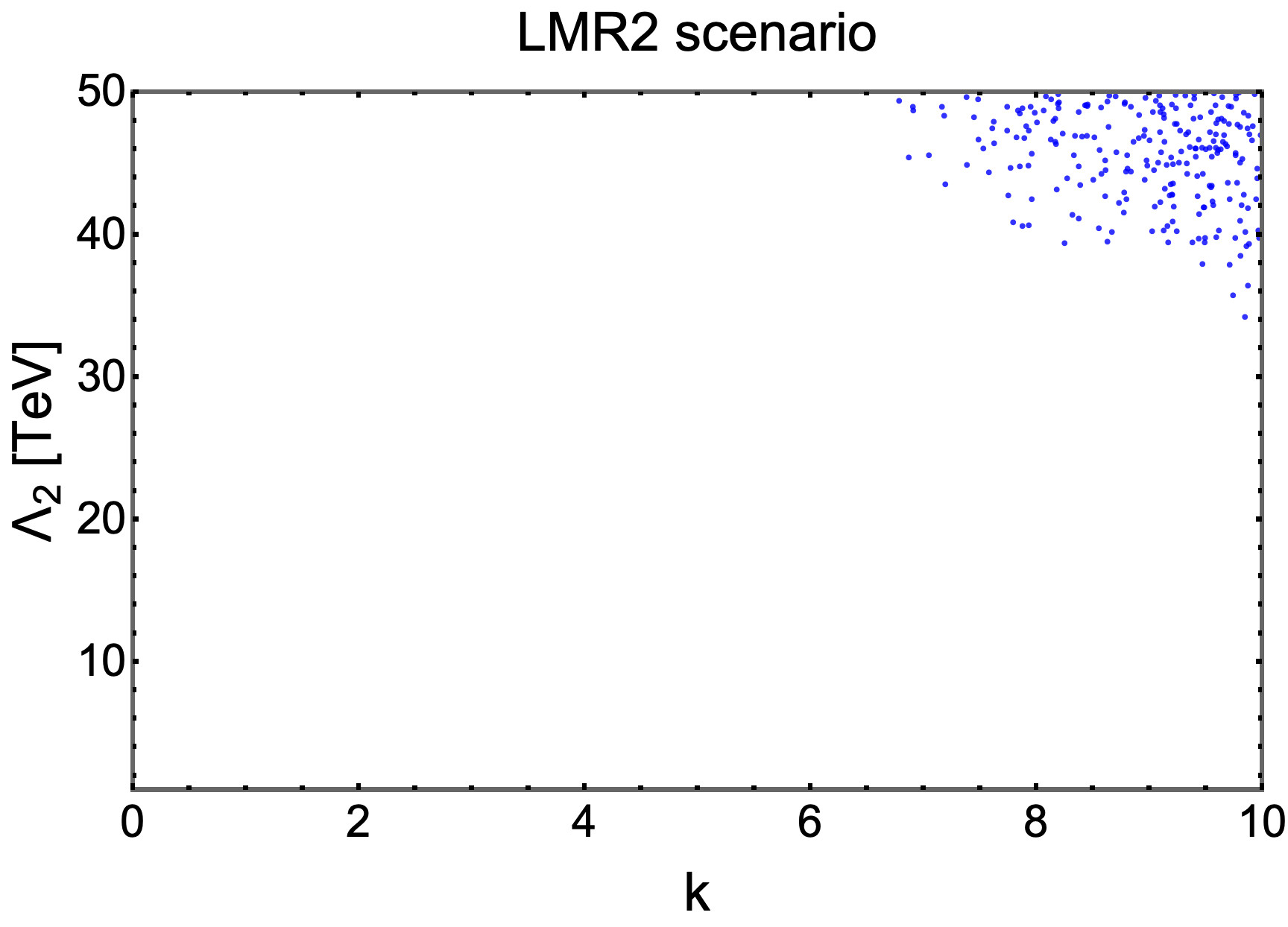}
	\end{subfigure}
	\caption{The correlations between the ratio $k=\La_1/\La_2$ with VEV $\Lambda_2$ for the four relation scenarios of lepton mixing angles. }
	\label{correlation_k_vs_lamda2}
\end{figure}

Furthermore, we obtain the correlation between the ratio $k=\La_1/\La_2$ and VEV $\La_2$ within four relation scenarios of lepton mixing angles, as respectively shown in four panels of Fig. \ref{correlation_k_vs_lamda2}. We see that the viable points in the four panels are distributed in the regions with high $k$ and $\La_2$ values, namely $k\gtrsim 7.42$ and $\La_2\gtrsim 39.77$ TeV for the LNR scenario, $k\gtrsim 6.87 (6.79)$ and $\La_2\gtrsim 34.21$ TeV for the LIR (LMR2) scenario, and $k\gtrsim 6.73$ and $\La_2\gtrsim 36.16$ TeV for the LMR1 scenario. Besides, the panels of LIR and LMR2 scenarios are almost similar. This result also occurred in Fig. \ref{correlation_s23_vs_z_Lamda2big}. Hence, we comment that the LIR and LMR2 scenarios give the same results, while the LNR and LMR1 scenarios do not change considerably. Therefore, in the following, we consider the model under only the LNR and LIR scenarios that satisfy the constraints from the lepton flavor violation process.

Now, we turn to the quark flavor phenomenologies. We randomly generate the free parameters $z,k$, $\La_2$, and $\theta$ similarly in the studies of leptonic flavor phenomenologies. Besides, the parameters $s_{12}^{d_R}$ and $\delta ^{d_R}$ are randomly extracted from ranges as
\be s_{12}^{d_R}\in [0,1], \hs \delta^{d_R}\in [0,2\pi],  \ee 
whereas the lepton mixing angles $\theta_{ij}^{e_R}$ are chosen in the LNR and LIR scenarios. 

\begin{figure}[h!]
	\begin{subfigure}{0.49\textwidth}
	\includegraphics[width=\textwidth]{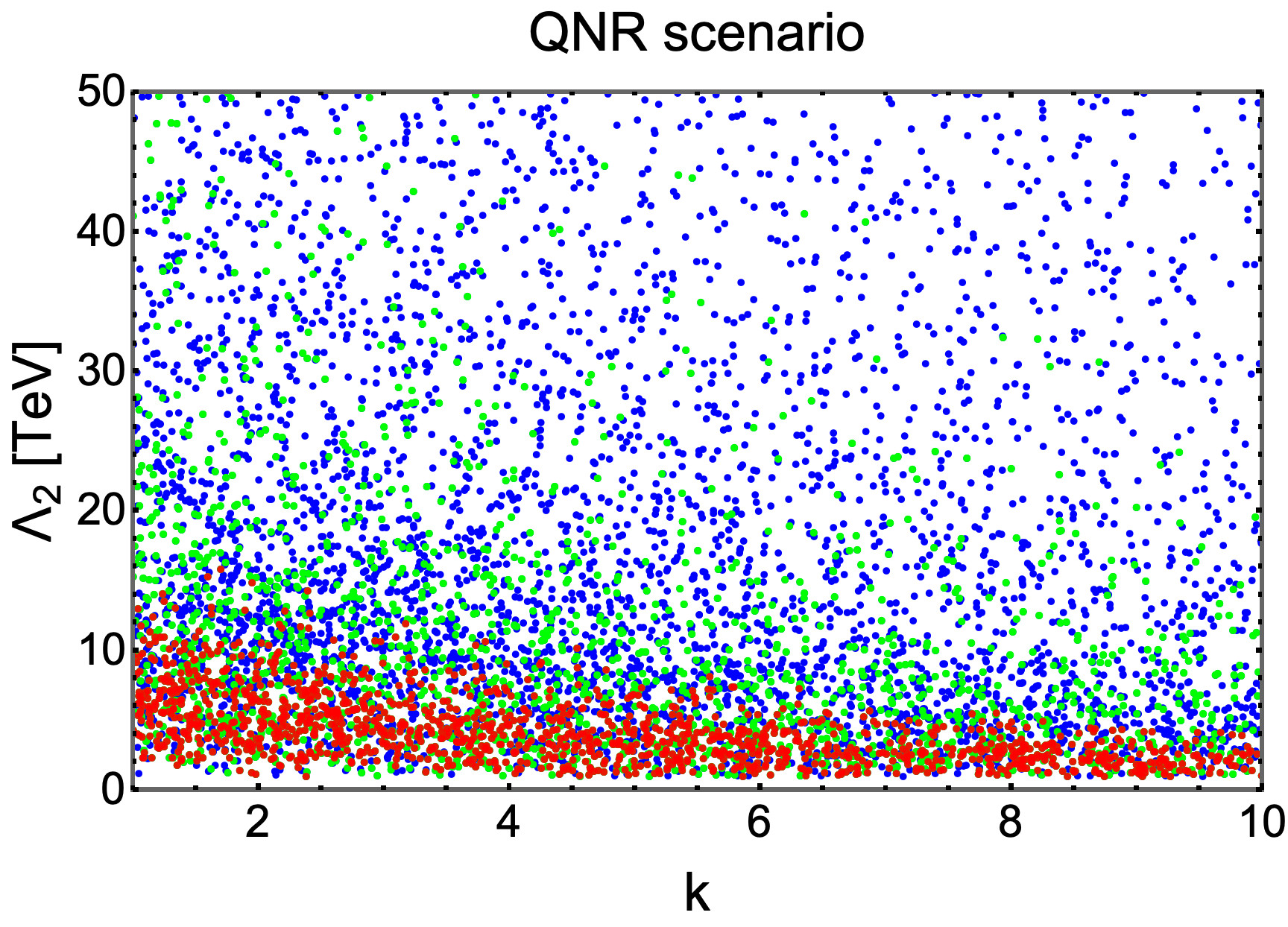}
	\end{subfigure}
	\hfill
	\begin{subfigure}{0.49\textwidth}
	\includegraphics[width=\textwidth]{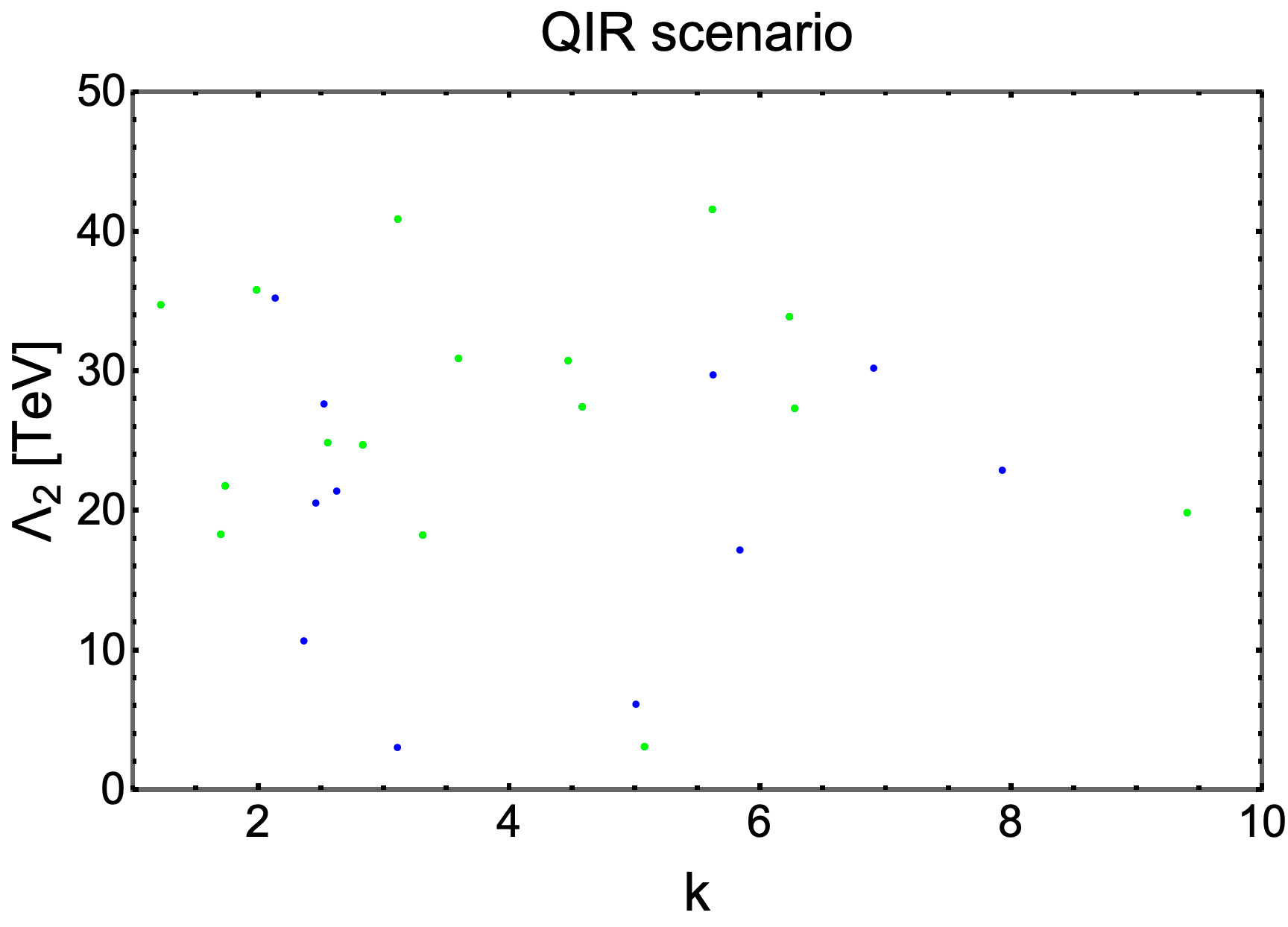}
	\end{subfigure}
	\medskip
	\begin{subfigure}{0.49\textwidth}
	\includegraphics[width=\textwidth]{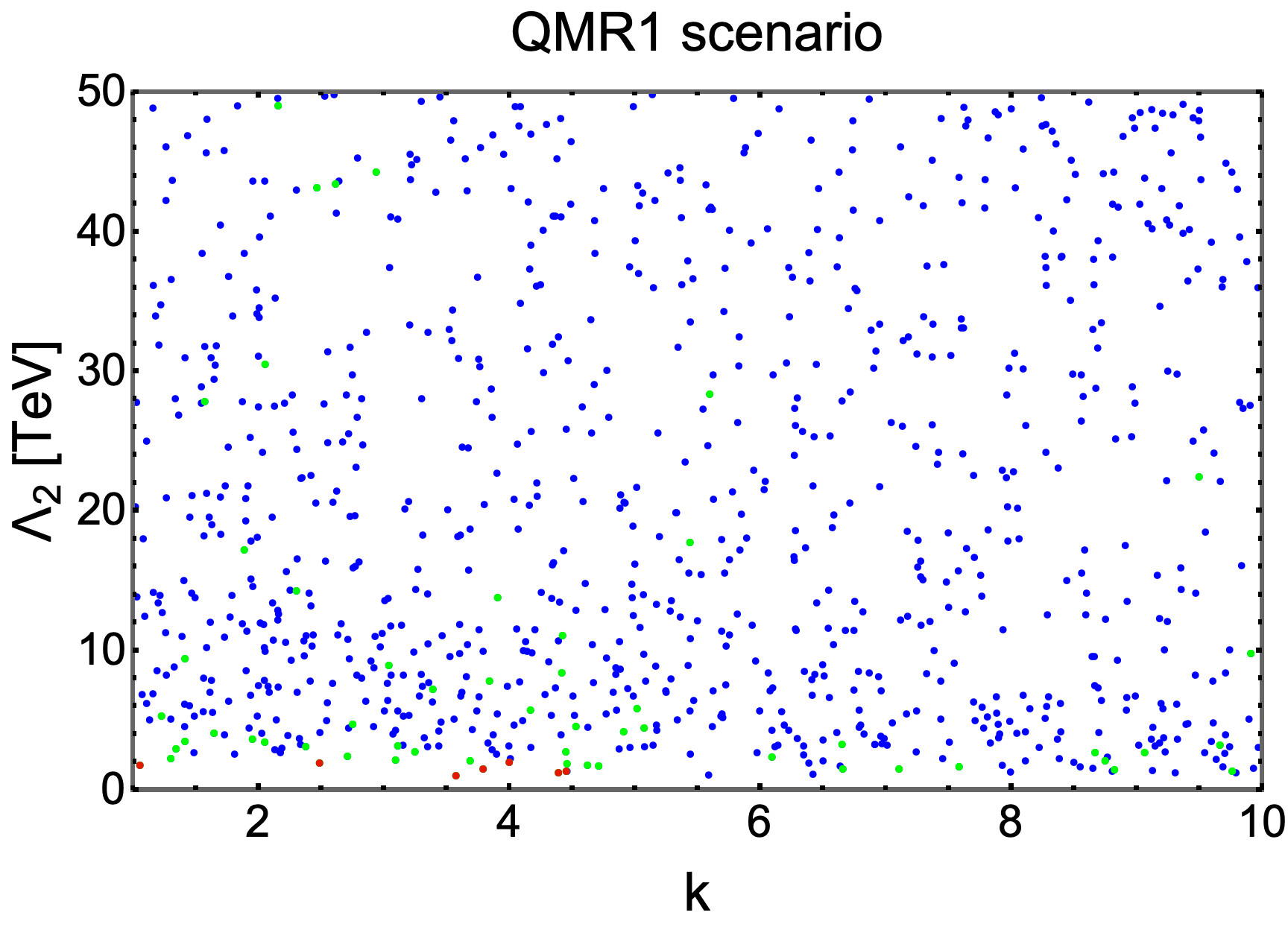}
	\end{subfigure}
	\hfill
	\begin{subfigure}{0.49\textwidth}
	\includegraphics[width=\textwidth]{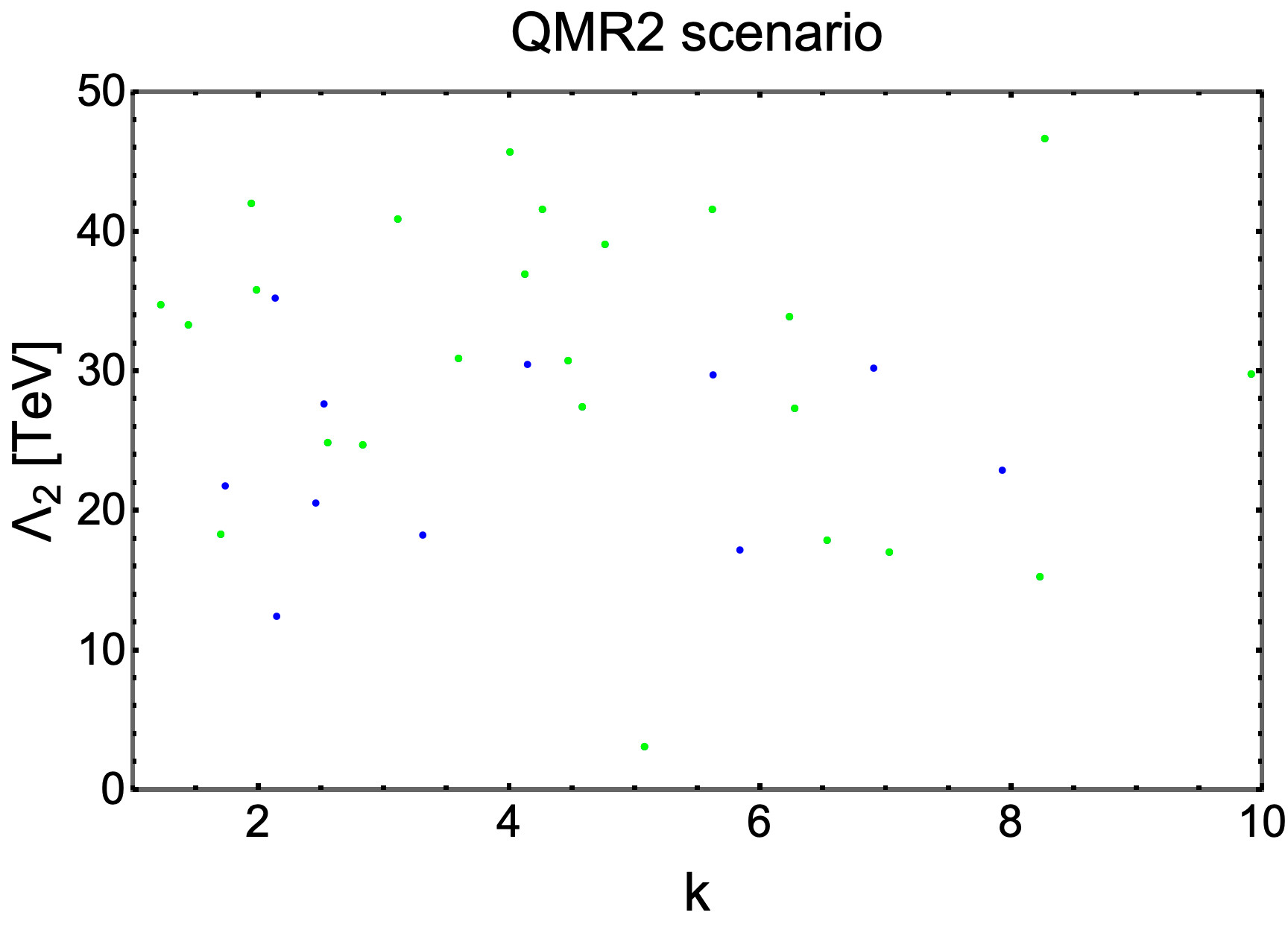}
	\end{subfigure}
	\caption{The correlations between the ratio $k=\La_1/\La_2$ with VEV $\Lambda_2$ for four relation scenarios of quark mixing angles with the LNR scenario of lepton mixing angles.}
	\label{La2_k_NHlepton}
\end{figure}

\begin{figure}[h!]
	\begin{subfigure}{0.49\textwidth}
	\includegraphics[width=\textwidth]{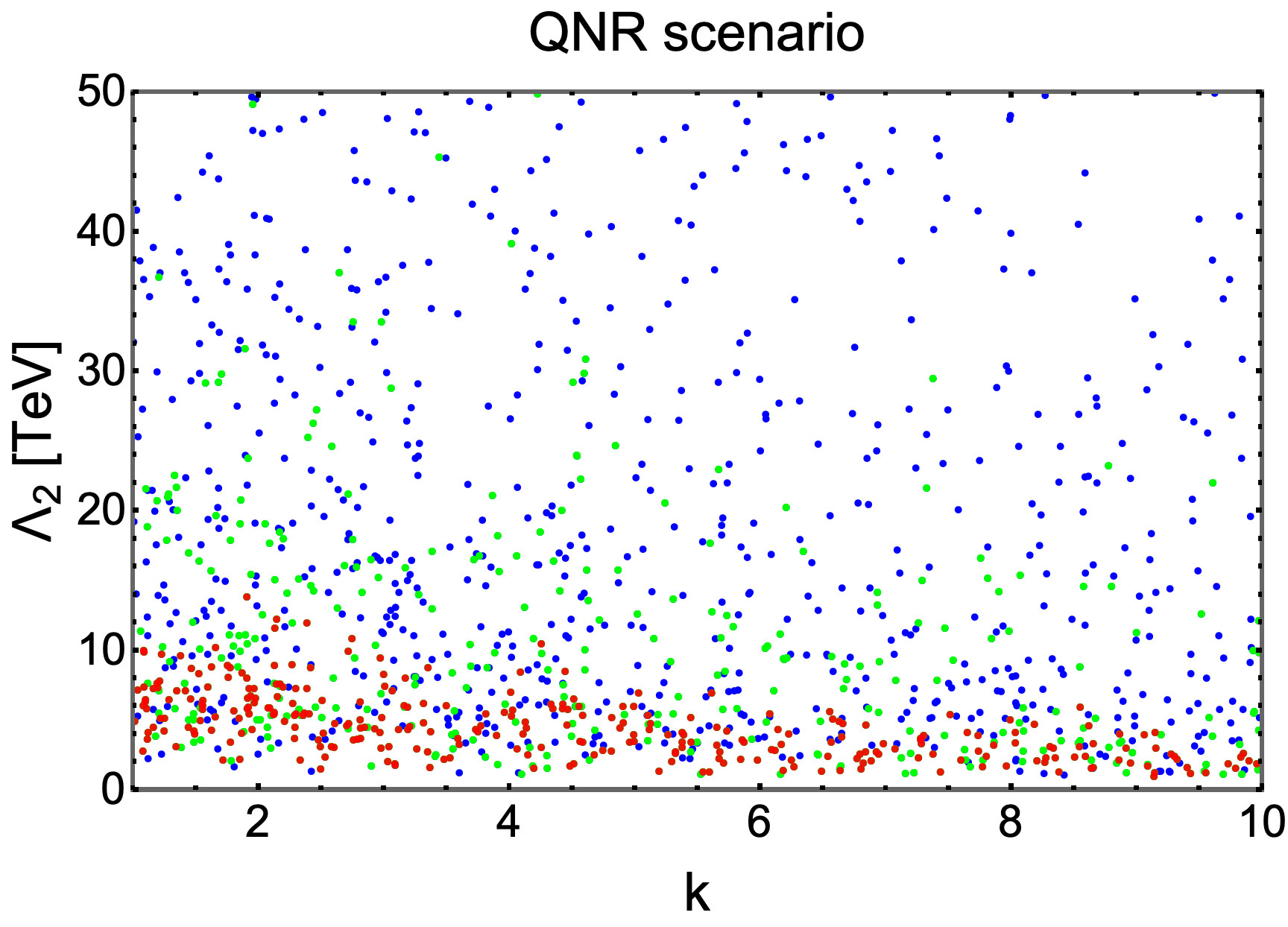}
	\end{subfigure}
	\hfill
	\begin{subfigure}{0.49\textwidth}
	\includegraphics[width=\textwidth]{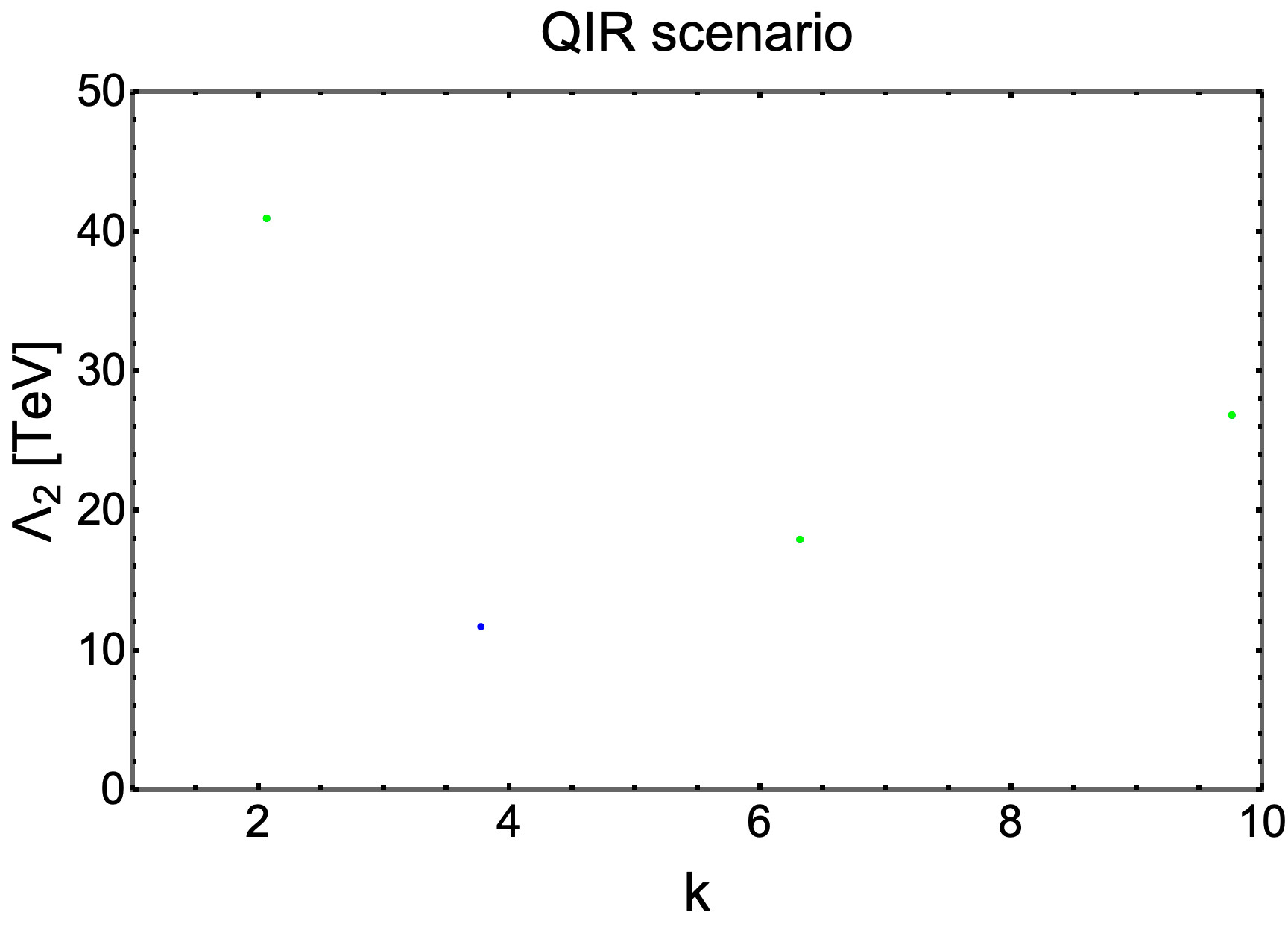}
	\end{subfigure}
	\medskip
	\begin{subfigure}{0.49\textwidth}
	\includegraphics[width=\textwidth]{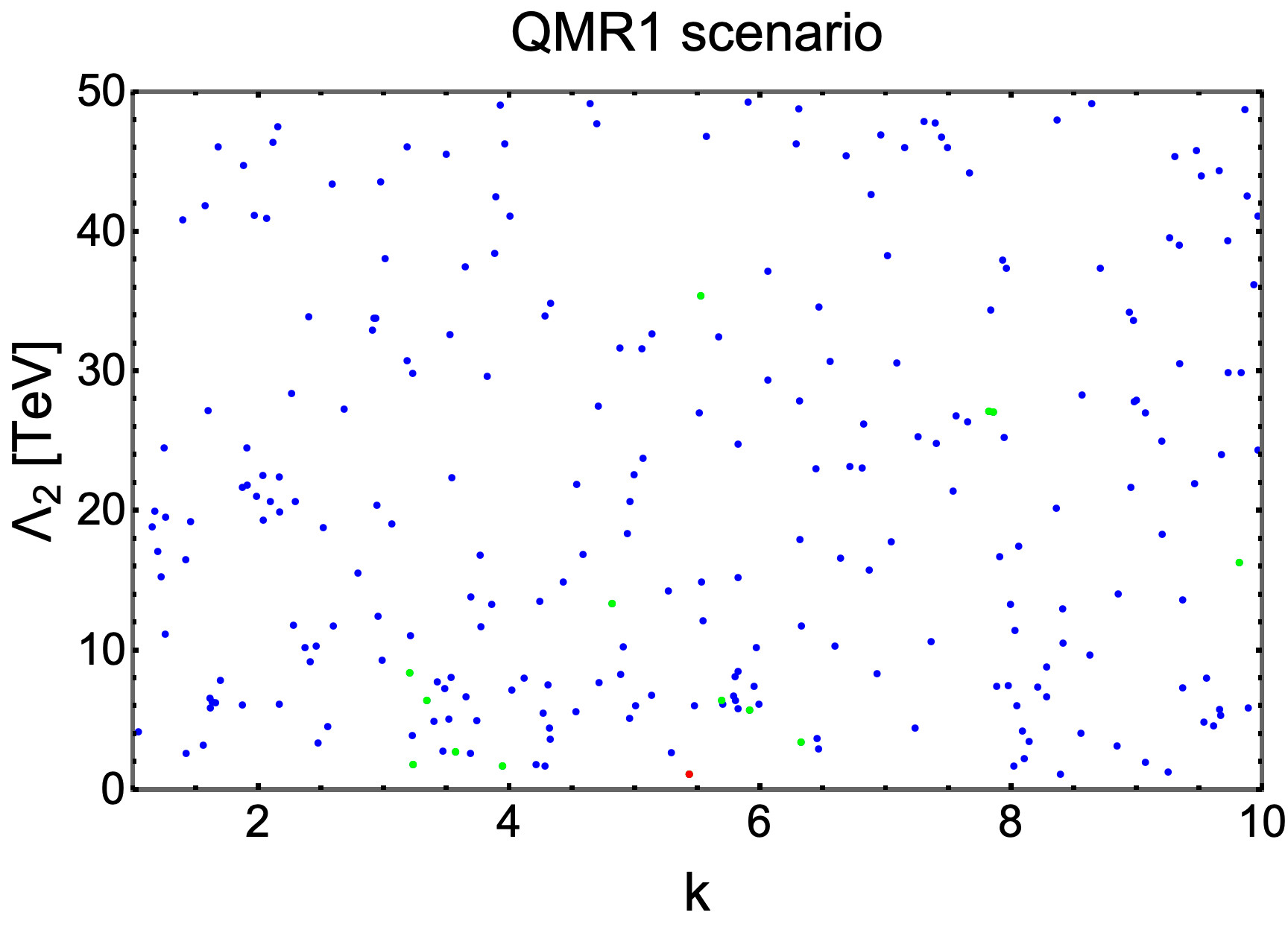}
	\end{subfigure}
	\hfill
	\begin{subfigure}{0.49\textwidth}
	\includegraphics[width=\textwidth]{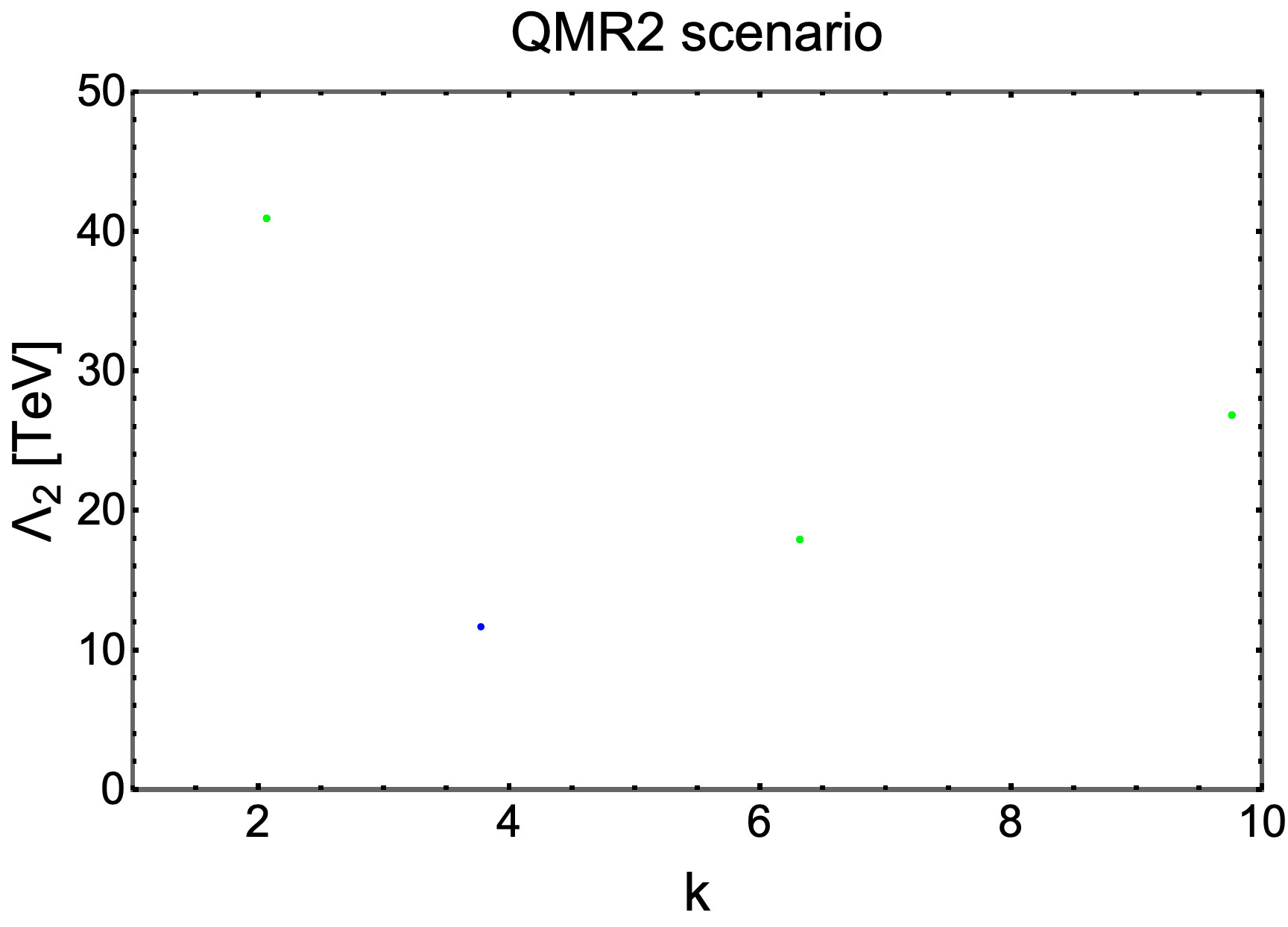}
	\end{subfigure}
	\caption{The correlations between the ratio $k=\La_1/\La_2$ with VEV $\Lambda_2$ for four relation scenarios of quark mixing angles with the LIR scenario of lepton mixing angles.}
	\label{La2_k_IHlepton}
\end{figure}

In Fig. \ref{La2_k_NHlepton}, we show the correlation points between the $k$ and $\La_2$ in four relation scenarios of quark mixing angles determined by Eqs. (\ref{quark_relation1})--(\ref{quark_relation2}), while lepton mixing angle is taken in the LNR scenario. All these points fulfill the constraints of $\Delta m_K$, BR$(B_s\to \mu^+\mu^-)$, and BR$(\bar{B}\to X_s\gamma)$ respectively expressed in Eqs. (\ref{constraint1}), (\ref{Bsmm_constraint}), and (\ref{bsga_constraint}). In addition, the red, green, and blue points satisfy the latest experimental limits of $\Delta m_{B_s}$ and $\Delta m_{B_d}$ within $1\sigma$, $1.25\sigma$, and $1.5\sigma$, respectively \cite{HFLAV:2022pwe}. From here, we comment that the blue points that are distributed in the regions with high $k$ and $\La_2$ values not only satisfy the present constraints but also the constraints from the lepton flavor violation processes (see Fig. \ref{correlation_k_vs_lamda2}). Such points appear in the QNR and QMR1 scenarios but do not appear in the QIR and QMR2 scenarios. A similar result is also shown in Fig. \ref{La2_k_IHlepton} plotted with the LIR scenario. These two figures imply that the model under QNR and LNR scenarios is preferred. Therefore, the following numerical studies will focus on these relation scenarios.

\begin{figure}[h!]
	\begin{subfigure}{0.49\textwidth}
	\includegraphics[width=\textwidth]{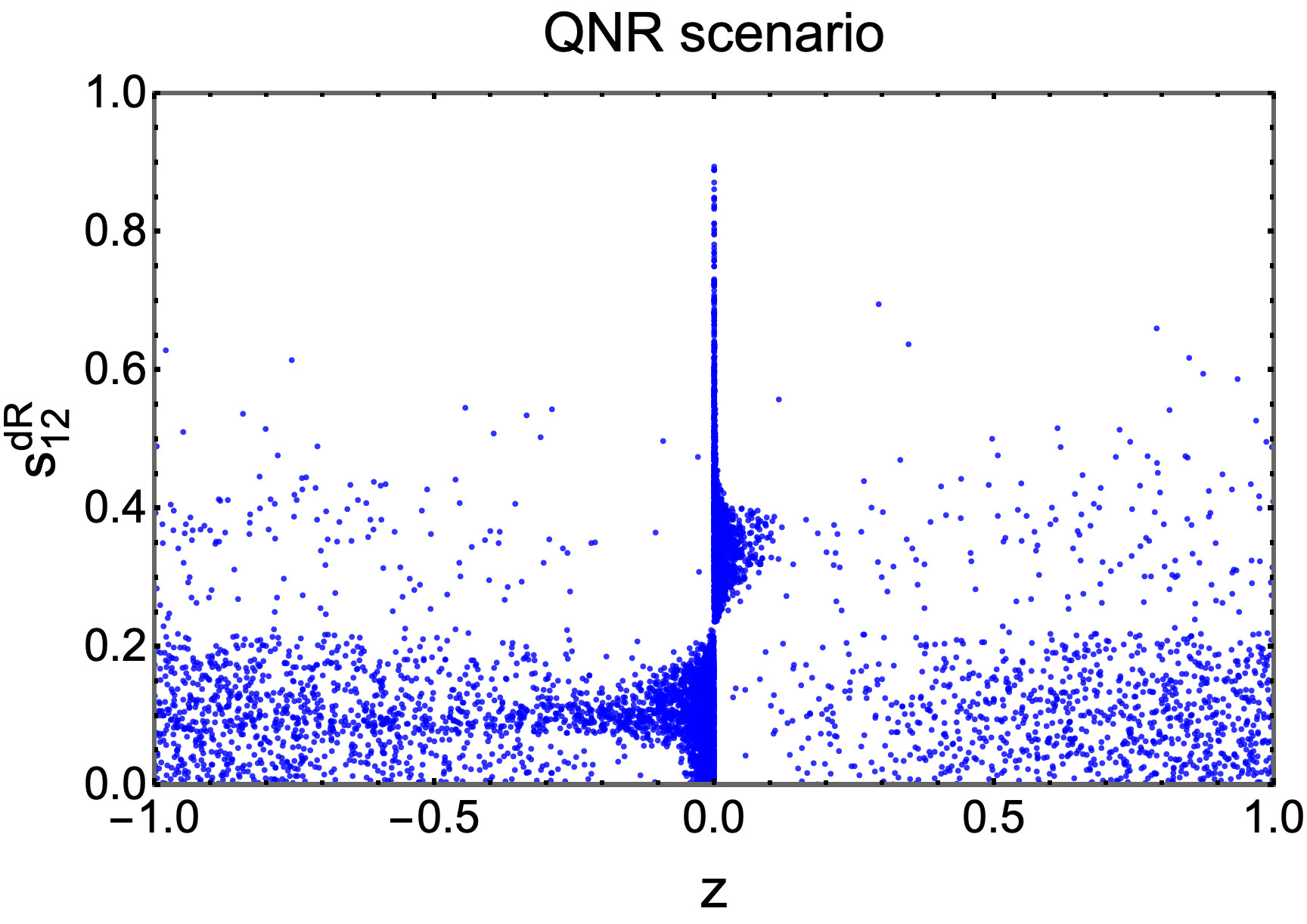}
	\end{subfigure}
	\hfill
	\begin{subfigure}{0.49\textwidth}
	\includegraphics[width=\textwidth]{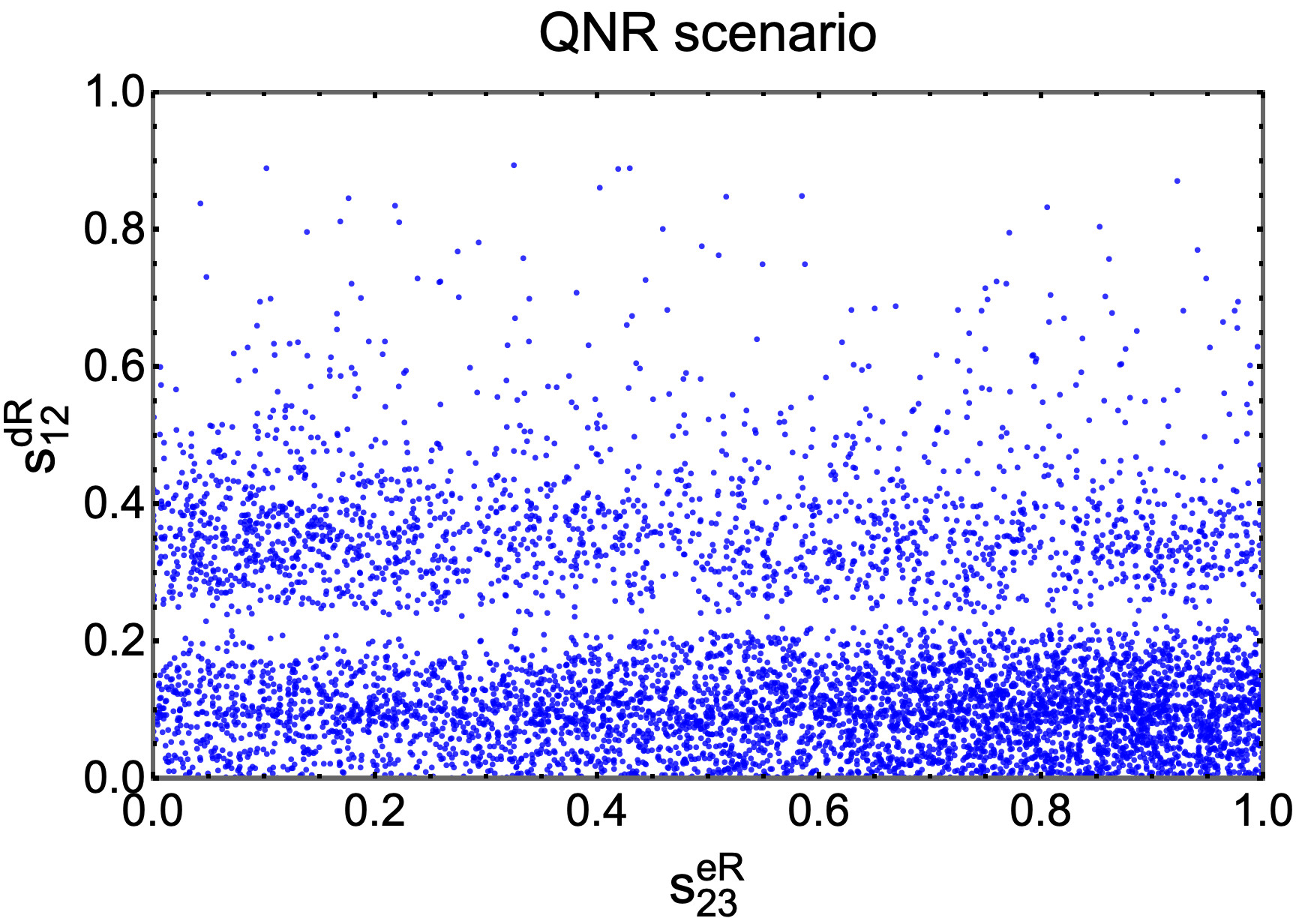}
	\end{subfigure}
	\begin{subfigure}{0.49\textwidth}
	\includegraphics[width=\textwidth]{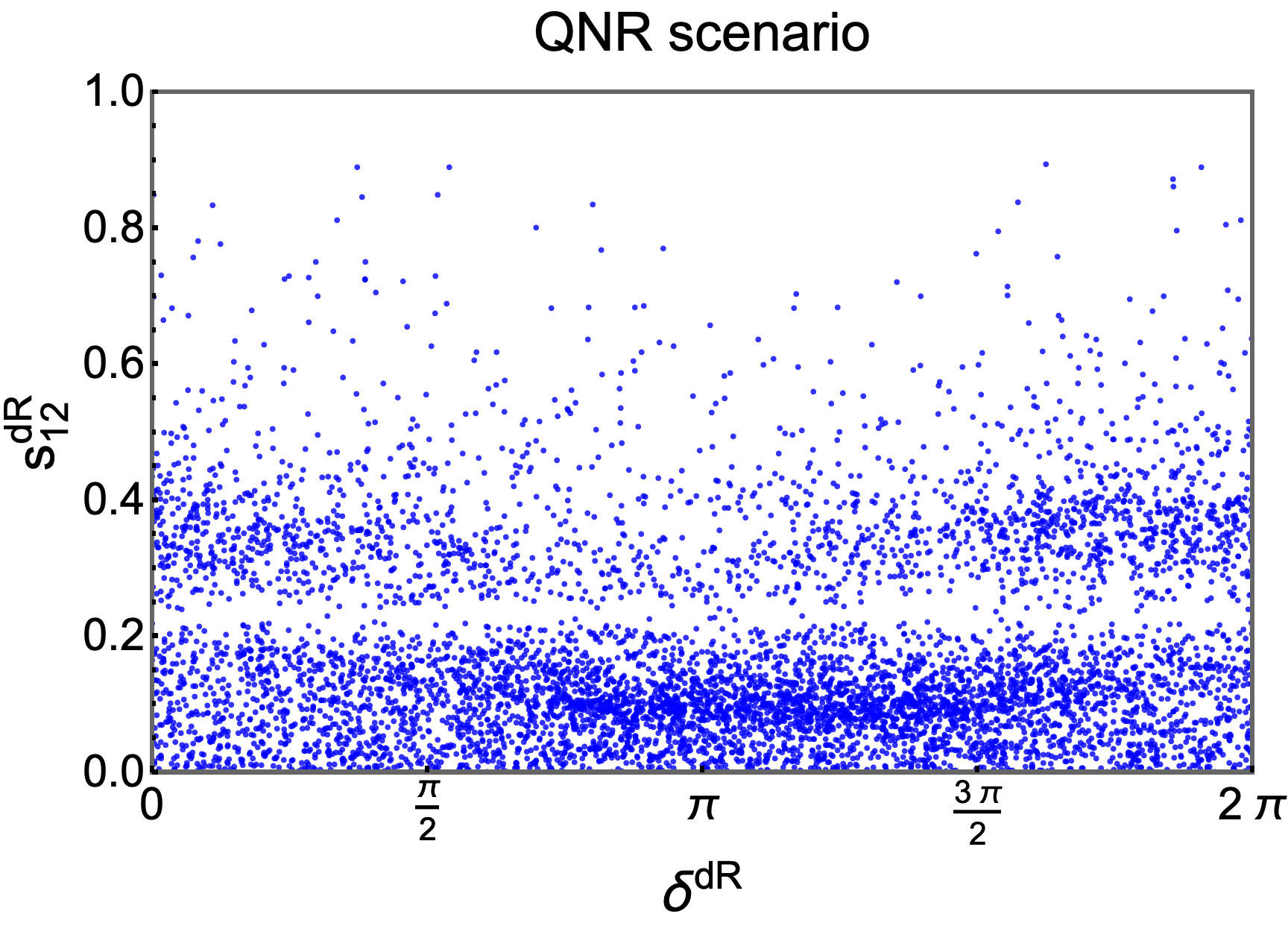}
	\end{subfigure}
	\caption{The correlation between parameter pairs, $s_{12}^{d_R}-z$, $s_{12}^{d_R}-s_{23}^{e_R}$, and $s_{12}^{d_R}-\delta^{d_R}$.}
	\label{correlations_quark}
	\end{figure}   

Now, we focus particularly on three correlations, $s_{12}^{d_R}-z$, $s_{12}^{d_R}-s_{23}^{e_R}$, and $s_{12}^{d_R}-\delta^{d_R}$, which are respectively shown in three panels of Fig. \ref{correlations_quark}. For the upper left panel, we see that the mixing angle $s_{12}^{d_R}$ is limited in a range $0\lesssim s_{12}^{d_R}\lesssim 0.2$ for $|z|\in[0.1,1]$. However, when $|z|$ decreases to less than 0.1, there are many points distributed in a wider range of $s_{12}^{d_R}$, i.e., $0\lesssim s_{12}^{d_R}\lesssim 0.4$. This can be interpreted by the following reason: when $|z|\sim 1$, the electroweak term proportional $v$ in $m_{Z'}$ will be much smaller than one relevant to the new physics contribution and can be ignored. As a result, $m_{Z'}$ now depends linearly on $z$, and then, several of the WCs are free of $z$ since it is canceled between numerator and denominator, such as the WCs in Eqs. (\ref{tree1})-(\ref{tree3}) and (\ref{loop1})-(\ref{pen2}). Therefore, the quark flavor observable is approximately independent of $z$. On the other hand, when $|z|$ is sufficiently small, the electroweak term significantly affects quark flavor processes. We would like to note that the upper limit of $s_{12}^{d_R}\sim 0.4$ is larger than the center value of $s_{12}^{\text{CKM}}=\la$ given in the Table \ref{input-par}. The upper right panel demonstrates that the mixing parameter $s_{12}^{d_R}$ of $V_{d_R}$ is independent of mixing parameter $s_{23}^{e_R}$ of $V_{e_R}$. The range of $s_{23}^{e_R}$ is not constrained tightly as the $s_{12}^{d_R}$ and whole range of $s_{23}^{e_R}$ satisfying the mentioned constraints. The correlation between $s_{12}^{d_R}$ with CP violation phase $\delta^{d_R}$ is displayed in the bottom panel. Here, the total range of $\delta^{d_R}$ fulfills the constraints from Eqs. (\ref{constraint2}) and (\ref{constraint1})--(\ref{bsga_constraint}). This also implies that the effect of $\delta^{d_R}$ on the quark flavor observables is negligible and can be ignored.  
        
\begin{figure}[h!]
\begin{subfigure}{0.49\textwidth}
	\includegraphics[width=\textwidth]{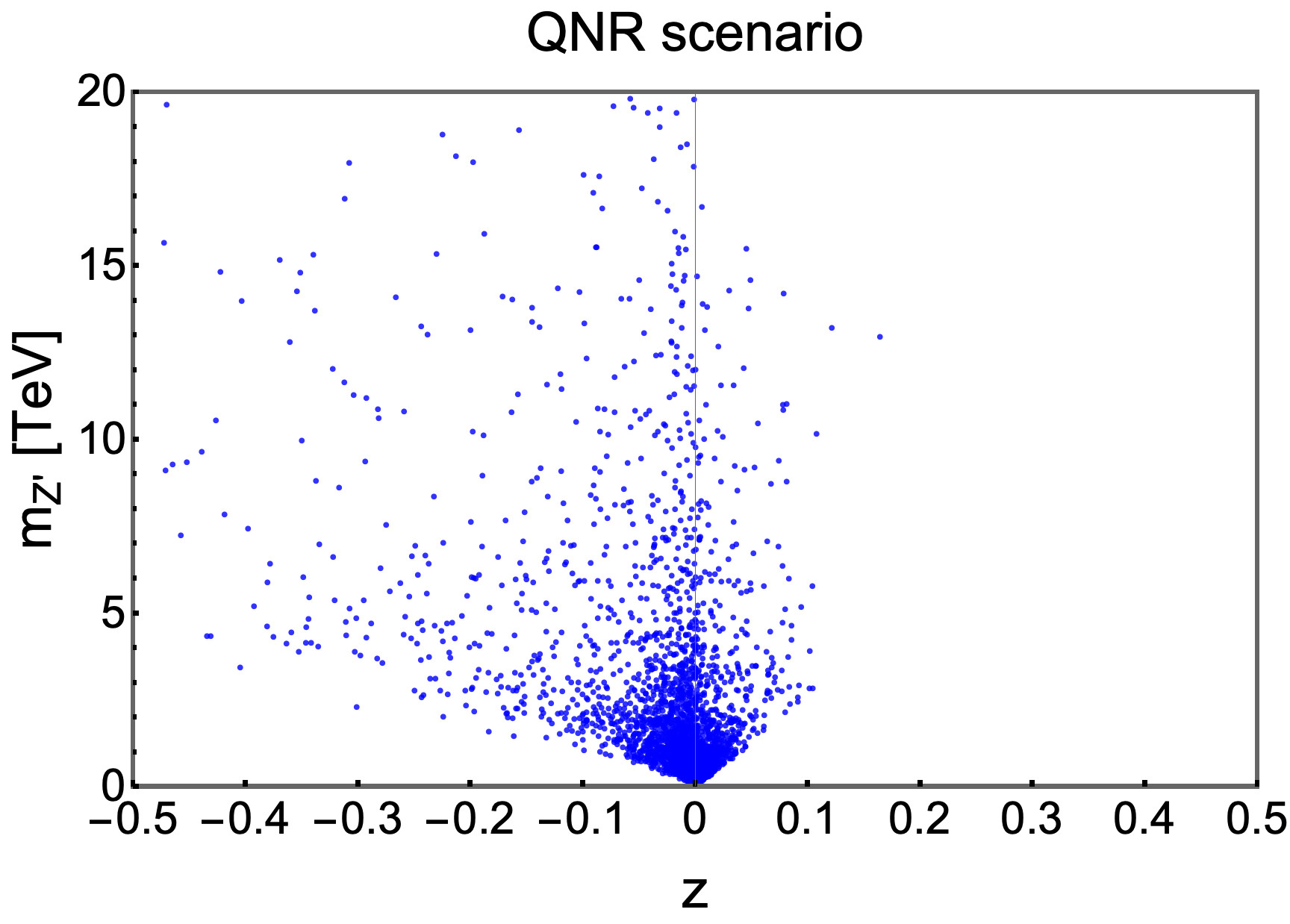}
	\end{subfigure}
	\hfill
	\begin{subfigure}{0.477\textwidth}
	\includegraphics[width=\textwidth]{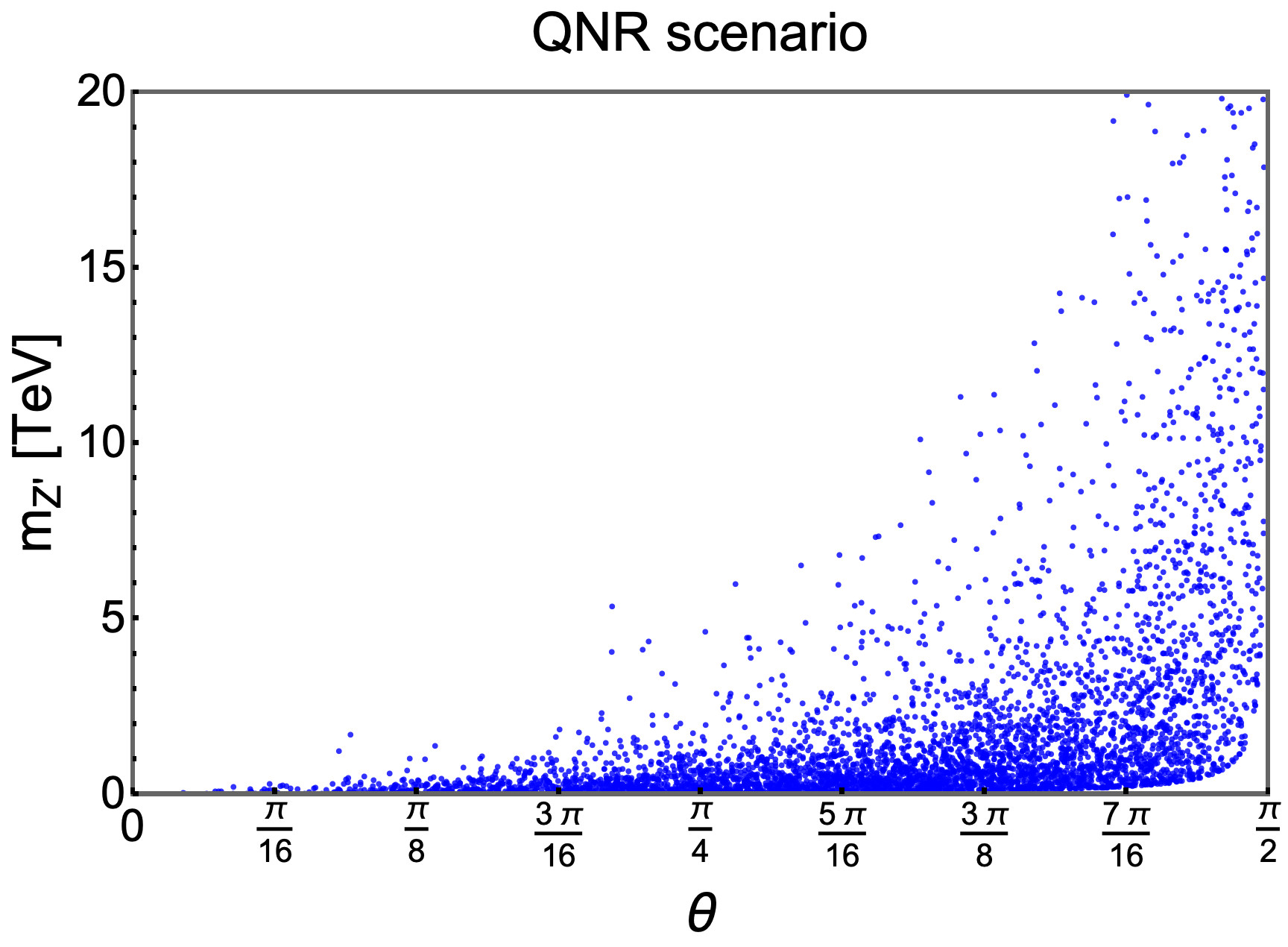}
	\end{subfigure}
	\caption{\label{correlation_z_mZp}The correlation between the mass of new gauge boson $m_{Z'}$ with the charge parameter $z$ (left panel), and with the mixing angle $\theta$ (right panel).}
\end{figure}

In Fig. \ref{correlation_z_mZp}, we show two correlations, $m_{Z'}-z$ (left panel) and $m_{Z'}-\theta$ (right panel). From here, we comment that the viable range of $z$ is $-0.5\lesssim z\lesssim 0.1$, whereas the whole range of $\theta$ is available. However, if $5\pi/16\lesssim\theta\lesssim\pi/2$ then $m_{Z'}\gtrsim 6$ TeV. This is consistent with the collider bounds (see below). Comparing with the results in Fig. \ref{correlation_s23_vs_z_Lamda2big}, Fig. \ref{correlation_k_vs_lamda2}, and Fig. \ref{correlations_quark}, we obtain the viable ranges for several parameters as 
\bea
&&0.1\gtrsim z\gtrsim 2.41\times 10^{-4}, \hs k\gtrsim 7.42, \hs \La_2\gtrsim 39.77\text{ TeV},\\
&&0.4 \gtrsim s_{12}^{d_R}\gtrsim 0.2, \hs \pi/2\gtrsim\theta\gtrsim 5\pi/16, \hs  1\gtrsim s_{23}^{e_R}\gtrsim 0,
\eea
while the effect of $\delta^{d_R}$ is insignificantly and it can be chosen arbitrarily.
  
\begin{figure}[h!]
	\centering
	\begin{tabular}{c} 
	\includegraphics[width=10.0cm]{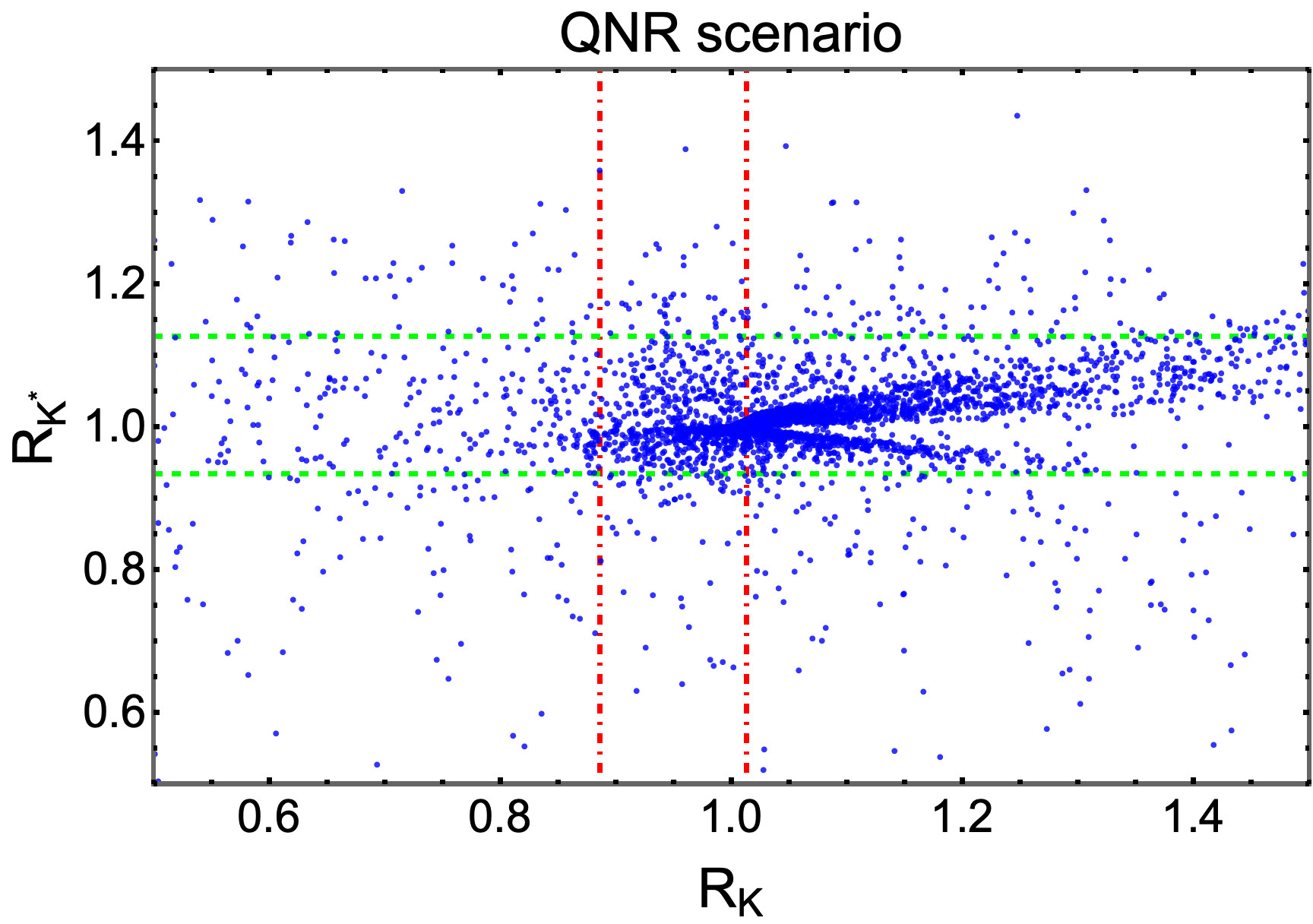}
	\end{tabular}
	\caption{\label{correlationRK_RKs}The correlation between the predicted $R_K$ and $R_{K^*}$. The dot-dashed red and green lines are correspondingly the current experimental limits of $R_{K}^{\text{exp}}$ and $R_{K^*}^{\text{exp}}$ \cite{LHCb:2021awg}. }
\end{figure}

Last but not least, we consider the LFUV ratios $R_{K}$ and $R_{K^*}$; their results are shown in Fig. \ref{correlationRK_RKs}. The blue points are plotted in which parameters satisfy all constraints given in Eqs. (\ref{constraint2}) and (\ref{constraint1})--(\ref{bsga_constraint}). We realize that the figure shows points concurrently meeting the measured results of both $R_{K}$ and $R_{K^*}$ given in the last column in Table \ref{SM_exp data}. Therefore, the model with the QNR scenario can explain several of quark flavor observables, including the meson oscillations $\Delta m_{K,B_s,B_d}$, BR$(\bar{B}\to X_s\gamma)$, BR$(B_s\to\mu^+\mu^-)$, and $R_{K^{(*)}}$. 

\section{\label{collider} Collider bounds}
The $Z'$ gauge boson in our model directly interacts with both ordinary quarks ($q$) and charged leptons ($l$), so it can be produced at the large electron-positron (LEP) experiments even the large hadron collider (LHC). In this section, we take the current negative search results reported by these experiments to impose a lower bound on the mass of $Z'$ boson \cite{Carena:2004xs,ALEPH:2005ab,ALEPH:2013dgf,ATLAS:2019erb,ATLAS:2019fgd,CMS:2021ctt}.
\subsection{LEP}
One of the processes searched at the LEP experiments is $e^+e^-\to f\bar{f}$, which generates a pair of ordinary charged leptons ($f=e,\mu,\tau$) through the exchange of $Z'$ boson. This process can be described by the following effective Lagrangian,
\be \mathcal{L}_{\text{eff}}=\frac{1}{1+\delta_{ef}}\frac{g^2}{c^2_Wm^2_{Z'}}\sum_{i,j=L,R}C_i^{Z'}(e)C_j^{Z'}(f)(\bar{e}\gamma_\mu P_ie)(\bar{f}\gamma^\mu P_jf), \ee
where $\delta_{ef}=1(0)$ for $f=e(\neq e)$, and the chiral gauge couplings are given by $C_{L,R}^{Z'}(f)=\fr 1 2 [g_V^{Z_2}(f)\pm g_A^{Z_2}(f)]$. LEP-II has probed all such effective contact interactions, and no significant evidence has been found for the existence of a $Z'$ boson. LEP-II also provided the lower limits of the scale of the contact interactions, $\La$, for all possible chiral structures and for various combinations of fermions \cite{ALEPH:2013dgf}. Consequently, the mass of $Z'$ boson is bounded by
\be m^2_{Z'}\gtrsim \frac{g^2}{4\pi c^2_W}|C_i^{Z'}(e)C_j^{Z'}(f)|[\La_{ij}^\pm(f)]^2, \ee
where $\La^+$ for $C_i^{Z'}(e)C_j^{Z'}(f)>0$ and $\La^-$ for $C_i^{Z'}(e)C_j^{Z'}(f)<0$. 

The strongest constraint for our model comes from the $e^+e^-\to \mu^+\mu^-,\tau^+\tau^-$ channel with $\La^+_{VV}=24.6$ TeV. It results in $m_{Z'}\gtrsim 5.9$ TeV for $z\simeq 0.05$ and $\theta\simeq 3\pi/8$.

\subsection{LHC}
At the LHC experiment, the $Z'$ neutral gauge boson can be resonantly produced in the new physics processes $pp\to Z'\to f\bar{f}$ for $f=q,l$. Additionally, the most significant decay channel of $Z'$ is given by $Z'\to l\bar{l}$ because of well-understood backgrounds \cite{ATLAS:2019erb,CMS:2021ctt} and that it signifies a boson $Z'$ having both couplings to lepton and quark like ours. The cross-section for the relevant process, in the narrow width approximation, takes the form \cite{Accomando:2010fz}
\be \sigma(pp\to Z'\to l\bar{l})\simeq \fr 1 3 \sum_q\frac{dL_{q\bar{q}}}{dm_{Z'}^2}\hat{\sigma}(q\bar{q}\to Z')\text{BR}(Z'\to l\bar{l}), \ee
where the parton luminosities $\frac{dL_{q\bar{q}}}{dm_{Z'}^2}$ can be found in Ref. \cite{Martin:2009iq}, while the peak cross-section is given by 
\be \hat{\sigma}(q\bar{q}\to Z')\simeq \frac{\pi g^2}{12 c_W^2}[(g_V^{Z_2}(q))^2+(g_A^{Z_2}(q))^2]. \ee
The branching ratio of $Z'$ decaying into the lepton pairs is $\text{BR}(Z'\to l\bar{l})=\Gamma(Z'\to l\bar{l})/\Gamma_{Z'}$, where the partial and total decay widths are respectively given by
\bea \Gamma(Z'\to l\bar{l})&\simeq&\frac{g^2m_{Z'}}{48\pi c_W^2}[(g_V^{Z_2}(l))^2+(g_A^{Z_2}(l))^2], \\
\Gamma_{Z'}&\simeq&\frac{g^2m_{Z'}}{48\pi c_W^2}\left\{\sum_f N_C(f)[(g_V^{Z_2}(f))^2+(g_A^{Z_2}(f))^2]+\frac{(C_{1L}^{Z_2})^2}{2}+(C_{2L}^{Z_2})^2\right\}\crn
&&+\frac{g^2m_{Z'}}{96\pi c_W^2}(C_R^{Z_2})^2\sum_{i=1}^3\left(1-\frac{4M_i^2}{m_{Z'}^2}\right)^{3/2}\Theta\left(\frac{m_{Z'}}{2}-M_i\right),\eea
assuming that $Z'$ is lighter than new Higgs bosons $\mathcal{H}_{1,2}$ and $\mathcal{A}$. Here, $f$ denotes the SM charged fermions, $N_C(f)$ is the color number of the fermion $f$, and $\Theta$ is the step function.

\begin{figure}[h!]
	\centering
	\begin{tabular}{c}
	\includegraphics[width=10.0cm]{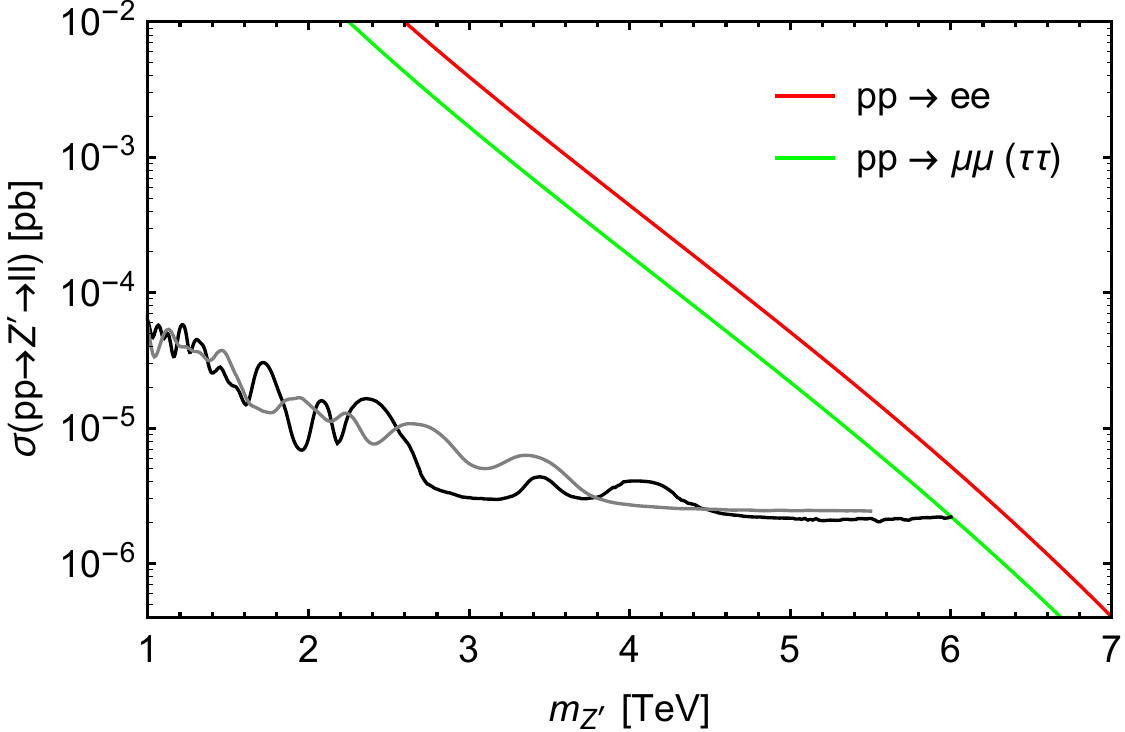}
	\end{tabular}
	\caption{\label{LHC}Dilepton production cross-section as a function of the new gauge boson mass $m_{Z'}$ for $z=0.05$ and $\theta=3\pi/8$. The black (gray) curve shows the upper bound on the cross-section obtained by the ATLAS 2019 results for $\Gamma/m=3\%$  \cite{ATLAS:2019erb} (the CMS 2019 results for $\Gamma/m=0.6\%$ \cite{CMS:2021ctt}).}
\end{figure}

Setting center-of-mass energy of $\sqrt{s}=13$ TeV and assuming $M_{1,2,3}=m_{Z'}/3$, in Fig. \ref{LHC}, we plot the cross section for the relevant processes as a function of the $Z'$ boson mass, given that $z=0.05$ and $\theta=3\pi/8$. Here, we also include the upper limits on the cross-section of these processes reported by ATLAS \cite{ATLAS:2019erb} and CMS \cite{CMS:2021ctt} experiments. We obtain a lower bound on the $Z'$ boson mass of $6$ TeV under the $\mu\mu~(\tau\tau)$ channel, while the $ee$ channel even implies a more strict constraint. Significantly, these signal strengths are separated, which can be used to approve or rule out the model under consideration.

We would like to note that the dijet signals also can provide a lower bound for the $Z'$ boson mass \cite{ATLAS:2019fgd}. However, in the present model, the coupling strengths between $Z'$ and the charged leptons are approximately equal to those of $Z'$ with the quarks, while the current bound on dijet signals is less sensitive than the lepton one, so the lower bounds implied by the dijet search are quite smaller than those result from the dilepton. In other words, in the present model, the dijet bounds for the $Z'$ boson mass are not significant.

\section{\label{conclusion}Conclusion}
In this work, we have proposed a model that is based on the gauge symmetry $SU(3)_C\otimes SU(2)_L\otimes U(1)_X\otimes U(1)_N$. This model is general and flavor-dependent but constructed minimally. The new charges $X$ and $N$ of the light fermions differ from those of the remaining fermions but determine the hypercharge $Y$ as $Y=X+N$, which conforms to the observables. We have shown that our model can provide a possible solution to several puzzles of the SM, including the observed number of fermion generations, the neutrino mass generation mechanism, and the flavor anomalies in both the quark and lepton sectors. The new physics effects at the LEP and LHC experiments have also been examined.  

With the relevant assumptions, the model leaves only six free parameters, including the charge parameter $z$, the new physics scale $\La_2$, the ratio $k=\La_1/\La_2$, the mixing angles $\theta_{12}^{d_R}$, $\theta_{23}^{e_R}$ and $\theta$. We have identified the allowed parameter space of the model, which is consistent with the experimental constraints, namely $0.1\gtrsim z\gtrsim 2.41\times 10^{-4}$, $k\gtrsim 7.42$, $\La_2\gtrsim 39.77 \text{ TeV}$, $0.4\gtrsim \sin\theta_{12}^{d_R}\gtrsim 0.2$, $1\gtrsim\sin\theta_{23}^{e_R}\gtrsim 0$, and $\pi/2\gtrsim\theta\gtrsim 5\pi/16$.

\section*{\label{acknowledgement}Acknowledgement}
N.T. Duy was funded by the Postdoctoral Scholarship Programme of Vingroup Innovation Foundation (VINIF), code VINIF.2023.STS.65. 

\bibliographystyle{JHEP}

\bibliography{combine}

\end{document}